\documentclass[preprint, 12p]{elsarticle}

\usepackage[OT1]{fontenc}
\usepackage{hyperref}
\usepackage{amsmath,amsthm,amssymb}     

\usepackage[english]{babel}
\usepackage[utf8]{inputenc}

\usepackage[caption=false]{subfig}
\usepackage[export]{adjustbox}
\usepackage{graphicx}
\usepackage{tabularx}

\usepackage{booktabs}

\usepackage[labelfont=bf]{caption}
\usepackage{color}

\newcommand{\cfuture}[1]{#1}

\newdefinition{definition}[theorem]{Definition}

\begin{document}
	
\begin{frontmatter}
\title{Community detection and Social Network analysis based on the Italian wars of the 15th century}

\author[pamplona]{J. Fumanal-Idocin}
\ead{javier.fumanal@unavarra.es}

\author[coruna]{A.Alonso-Betanzos}
\ead{amparo.alonso.betanzos@udc.es}

\author[granada]{O. Cordón}
\ead{ocordon@decsai.ugr.es}

\author[pamplona,arabia]{H. Bustince}
\ead{bustince@unavarra.es}

\author[bratislava]{M. Min\'arov\'a}
\ead{maria.minarova@stuba.sk}

\address[pamplona]{Public University of Navarra and Institute of Smart Cities, Campus Arrosadia s/n, 31006 Pamplona, Spain}
\address[arabia]{Faculty of Computing and Information Technology, King Abdulaziz University, Jeddah, Saudi Arabia}
\address[bratislava]{Slovak University of Technology in Bratislava,Radlinsk\'eho 11, 81005 Bratislava, Slovakia}
\address[coruna]{CITIC. University of A Coruña, 15071 A Coruña. Spain}
\address[granada]{Andalusian Research Institute DaSCI, "Data Science and Computational Intelligence", University of Granada, 18071 Granada, Spain}

\begin{abstract}
In \cfuture{this contribution} we study social network modelling \cfuture{by} using human interaction as a basis. To do so, we propose a new set of functions, affinities, designed to capture the nature of the local interactions among each pair of actors in a network. \cfuture{By} using these functions, we develop a new community detection algorithm, the Borgia Clustering, where communities naturally arise from the multi-agent interaction in the network. We also discuss the effects of size and scale for communities regarding this case, \cfuture{as well as how} we cope with the additional complexity present when big communities arise. Finally, we compare our community detection solution with other representative algorithms, finding favourable results.
\end{abstract}


\end{frontmatter}
\section{Introduction} \label{sec:introduccion}

Network analysis has become an important tool to study systems composed of interacting agents, such as proteins or human societies\cite{scott1988social, wasserman1994social, borgatti2009network, horvath2011weighted, BENITEZANDRADES2020154}. One of the key ideas in social sciences \cfuture{is the one that} we, human beings, are embedded by our own social nature in a complex web of social relations and interactions. Traditionally, this hierarchy that we have formed has been modelled as a network, where each person is represented as a node that is connected to others according to some criteria. Social networks \cfuture{analysis stands} as an appropriate tool to understand many characteristics of the human behaviour, as it seems that many of us are deeply affected by the social structure in which we take part \cite{fischer1982dwell}. Adjacency matrices are the most common form of network representation \cite{wellman1988social}. However, if there is more data available to construct the network, a more complex model can be used \cite{stopczynski2014measuring}. Depending on the context, different models have been developed to mimic human behaviour: in an educational organization \cite{sanchezmar2015} and when dealing with conversations \cite{2019cascades}.

There are many problems related to social networks, such as information diffusion \cite{lobel2015information}, social circles detection \cite{McAuley2014DiscoveringSC}, coalition formation \cite{SLESS2018217}, recommendation systems \cite{YU2018312} or user behaviour prediction \cite{LUO20191023}. Social networks can also be seen as multi-agent systems, and it is possible to study their emergent properties \cite{DELGADO2002171}.  One key step to analyse a network is to identify its community structure: the groups of nodes that can be identified as a functional sub-partition of the graph\cite{Girvan2002Jun, zhou2003distance} e.g a group of friends or a protein complex\cite{palla2005uncovering}. Communities are important because we can infere significant knowledge from a node or a set of nodes if we know whether or not they share the same community and what kind of community it is.  One classical method \cfuture{to develop} community detection is hierarchical clustering \cite{girvan2002community}, although there are many algorithms \cfuture{performing} community detection in a social network that improve the results obtained by this method. Authors in \cite{pizzuti2008ga} use a genetic algorithm to identify densely connected groups of nodes. The algorithm in \cite{MA2020533} performs an initial community detection in the most important nodes in the network, and then labels the rest of them. 

\cfuture{Modularity is a measure that quantifies the quality of a graph partition into different modules. Networks with high modularity have a high number of edges between the nodes within modules, and a low number of them between nodes in different modules \cite{newman2006modularity}. There are many modularity-based methods to perform community detection.} The proposal in \cite{newman2004fast} uses the idea of modularity and it is much faster than previous algorithms.  Authors in \cite{Blondel_2008} propose a modularity optimization approach to work on large networks, with this being one of the most extensively used community detection methods.

In this paper, our goal is to solve some of the social network analysis problems that we have identified \cfuture{both in} actor interaction and community detection. First, current community detection algorithms \cfuture{do not take into account neither the different nature of the involved human beings when applied to social networks, nor the impact to its structure}. We have also noted how scale in social networks can alter its dynamics. Besides, when it comes to community detection, many of the existing algorithms have problems in densely connected graphs and the appropriate algorithms are often awfully time consuming \cite{danon2005comparing}.

To solve these problems, we studied how human-inspired algorithms can lead to a better understanding of the structure of a social network and its communities. In contrast to existing literature, which uses an adjacency matrix to model a network\cite{newman2004finding}, we propose a set of functions \cfuture{for the sake of better capturing} the relationships between each pair of actors \cfuture{in the social network}. We model an algorithm that forms communities in \cfuture{a} similar fashion to real social networks. \cfuture{Finally, we propose} a new representation space for these graphs, called ``affinities", that can be calculated from the original adjacency graph. These affinities model how strong the relationship between two nodes according to different criteria \cfuture{is}. Based on these functions, we have developed a new algorithm \cfuture{based on} the gravitational algorithm described in \cite{Wright}. \cfuture{We have called our new algorithm to perform community detection the Borgia Clustering.}


Using the affinity functions and the Borgia Clustering, we aim at having a better understanding of local interactions and how they can affect global dynamics. Regarding community detection, our target is to obtain a state-of-the-art algorithm to perform this task. \cfuture{Such algorithm would be able to generate a dendrogram faithfully reflecting} the evolution of the network in the clustering process and choose the right configuration inside it.

To test the quality of our proposals, we have studied how the new affinities affect the interaction between actors in some real-world datasets. We have also tested our community detection method in four different networks.

The rest of the paper is as follows. In section \ref{sec:pre} we explain the basics of the gravitational algorithm and graph theory and we explain what is a T-Norm and an Overlap.  In section \ref{sec:affinities} we explain the new representation space for social networks based on the traditional adjacency matrix used for graph representation. In section \ref{sec:core} we introduce the Borgia Clustering algorithm, the historical moment that inspired it, and how it did so. In section \ref{sec:scale} we discuss some issues that we found working \cfuture{with} communities of different sizes. In section \ref{sec:results} we test our algorithm on three real world datasets and in section \ref{sec:comparison} we compare the quality of our solution against other representative algorithms. Finally, in section \ref{sec:conclusions} we summarize the whole work and state some future guidelines.

\section{Preliminaries} \label{sec:pre}

In this section we will briefly explain some of the already-existing concepts related to some of the new proposals in this paper:

\begin{itemize}
	\item The gravitational clustering algorithm.
	\item Aggregation functions.
	\item Graph theory.
\end{itemize}

\subsection{Algorithm of Gravitational Clustering}

The algorithm of Gravitational Clustering \cite{Wright} employs the Newton gravitational law within the process of clustering. In this algorithm, each observation is a particle that attracts the others according to \cfuture{their distances and masses}. When two particles are closer than the collision distance, they are merged into a single one. Its mass is the sum of those particles and \cfuture{its position is their center of masses}. This process repeats until only one particle exists. This algorithm results in a dendrogram containing each particle fusion. \cfuture{Finally the most stable configuration (with the largest lifespan) is taken as the resulting one.} The scheme is as follows:

We suppose that we have $n$ particles $p_1,\dots,p_n$, with their positions  $s_1,\dots,s_n \in {R}^n$. We also have two parameters: $\epsilon$, which establishes the collision distance for two particles, and $\delta$ which determines the movement for the fastest particle in each iteration.
\begin{enumerate}
	
	\item Initially we assign a mass ($m_i$) 1 to each particle $p_i$.
	
	\item We fix real positive parameters  $\epsilon$ and $\delta$
	
	\begin{itemize}
		\item We utilize $\delta$ for determining the actual time step longitude\cfuture{,} $dt$. \cfuture{It is the time in which the fastest particle moves.}. 
		
		\item If in a moment two particles find themselves in a distance less than $\epsilon$ we unify them in one particle\cfuture{. The mass of the new resulting particle is the sum of both masses and its position is their centre of masses. Likewise in the case of three or more particles.}
	\end{itemize}
	
	\item Initial time is set to $t=0$.
	
	\item We repeat the following steps (i)-(iv) until a single particle remains.
	\begin{itemize}
		\item[(i)] In each time interval 
		$[t,t+dt]$, for each particle $p_i$ we compute its movement influencing function:
		
		\begin{equation}\label{eq:clasico}
		g(i,t,dt)=\frac12 G \sum _{j\neq i} \frac{m_i(t) m_j(t)}{m_i(t)}\frac {s_j(t)-s_i(t)}{\vert s_j(t) - s_i(t)\vert^3}dt^2
		\end{equation}
		where $G$ is a positive constant.
		\item[(ii)] 
		For each particle $i$, its new position is:
		\[
		s_i(t+dt)=s_i(t)+  g(i,t,dt)
		\]
		\item[(iii)] We increment $t$ to $t+dt$.
		\item[(iv)]  If two particles $i$ and $j$ are closer than $\epsilon$, they are fused as explained above.
	\end{itemize}
\end{enumerate}
After the algorithm ends, we have just one particle. The duration of the entire process is denoted by $T$. \cfuture{We measured the duration of each iteration, as well, with the aim of being able to detect afterwards the most stable configuration. The final result is the configuration that lasted the longest period of the simulated time.}

\subsection{Aggregation functions}
\begin{definition}
	Aggregation functions are used to combine the information of multiple numerical sources into a single one.
	A function $A:[0,1]^n \rightarrow [0,1]$ is said to be n-ary aggregation function is said to be a n-ary aggregation function if the following conditions hold \cfuture{\cite{beliakov2016practical}}:
	\begin{itemize}
		\item A is increasing in each argument: $\forall i \in \{1,\dots,n\}, i < y, A(x_1,\dots,i,\dots, x_n) \le A(x_1,\dots,y,\dots, x_n)$
		\item $A(0,\dots,0) = 0$
		\item $A(1,\dots,1) = 1$
	\end{itemize}
\end{definition}

\begin{definition}
	A bivariate aggregation function T:$[0,1]^2\rightarrow[0,1]$ is a t-norm if, $\forall x, y, z \in[0,1]$, it satisfies the following properties:
	\begin{itemize}
		\item Commutativity: $T(x,y)$=$T(y, x)$
		\item Associativity: $T(x, T(y, z))$ = $T(T(x, y),z)$.
		\item Boundary condition: $T(x,1) =x$
	\end{itemize}
\end{definition}

Two prototypical examples of T-norms are the minimum and the product.

\subsection{Graphs}

A graph $G$ is represented as $G(V, E)$ where $V$ is a set of vertices and $E$ is a set of edges that connect some pairs of vertices in the graph G. 

There are different kinds of graphs, depending on the information related to each edge. In case we have some information regarding the strength of the relationships, we call \cfuture{it} a weighted graph. If we \cfuture{do not have such information}, it is called an unweighted graph. Edges can also be undirected, when an edge between $V_i \leftrightarrow V_j$ represents a bidirectional relationship, or directed, when the edge between $V_i \rightarrow V_j$ can be different from the edge $V_j \rightarrow V_i$.

Graphs can be characterized by using many statistics such as the average number of connections per node, the average path lengths between nodes, etc. 

%

Graphs can be modelled by using different representations. The most common ones are the adjacency or connectivity matrix and the adjacency list. 

\begin{figure}
	\centering
	\subfloat[]{
		\adjustbox{valign=b}{$
			\begin{bmatrix}
			0 & 5 & 0 & 0 \\
			0 & 0 & 3 & 0 \\
			0 & 0 & 0 & 0 \\
			0 & 1 & 7 & 0 
			\end{bmatrix}$
	}}
	\quad 
	\subfloat[]{
		\adjustbox{valign=b}{\includegraphics[width=0.3\linewidth]{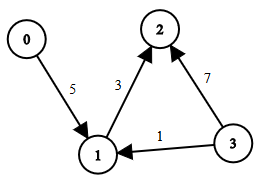}}}
	\caption{\textbf{Directed and weighted graph.} \textbf{a}. Adjacency matrix of the graph. \textbf{b}. Visual representation of the graph.}
	\label{fig:visual_matrix_example}
\end{figure}

The adjacency matrix $A$ of a graph $G$ is a $N\times N$ matrix where N is the number of nodes \cfuture{in G}. Each \cfuture{entry $A_{ij}$} in this matrix corresponds to the value associated with the $V_i \rightarrow V_j$ edge. If the graph is unweighed, those values will be 0 or 1, while if the graph is weighted, those values will be the corresponding weights for each edge (Fig. \ref{fig:visual_matrix_example}). The adjacency list is similar to the adjacency matrix, but instead of storing $N \times N$ elements, we store a list for each vertex containing the rest of nodes which are communicated with it.

\section{Affinity functions as an actor representation} 
\label{sec:affinities}

We define ``Affinities" as a set of functions \cfuture{of two different actors, Actor$_x$ and Actor$_y$, establishing their mutual relation using} $C$: 
\[
A_C:[Actor_x, Actor_y] \rightarrow [0,1]
\]
\cfuture{Usually, this $C$ is the adjacency matrix that quantifies the relationships in each pair of actors, although $C$ can be, for example, another affinity matrix, or a list of them}. The affinity between two actors shows how strongly these two are connected. \cfuture{Since affinities are not necessarily symmetrical, the strength of this interaction depends on who the sender and receiver are}, as happens in human interaction e.g. unrequited love. 

\renewcommand{\arraystretch}{2}
\begin{table}[!htb]
	\begin{tabularx}{\textwidth}{|l|X|}
		\hline 
		$Affinity_{x,y}$ & \textit{Formula} \\ 
		\hline 
		Best friend &  $A_C(x,y)=\frac{C_{x,y}}{ \sum_{a=1}^{N}C_{x,a}}$\\
		\hline
		Best Common friend & 		\cfuture{$A_C(x,y)=Max\{Min(C_{x, z}, C_{y, z})\}/\sum_{a=1}^{N}C_{x,a}$}\\
		\hline 
		Friends forever &  $A_C(x,y)=\sum(\frac{C_{x,y}(t)}{ \sum_{a=1}^{N}C_{x,a}(t)})\frac{1}{|T|},$ \cfuture{$\forall t \in T$}\\   
		\hline 
		Social networking &  $A_C(x,y)=Mean(A_C'(x',y)) $\newline $\forall x'$, \cfuture{such that} $ A_C'(x,x') > 0$\\ 
		\hline 
		Machiavelli &  $A_C(x,y)=1 - \frac{abs(I_x-I_y)}{Max(I_x, I_y)},$  $ I_a = Sum(Degree(x')) $ \newline $\forall x$ \cfuture{such that} $C(a,x') > 0$\\
		\hline
	\end{tabularx} 
	\caption{\textbf{Formula proposed for each of the affinities.} $C$ is the adjacency matrix and 		\cfuture{$N$ is the total number of actors in $C$.} }
	\label{tab:formulas}
\end{table}
\renewcommand{\arraystretch}{1}

We proceed to list some affinity functions:

\begin{itemize}
	\item \textbf{Best Friend affinity \cfuture{(BF)}}: the affinity of the actor $Actor_x$ over the $Actor_y$ is defined as the percentage of the total connectivity of Y that corresponds to $C_{x,y}$.
	\item \cfuture{\textbf{Best Common Friend affinity \cfuture{(BCF)}}: the affinity between two actors is defined as the biggest affinity common to the both of them. It can be computed using both the adjacency matrix or another previously calculated affinity.}
	\item \textbf{Friends Forever affinity \cfuture{(FF)}}: the affinity of two actors \cfuture{reflects the durability of the relation in time.}
	\item \textbf{Social Networking affinity \cfuture{(SN)}}: \cfuture{the affinity between two actors, Actor$_x$ and Actor$_y$, is based on the affinities of the actors connected to Actor$_x$ with respect to Actor$_y$}. 
	\item \textbf{Machiavelli Affinity \cfuture{(MA)}}: the affinity between two actors is based on the social structure that is built around the two of them.
\end{itemize}

All but the Machiavelli affinity are personal affinities. A formula to calculate each affinity function is in Table \ref{tab:formulas}.

There are, mainly, two types of affinity functions: personal affinities and structural affinities. Personal affinities stablish the strength of a interpersonal connection $Actor_x \rightarrow Actor_y$ using  their respective connections and shared friends. \cfuture{Structural affinities quantify the relationship of a pair of actors based on the properties of their nodes, such as their degree or betweenness.}

\cfuture{One particular difference between these two types is that the expected value of a personal affinity is affected by degree of an actor, but this does not necessarily happen in structural affinities. This is because the personal affinity functions behave like a zero-sum game. It can be easily seen in the best friend affinity, where the higher the number of connections the higher the denominator value in the expression.}

\subsection{Effects of different affinity functions in Plato's \textit{Republic}}
\begin{figure}
	\centering
	
	\subfloat[width=0.45\textwidth][]{\includegraphics[width=0.40\textwidth]{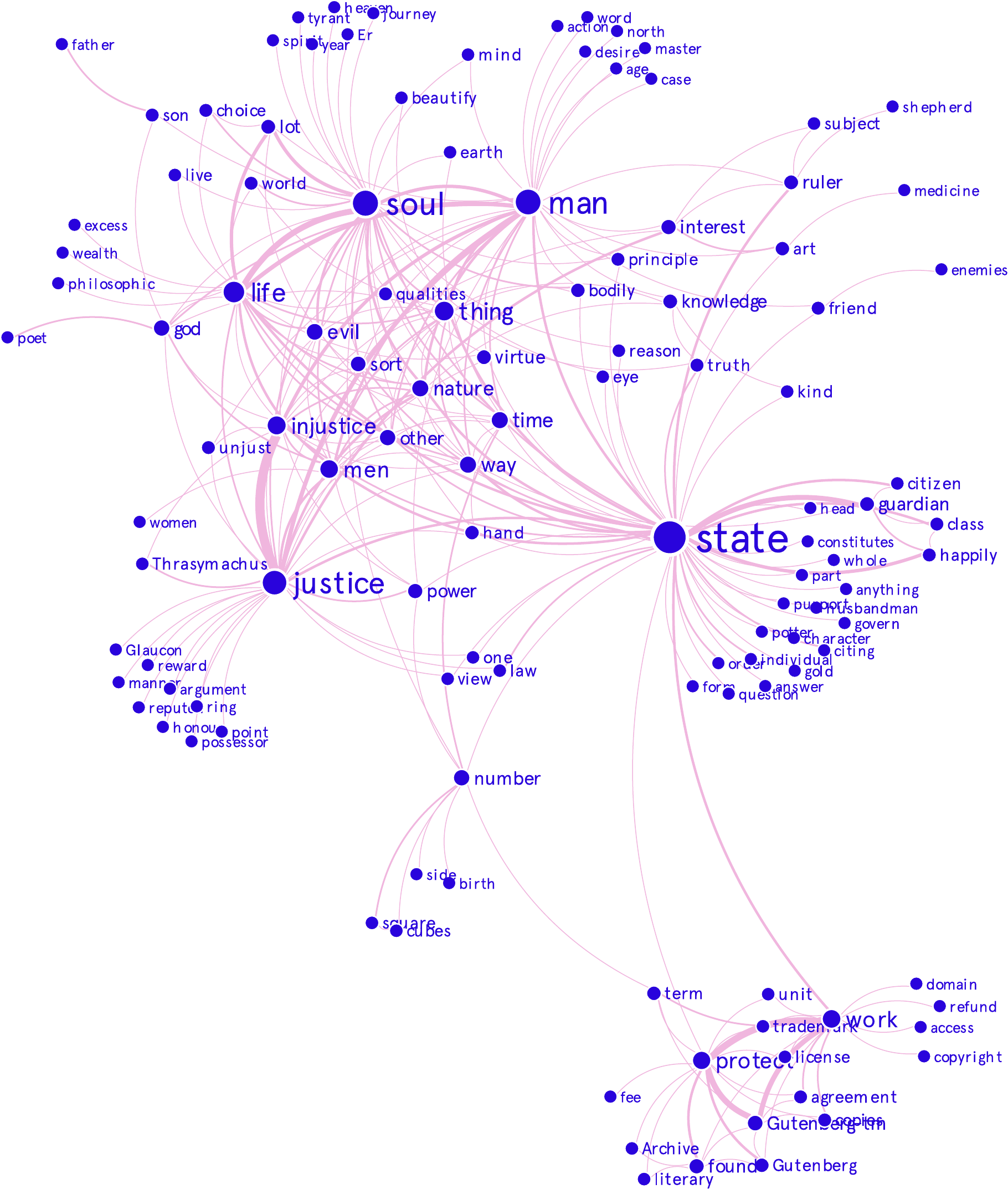}}
	\qquad
	\subfloat[width=0.45\textwidth][]{\includegraphics[width=0.47\textwidth]{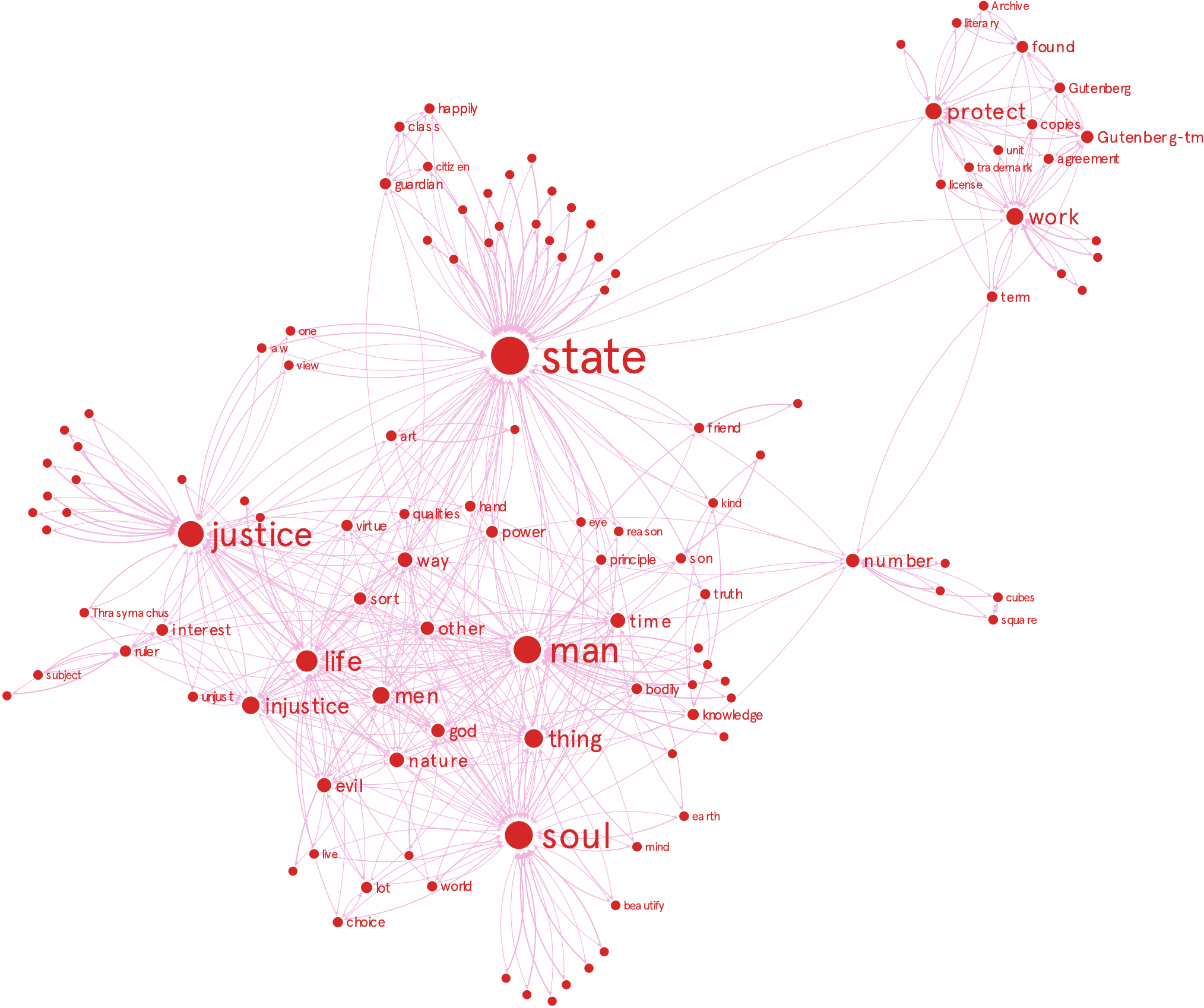}}	\\
	
	\subfloat[width=0.45\textwidth][]{\includegraphics[width=0.43\textwidth]{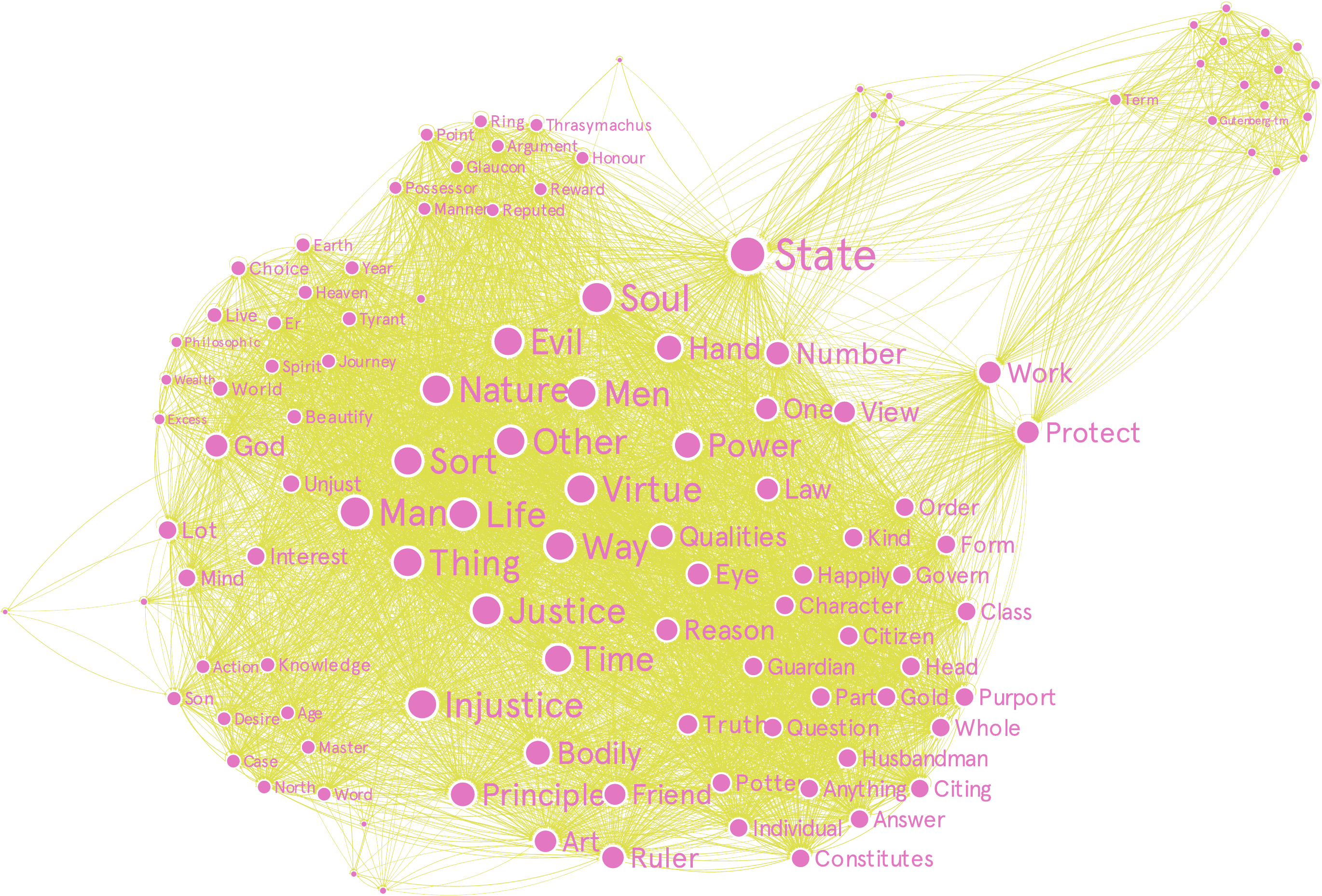}}
	\subfloat[width=0.45\textwidth][]{\includegraphics[width=0.43\textwidth]{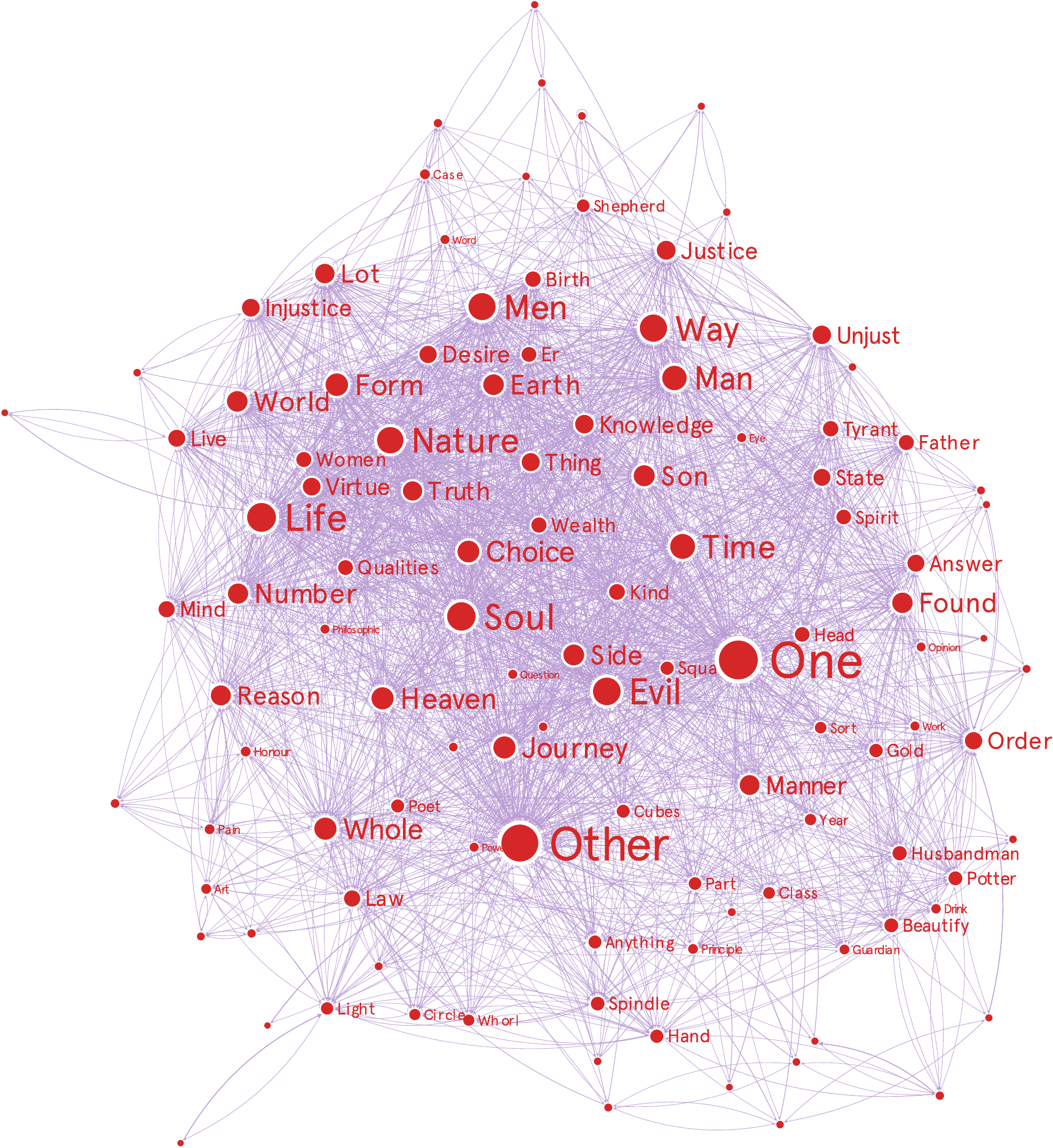}} \\
	
	\subfloat[width=0.45\textwidth][]{\includegraphics[width=0.43\textwidth]{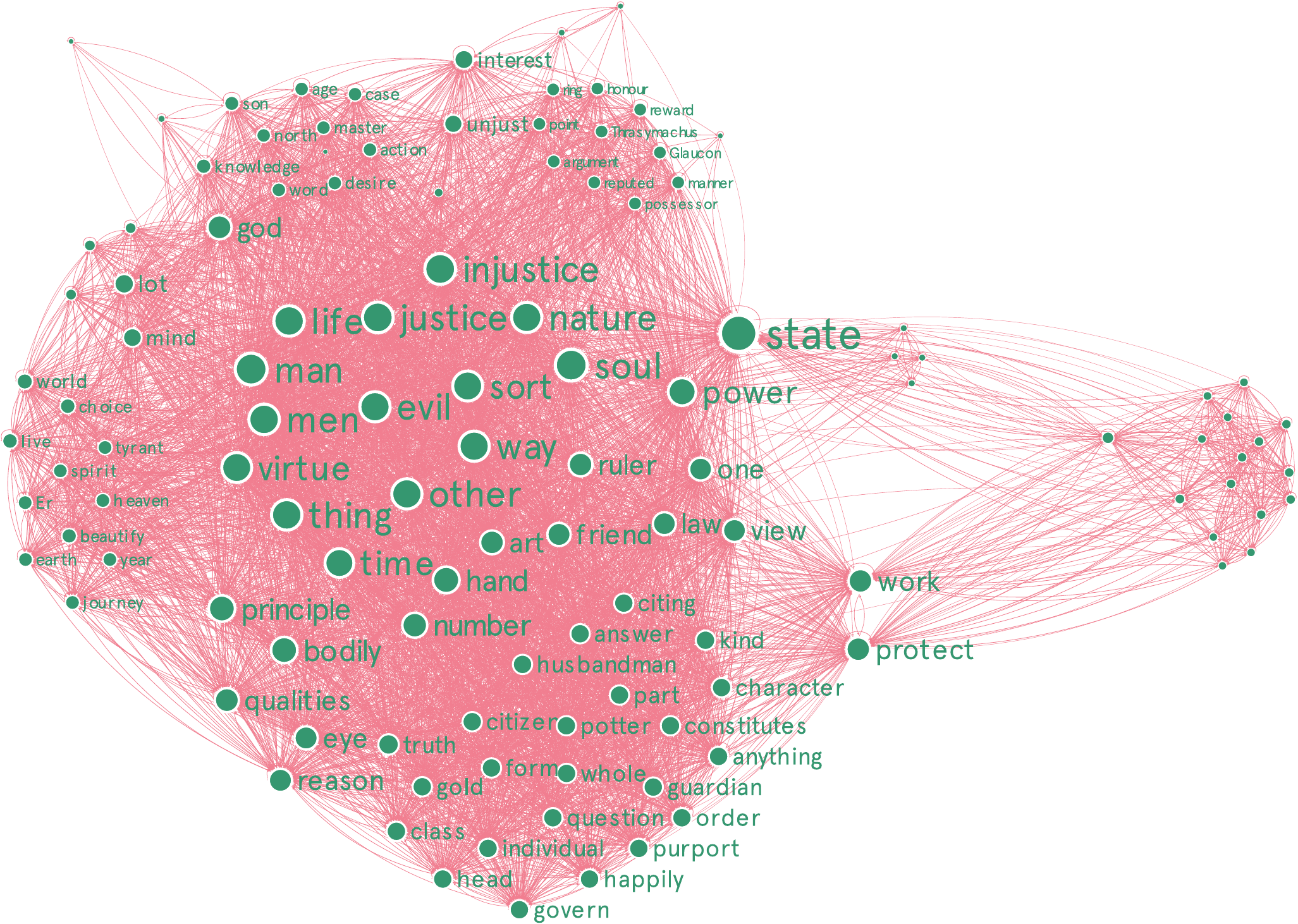}}
	\subfloat[width=0.45\textwidth][]{\includegraphics[width=0.43\textwidth]{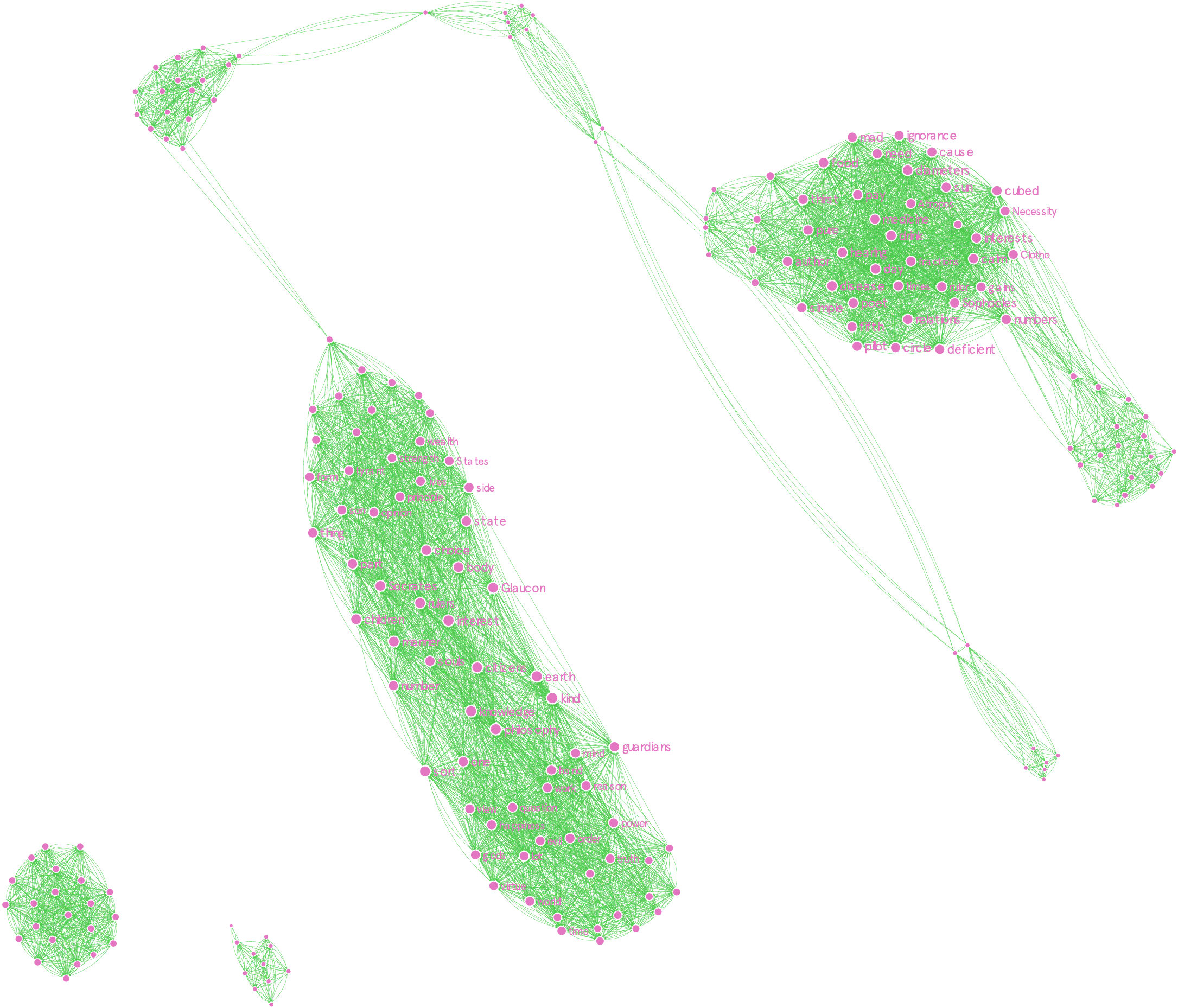}}
	\caption[Network representation for the social network for Plato's \textit{Republic}.]
	{\cfuture{\textbf{Network representation of the social network for the Plato's \textit{Republic}.} \textbf{a}. Co-appearances network. \textbf{b}. Best friend affinity network. \textbf{c}. Best common friend affinity network.  \textbf{d}. Friends forever affinity network. \textbf{e}. Social network affinity network. \textbf{f}. Machiavelli affinity network. All of them have been drawn using the Force Atlas 2 algorithm \cite{hu2005efficient}}}
	\label{fig:affinities_plato}
\end{figure}

\begin{figure}
	\centering
	\subfloat[width=0.45\textwidth][]{{\includegraphics[width=0.35\textwidth]{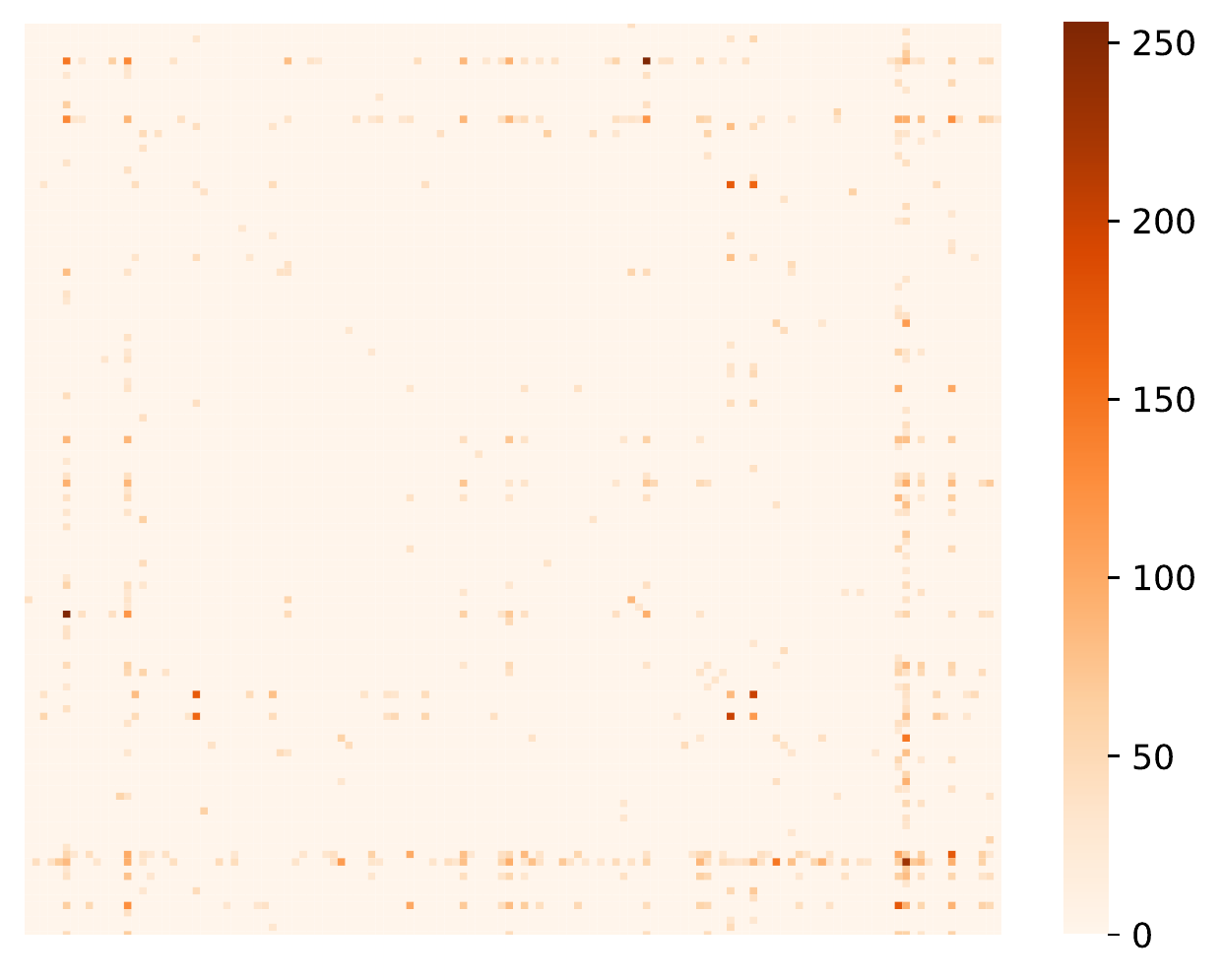}}}
	\qquad
	\subfloat[width=0.45\textwidth][]{\includegraphics[width=0.35\textwidth]{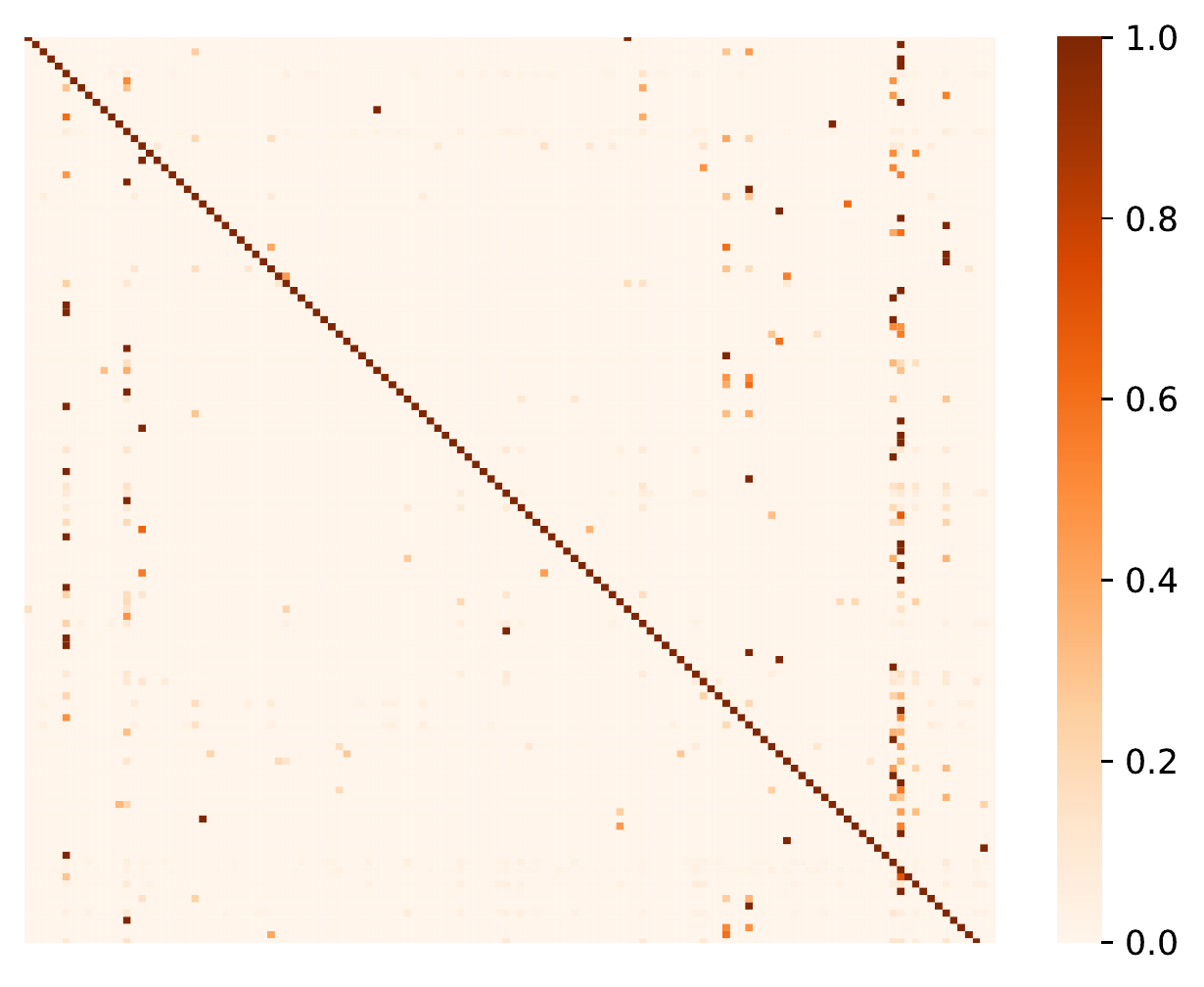}}
	\\
	\subfloat[width=0.45\textwidth][]{\includegraphics[width=0.35\textwidth]{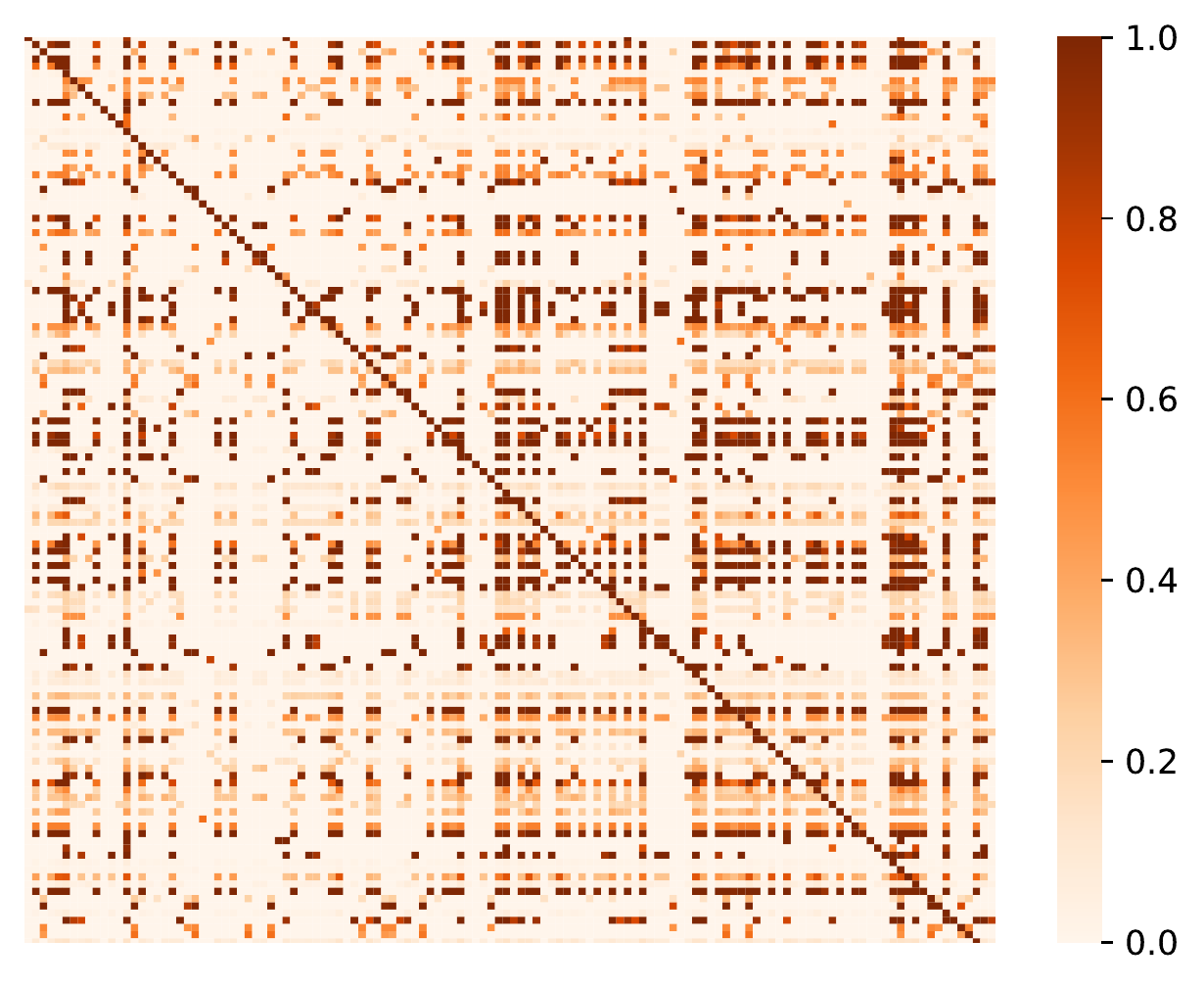}}
	\qquad
	\subfloat[width=0.45\textwidth][]{\includegraphics[width=0.35\textwidth]{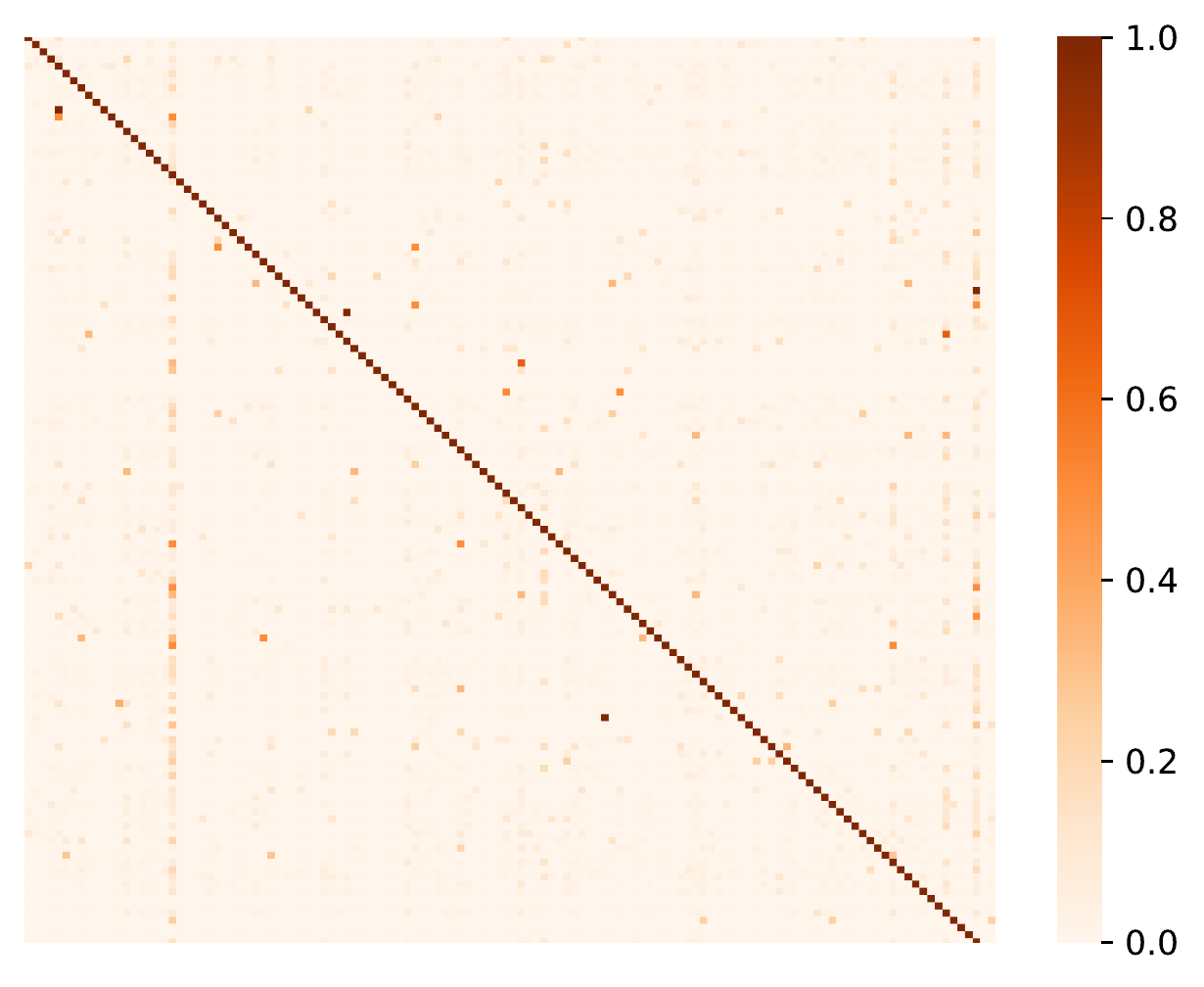}}
	\\
	\subfloat[width=0.45\textwidth][]{\includegraphics[width=0.35\textwidth]{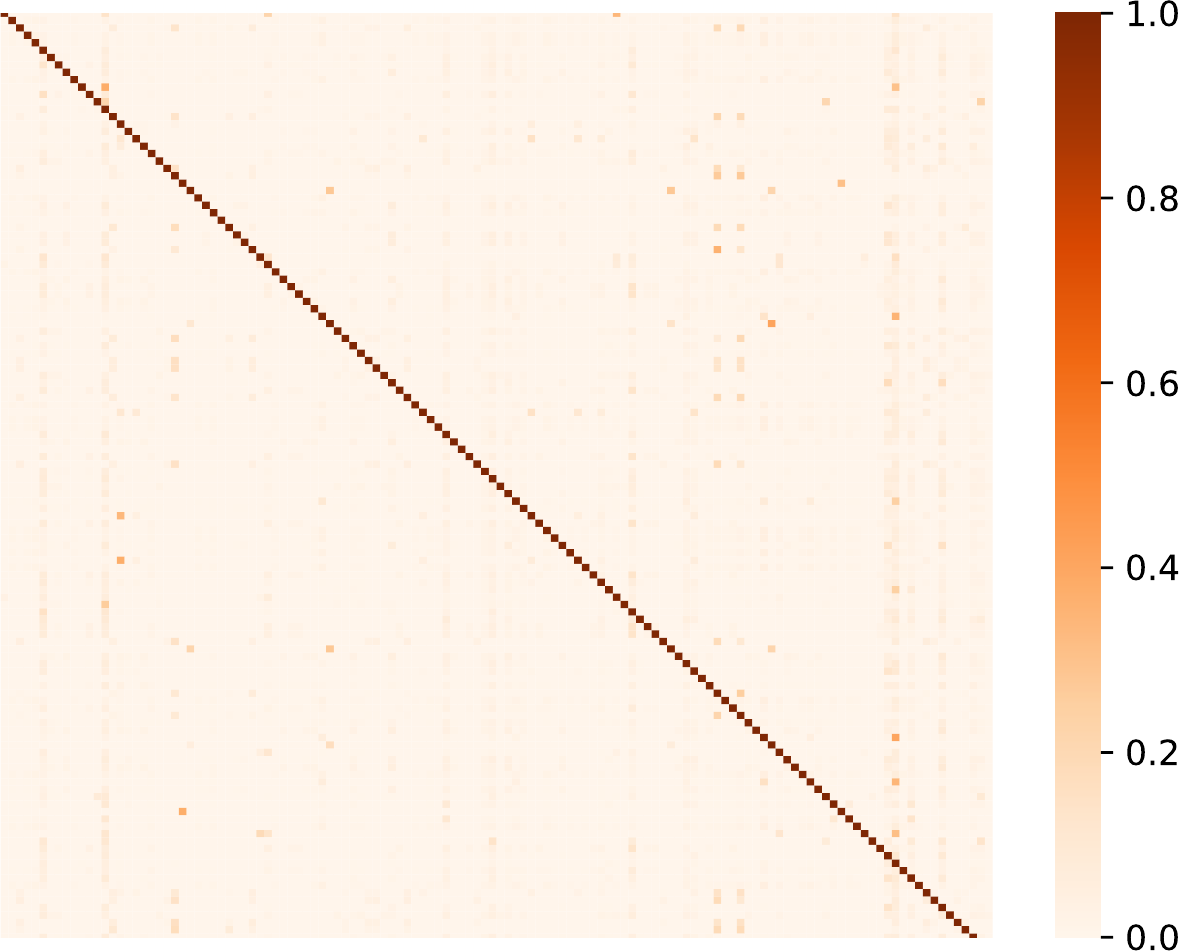}}	
	\qquad
	\subfloat[width=0.45\textwidth][]{\includegraphics[width=0.35\textwidth]{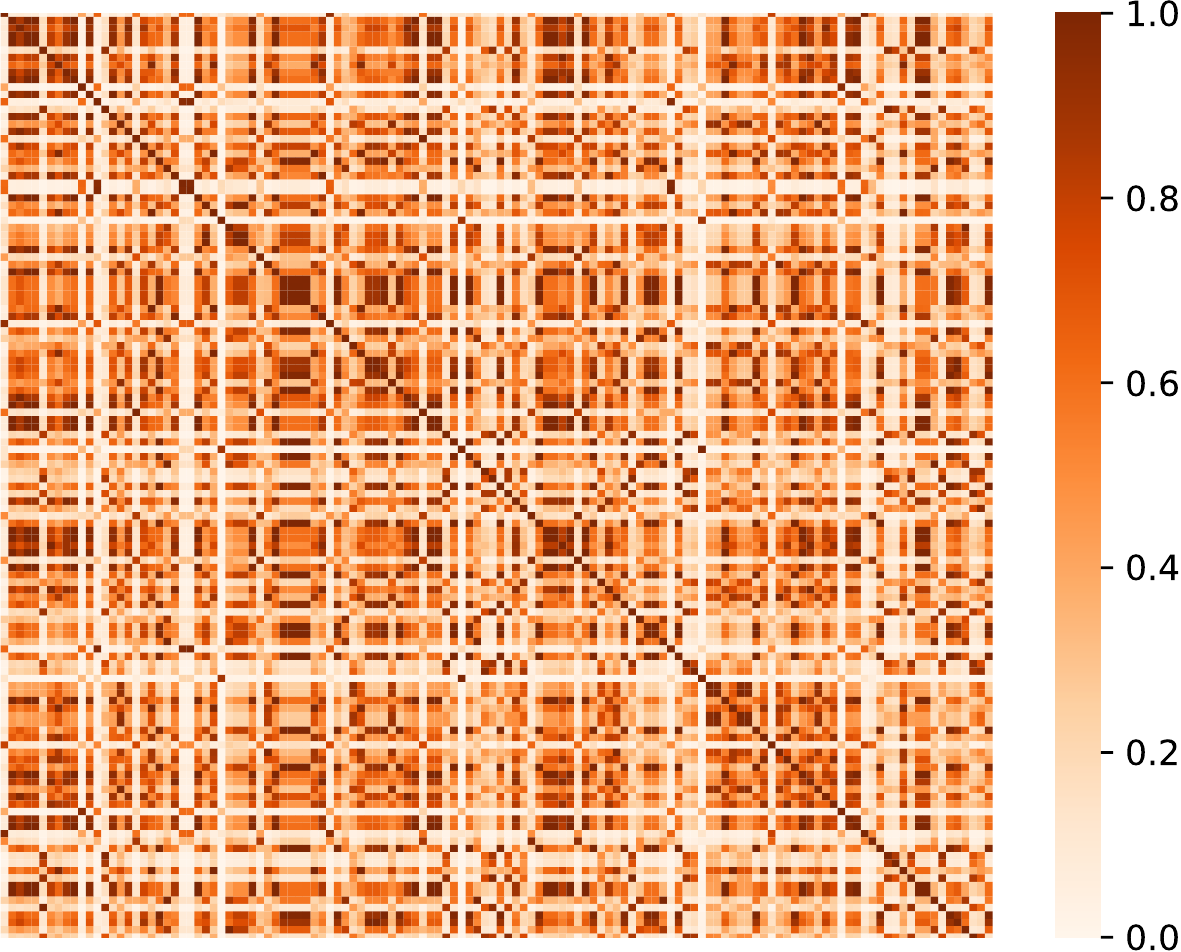}}
	\caption[Heatmap representation for the social network for the Plato's The Republic.]
	{\cfuture{\textbf{Heatmap representation of the social network in Plato's \textit{Republic}.} \textbf{a} The original matrix of co-occurrence of each pair of words. \textbf{b} The affinity matrix for the Best friend affinity. \textbf{c}. The affinity matrix for the Best Common Friend. \textbf{d}. The affinity matrix for the Friend Forever affinity. \textbf{e}. The affinity matrix for the Social Networking affinity \textbf{f}. The affinity matrix for the Machiavelli affinity.}}
	\label{fig:heatmaps_plato}
\end{figure}

\cfuture{For} illustration, we consider the network of word association in classical literature. In particular, we take \textit{Republic}, by Plato. We visualize the resulting network (Figure \ref{fig:affinities_plato}) and the heatmap for the original network (Figure \ref{fig:heatmaps_plato}) for each different affinity. \cfuture{In these networks, each node corresponds to a different word in the original work, and its size is directly proportional to its degree. In the co-occurrences network (Figure \ref{fig:heatmaps_plato}a), each edge represents the number of times when two words appear together in a paragraph, and in the affinity graphs  (Figure \ref{fig:heatmaps_plato}b-f), each edge is the affinity value for each pair of nodes. For the sake of clarity, only the 130 most frequent words are present in each network. }

We have computed all the different affinities \cfuture{provided} in Table \ref{tab:formulas}. \cfuture{Depending on the affinity function used, the resulting edges and their weights can be very different. In Table \ref{tab:aff_effects} we have computed the five affinities for the entity ``Man", and we proceed to discuss the results obtained:}

\begin{enumerate}
	\item \cfuture{Best friend affinity: we obtained a network with the same edges as the original adjacency matrix, but with a weight for each edge. For example, let us consider the ``Man" actor entity. In Table \ref{tab:aff_effects} there are the top incoming and outgoing BF affinities for the ``Man" entity. Outgoing entities are the same as in the adjacency matrix. However, the incoming edges are terms that associate exclusively or almost exclusively with this actor. This happens because ``Man" is an actor that appears constantly in the text, and so it appears mostly with other entities that arise frequently, such as ``Justice" or  ``State". However, some less important actors appear almost entirely associated with ``Man", like ``Desire" or ``Master". So, using this affinity we can easily observe which are the concepts semantically closer to an actor.}
	\item \cfuture{Best common friend affinity: this affinity is capable of ``deducing" edges based on the already existing ones in the adjacency matrix. This leads to a noticeable increment in the number of edges in the network. The density of the network corresponding to the original adjacency matrix is 0.0346, while the best common friend shows a density of 0.3437. This is a consequence of the small world problem \cite{Milgram1967TheSW}. Usually, small nodes are connected to a high-degree node (hub) and to some other small nodes, so many pairs of nodes without an edge between them do have a common connection. Only four actors that were connected in the adjacency matrix did not share any common association. If we look at Table \ref{tab:aff_effects}, the column BCF shows the result of this affinity for the ``Man" entity. The outgoing edges are the same as in the rest of the affinities and the incoming edges are actors that were very affine to one of the top outgoing affinities of ``Man". For example, the ``Tyrant" actor in the adjacency matrix is only connected to the ``Soul" actor, which is one the top connections of ``Man". This results in a high BCF affinity value from ``Tyrant" to ``Man", as the only connection of ``Tyrant" is also a very important connection for ``Man".}
	\item \cfuture{Friends forever affinity: the friends forever affinity is computed by using the ten different chapters of the book as a time unit. This network is very different of the other ones, as each pair of actors needs to repeatedly appear over the whole book in order to have a high affinity value, which can be more revealing to the nature of the original material than the rest of affinities. As the book changes the topic in each chapter, the associations and words not linked to any particular subject are favoured here. For example, taking the ``Man" entity in Table \ref{tab:aff_effects}, we can see in the column FF the concept ``Evil", which does not appear in any other of the affinity functions. The association of the concepts ``Evil" and ``Man" does not appear as many times as other important associations, but it is repeatedly discussed in all of the different chapters of the books, which resulted in a high FF affinity value.}
	\item \cfuture{Social networking affinity: we obtained again a network with the same number of edges as in the best common friend affinity, due to the same reasons. High values of this affinity reveal local social groups, because in order to have a high affinity between Actor$_x$ and Actor$_y$, the majority of connections of Actor$_x$ should have a high previous affinity value with respect to Actor$_y$ (Ac' in the formulation in Table \ref{tab:formulas}). If we look at the SN column in Table \ref{tab:aff_effects}, we can see that the outgoing edges are quite similar to the ones in other affinities, but the incoming edges change significantly. These edges arise from concepts that are semantically very close to the ``Man" actor. This happens because Actor$_x$, in order to have a high affinity value with Actor$_y$, needs to be connected with other actors connected to Actor$_y$. This results in that in order to have a high affinity with ``Man", an actor needs to have a high affinity with other actors that also have a high affinity with ``Man".}
	\item \cfuture{Machiavelli affinity: nodes are more affine to each other depending on their importance in the network, so equally important actors in the book, e.g. Life and Justice, appear very close to each other. The structure of this network is particularly different from the rest of them. This can be clearly seen in Fig. \ref{fig:affinities_plato}f. The bigger mass contains all the most important actors in the network, while the other masses refer to less important concepts or characters that appear in the book. If we look at incoming and outgoing edges in Table \ref{tab:aff_effects}, we can see that they are very similar to those in the adjacency matrix. This is because ``Man" is a key concept of this book, and so are ``Life", ``Soul" or ``Justice", which are its higher Machiavelli affinities. Actors with similar Eigenvector centrality value \cite{friedkin1991theoretical} result in a high Machiavelli affinity.}
\end{enumerate}

\begin{table}
	\centering
	\cfuture{\begin{tabular}{lcccccc}
			\toprule
			Outgoing & Adj. & BF & BCF & FF &  SN & MA  \\ 
			\midrule
			Top 1              & Justice   & Justice     & Justice      &   Life     & Soul       & Life               \\
			Top 2              & Life      & Life        & Injustice    &  Other     & State      & Justice              \\
			Top 3              & Injustice & Injustice   & Soul         &  One       & Justice    & State          \\
			Top 4              & Soul      & Soul        & Life         &  Soul      & Life       & Men             \\
			Top 5              & State     & State       & State        &   Evil     & Injustice  & Soul            \\ 
			\bottomrule
			Incoming &&&&&& \\ 
			\midrule
			Top 1          & Justice     &    Master     &    Tyrant     &  Reward  &  Shepherd   &  Soul        \\
			Top 2          & Life        &    Desire     &    Desire     &  God     &  Medicine   &  Justice    \\
			Top 3          & Injustice   &    Action     &    Spirit     &  Gold    &  Father     &  Life       \\
			Top 4          & Soul        &    Word       &    Journey    &  Work    &  Age        &  State            \\
			Top 5          & State       &    Case       &    Master     &  Protect &  Enemies    &  Men           \\ 
			\bottomrule
	\end{tabular}}
	\caption{\cfuture{\textbf{The effect of different affinity calculations for the ``Man'' actor in Plato's \textit{Republic}}. Each column shows the top values for the adjacency matrix and different affinity functions for each edge of the ``Man" actor.}}
	\label{tab:aff_effects}
\end{table}

\section{Borgia Clustering} \label{sec:core}

\begin{figure}
	\centering
	\subfloat[width=.45\textwidth][]{\includegraphics[width=.44\textwidth]{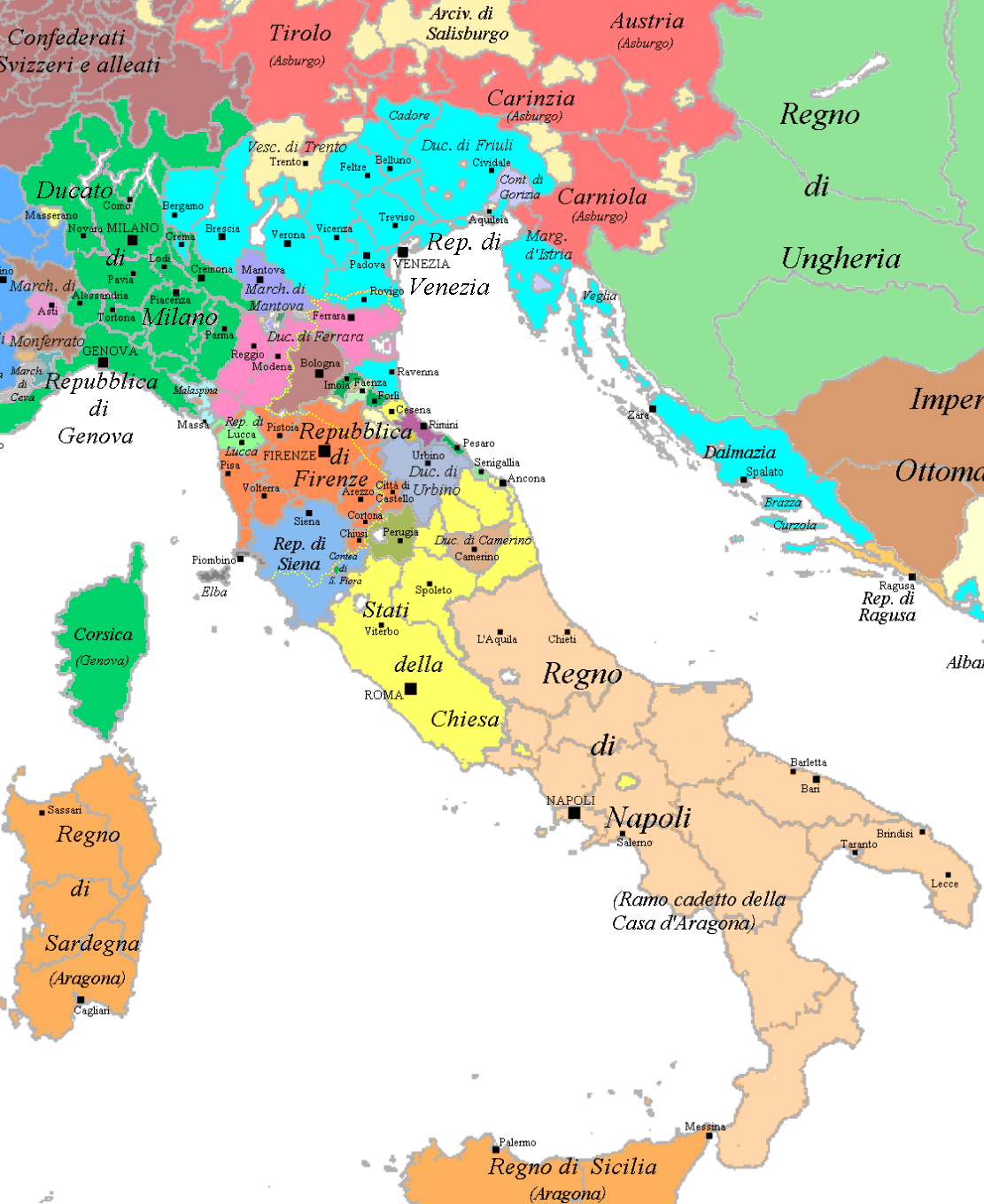}}
	\qquad
	\subfloat[width=.45\textwidth][]{\includegraphics[width=0.44\linewidth]{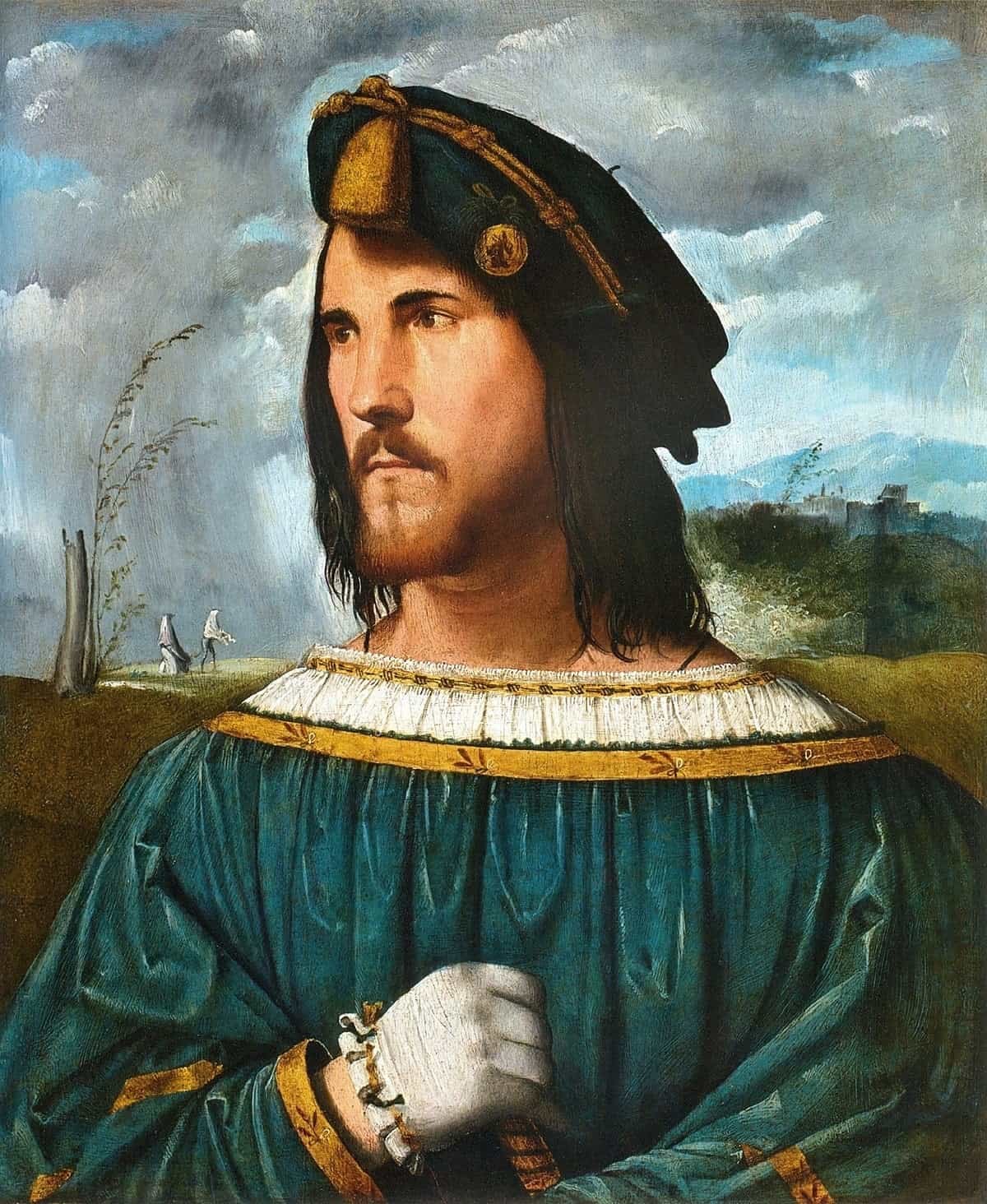}}
	\caption[Italy in times of the Borgias.]
	{\textbf{Italy in times of the Borgias.} \textbf{a}. Italy's lordships before Cesare Borgia's campaign as Commander in Chief of the Papal Army. Each colour represents a different faction. There is a large amount of independent and small counties in central Italy, more precisely, under \textit{de iure} territories of the Papal States\cite{italy_before}. \textbf{b.} A portrait of Cesare Borgia circa 1500.}
	\label{fig:italy}
\end{figure}

Using the affinity functions, we have developed a community detection algorithm, the Borgia Clustering, based on an important chapter in the European history: the 15th-century Italian Wars. In 1497, Cesare Borgia, under the command of his father, Alexander VI, and as Commander in Chief of the Papal Army, marched through the centre of Italy, conquering all the territories that have been traditionally linked to the Papal States \cite{catalan2008principe} (Fig. \ref{fig:italy}). This thrilling moment in Renaissance history provided us with not only memorable moments of unparalleled initiative and wit, but also with an excellent example regarding human interaction in both personal and communitarian levels.

This algorithm is based on the classical Gravitational Clustering algorithm \cite{Wright}, \cfuture{which has not previously been applied for community detection}. Each actor starts as a different community that gets closer to the others due to the effect of an attraction force. Our contention is that the Borgia Clustering force and particle movement generates communities emulating the dynamics seen during the XV Italian wars:

\begin{itemize}
	\item All countries aim to grow. Although it may seem a trivial thing, this is not always the case. Sometimes states prefer to create and maintain a balance rather than breaking it in its own favour. This means that all the attractive forces must be \cfuture{non negative}, and there must always be at least one bigger than zero.
	\item Some parts of Italy are culturally more similar to different countries. Naples, for example, is much more influenced by the Spanish culture than Milan, and future dynastic unions and conquests will make evident these differences. So, the attraction force for a pair of particles must grow with their affinity.
	\item The differences in size and power make some alliances more valuable than others. The Italian republics look for alliances not only in their peninsula but on the lands of more powerful nations. This behaviour favours the creation of opposing sides led by great powers, as was the case with the many Franco-Spanish wars that occurred in the fifteenth and sixteenth centuries. The attractive force of an actor must grow with their inherent value and influence in the net.
	\item There were many countries of different size and importance involved in that historical event. France and Spain were the biggest ones, followed by  Milan, Venice, Florence and the Kingdom of Naples. There were also the Papal States and the ``independent lords" in central Italy. Contrary to naive thinking, the smallest counties were not conquered by the biggest countries, which are the ones with bigger armies, but by the Papal States. So, even though more size implies more attractive force, \cfuture{this} must not be the only factor, and the actors size and scale must be also taken into account.
\end{itemize}

To represent each actor, we use a combination of the best friend affinity matrix and the best common friend affinity matrix, and an influence matrix, $S$, based on the former. By using the best friend affinity, we favour strong pair-wise interactions, and with the best common friend affinity, we also favour the formation of communities whose members share a high number of friends. Also, each actor has a ``social value", equal to its number of connections, that reflects its popularity. 

The actors attract to each other in a simulation of gravitational-like force, depending on their social value, affinity and distance. We consider that two actors $a, b$ collide when the value of $S(a, b)$ is bigger than $S(b, b)$. Then, we interpret that the actor $a$ has as much influence \cfuture{over} the actor $b$ as $b$ has over itself. In that moment they are fused to form a new community. As a result of this, the \cfuture{most} affine pairs of actors will naturally join first and start forming communities.

We have decided to construct the Borgia Clustering algorithm by modifying the classical gravitational algorithm. because it can be easily modified to apply our three ideas to effectively model our wanted dynamics and still keep the physical interpretation. To obtain the desired behaviour, we have performed the following modifications:

\begin{enumerate}
	\item Particles in the original algorithm have been substituted by actors.
	\item We have revamped the attraction \cfuture{force in a way that now it takes the size, the distance and social value of each actor into account.}
	\item The collision condition and fusion procedure for two particles has been adapted to actors.
	\item We have \cfuture{replaced} the idea of position \cfuture{by} the idea of influence. The position matrix has been substituted by an influence matrix, $S$.
	\item Besides the influence matrix, we keep an Affinity matrix, $A_C$, that contains the affinity for each pair of actors and/or communities alongside the execution of the algorithm.
\end{enumerate}

Finally, we have also studied how to optimize the delta parameter, which controls the maximal movement of each particle for each iteration, to speed up the computation.

\subsection{Particle Modification}

In social networks, each particle is not a point in the space, but an actor. Actors have a set of additional properties, usually related to the semantic information available for each one. The most important one is the connectivity, which allows us to compute the best friend and best common friend affinity.

Actors have an initial social value, $m$, that initially corresponds to its degree, that is similar to the concept of particle mass in the original algorithm. This allows us to display two well-known social behaviours: popular people are more socially attractive than people with fewer friends. \cfuture{This fact is} called cummulative advantage in network theory \cite{merton1968matthew} (from the physical point \cfuture{of} view, a greater mass implies a greater attractive force). \cfuture{Furthermore}, people with few friends try to hang out with popular people, but not the reverse (greater mass also implies less movement).

The concept of particle position is substituted for the actor pairwise-influence, denoted as the matrix $S$, that is \cfuture{built up by using} an affinity matrix. The \cfuture{value} $s_{ij}$ reflects the influence that the actor $j$ has over the actor $i$. The self-influences ($s_{i,i}$) are set to 1. Actors will interact with each other along time and they will move closer as time passes due to this interaction. Besides, self-influences will decrease during the execution of the algorithm.

The matrix \cfuture{$A_C$} is initially the same as the matrix $S$, but $A_C$ only changes its values when two actors collide and form a new community.

\subsection{Actor fusion}

The condition \cfuture{for checking} whether two actors have collided using the Euclidean distance is not good enough in this context due to the curse of dimensionality. Thus, we have used the idea of influence instead of position to check whether two actors should be fused or not. 

Each actor starts as being completely influential over itself. During the execution of the algorithm, actors attract \cfuture{to} each other, causing a reduction in their own self-influence, and augmenting the influence other actors have over them. When the influence of one actor over another is greater than the self-influence of that actor, those actors will collapse into a new community.

The fusion of two actors $a,b$ results in the emergence of a new actor whose social value is the sum of $m_a$ and $m_b$, whose position is their centre of masses of $s_a$ and $s_b$, and whose affinities in the affinity matrix are calculated using formula (\ref{eq:new_af}).

\begin{equation}\label{eq:new_af}
A_C(ab) = \frac{m_aA_C(a) + m_bA_C(b)}{m_a + m_b}
\end{equation}

\subsection{New attraction formula}
\label{subsec:attraction_formula}

We have modified the attraction formula to take into account not only the mass and distance, but the whole set of characteristics present in each actor. 

We take into account the bilateral relationships using the $A_C$ and $S$ matrices. The $A_C$ matrix returns the affinity for a pair of actors/communities, and we use the $S$ to compute the \cfuture{Euclidean} distances between them, so that the more common or similar affinities between two actors \cfuture{are}, the higher the attraction force will be.

We also need to take into account the size of the actor/community. To do so, we add a ``greedy expanse" penalization parameter, $p$. This $p$ penalizes the size of the actor in a non-linear way, so that the bigger it gets, the more difficult to move. The penalization is computed as $\frac{1}{m_x^p}$ for each actor $x$. This idea used here \cfuture{for the sake of coping with} size is what we have called the ``Early Roman policy",  in section \ref{sec:scale}.

\cfuture{The product of the masses} and the $A_C$ are aggregated \cfuture{by} using a T-norm as aggregation function \cite{GUPTA1991431}. Using a T-norm is important for computational speed. \cfuture{T-norms} are always \cfuture{less than or equal to} the minimum function. If the $A_C(x,y)=0$, we do not need to calculate the attraction force between $x$ and $y$ because we know that it must be 0. 

The computational cost of the original gravitational algorithm is $O(kn^2)$, where $k$ refers to the number of iterations and $n$ to the number of particles. Because of the \cfuture{T-norm aggregation}, in our case, the cost is $O(kl)$, $l$ representing the number of edges in the graph. This is a significant result since the maximum possible value for $l$ in a undirected network is $n^2$ and real networks are usually sparse. Therefore, the complexity order reduction is actually very large.

As a result, the final formula for \cfuture{the} attraction force is the following:
\begin{equation}\label{eq:force}
\textbf{F}_{xy}=\frac{T((m_xm_y)^c,A_C(x,y))}{m_x^{p}},\frac{\textbf{s}_x-\textbf{s}_y}{\vert \textbf{s}_x-\textbf{s}_y \vert^3}dt
\end{equation}

In it, $T$ stands for a t-norm function, $A_C$ for the chosen affinity function, $\textbf{s}_x$  for the influence vector of actor x.

\subsection{Choosing a configuration}

In the original gravitational algorithm, the configuration \cfuture{with the longest life} in simulated time is chosen as the final one. This criterion is extremely fast to compute but tends to return a very small number of communities. This happens because in the last steps of the algorithm we only have a reduced number of particles that move very \cfuture{slowly} because they are very heavy.

\cfuture{There} is no problem if the desired number of communities is low ($< 5$) but in case we want more, it is necessary to add an extra term to measure the quality, $R$, of each configuration, $Z$:
\begin{equation} \label{eq:cut_dendro}
R(Z) = SimulatedTime(Z) * log(NumCommunities(Z))
\end{equation}
where q is a partition of the graph. \cfuture{By} using this formula, we reward both stability in time and a higher number of different communities.

It is also possible to specify the exact number of communities wanted.

\subsection{Formulation of the algorithm}

To sum up, the formulation of the Borgia Clustering algorithm is the following one:

\begin{enumerate}
	\item We assign a social value to each  actor equal to its degree in the original network.
	\item \cfuture{We set t as 0 and the parameter $\delta$. The $\delta$ parameter restricts the shift magnitude in each iteration for the fastest particle. Then, the $dt$ of each particular iteration is computed based on the movement of the fastest particle. The rest of particles' movement is computed based on this $dt$ value.}
	\item We compute the affinity matrix, $A_C$. Then, we set the initial influence matrix, $S$, equal to $A_C$. 
	\item Next, clustering and attracting steps alternate alongside with the  repetition of a) - d) steps until all actors are fused into one.
	\begin{itemize}
		\item[a)] The function for driving the movement of each actor $i$ in a time interval $[t,t+dt]$ is:
		
		\begin{equation}\label{eq:accel}
		\textbf{g}_i(t)=\frac{1}{m_i^p} \sum _{j\neq i} T(\cfuture{(m_xm_y)^c},A_C(i,j)) \frac {\textbf{s}_j(t)-\textbf{s}_i(t)}{\vert \textbf{s}_j(t)-\textbf{s}_i(t)\vert^3}dt
		\end{equation}
		
		\item[b)] The fastest actor is indexed as $F$
		\[
		F=arg(\max_i\{\vert\textbf{g}_i\vert(t)\}),
		\]
		$dt(t)$ for the next step is computed from $\vert\textbf{g}_F(t)\vert=\delta$:
		$$\delta=\frac{1}{m_F^p} \sum _{j\neq F} T(\cfuture{(m_xm_y)^c},A_C(F,j)) {\vert\frac {\textbf{s}_j(t)-\textbf{s}_F(t)}{\vert \textbf{s}_j(t)-\textbf{s}_F(t)\vert^3}\vert} dt\Rightarrow$$
		\begin{equation}\label{eq:dt(t)}
		\Rightarrow  dt(t)=\frac{\delta m_F^p}{\sum _{j\neq F} T(\cfuture{(m_xm_y)^c},A_C(F,j))\vert\frac {\textbf{s}_j(t)-\textbf{s}_F(t)}{\vert \textbf{s}_j(t)-\textbf{s}_F(t)\vert^3}\vert}
		\end{equation}
		\cfuture{It is apparent} that for each $t$, $dt(t)$ is positive. The influence vector of each actor $i$ is set:
		\begin{equation}\label{eq:positionUpgr}
		\textbf{s}_i(t+dt(t))=\textbf{s}_i(t)+\frac{\textbf{g}_i(t)}{m_i}
		\end{equation}
		\item[c)] $t\leftarrow t+dt(t)$.
		\item[d)]  The test is executed inspecting whether there are actors $i$ and $j$ that meet the collision condition ($s_{i, j} \geq s_{j,j}$). If so, they fuse into one with new mass and influence vector as described above. 
	\end{itemize}
\end{enumerate}

\subsection{About the particle system contraction}
Since a Markovian process is studied, where only the actual state of the system is taken into consideration, in order to prove the convergence of the algorithm it is enough to prove the contractiveness of the particles system. The authors in \cite{AGOP19} proved the convergence of the original gravitational algorithm using an overlap function, $G_s$, instead of the product, using this formula for the attraction force between particles $x$ and $y$ is:
\[
\textbf{F}(x, y) = G_s(m_xm_y) \frac {\textbf{s}_x-\textbf{s}_y}{\vert \textbf{s}_x-\textbf{s}_y\vert^3}
\]

In the case of the Borgia algorithm, the attraction force is the one described in Eq. \ref{eq:force}. All the statements for the original attraction force formula still hold true for the new one. So, the convergence proof for the Borgia Algorithm is analogous to that already present in \cite{AGOP19}.

\subsection{Parameter selection}
\label{sec:parameters}
The Borgia Clustering algorithm depends on a number of different parameters:

\begin{enumerate}
	\item Affinity function: we can choose one or a combination of different affinity functions. Depending on the chosen one, we obtain different behaviours \cfuture{e.g}. Best Friend affinity favours one-to-one interactions, while the  Social Networking favours local high-density groups to attract.
	\item Attraction force: we need to choose a T-norm and an exponent for the product of the masses. In the case of the T-norm, we can use the product as a default, and in case of the \cfuture{exponent, $c$, there are different alternatives: previous results in \cite{Wright} indicates that $c=0$ or $c=1$} gives the best results in clustering.
	\item Greedy expanse penalization factor: the bigger this factor, the harder it is for big masses to keep growing. Generally speaking, if we want to favour local interactions, we should set a high value ($+5$). If we want the big actors to take the lead in the process we should set a low value ($0$ or $1$).
\end{enumerate} 

\subsubsection{Affinity selection}
\label{sec:affinity_selection}

We use the affinity function in the Borgia Clustering algorithm to reflect local pair-wise interactions and local group-level interactions. We reflect both characteristics using the \cfuture{best friend} and the \cfuture{best common friend} affinities (using the formulas in Table \ref{tab:formulas}). By using the former, we can compute how affine \cfuture{two actors are.} Based on the relationship between them; and using the latter, we can compute their affinity according to the common connections they share. To take both affinity functions into account, we have used a convex combination of \cfuture{them}:
\[
A_C(x, y) = \alpha BestFriend_C(x, y) + (1 - \alpha)  BestCommonFriend_C(x, y)
\]
and then we choose and appropriate $\alpha$. \cfuture{We have tested the effects of different $\alpha$ using the famous Zachary network \cite{zachary1977information}, the co-occurrences character network of the \text{Game of Thrones} (GoT) novels \cite{mathbeveridge2017Aug} and the Eurovision voting phase \cite{BibEntry2019Jul}. A detailed description for these networks can be found in Sections \ref{sec:results} and \ref{sec:comparison}}.
	
\begin{figure}
	\centering
	\subfloat[width=0.30\textwidth][]{{\includegraphics[width=0.29\textwidth]{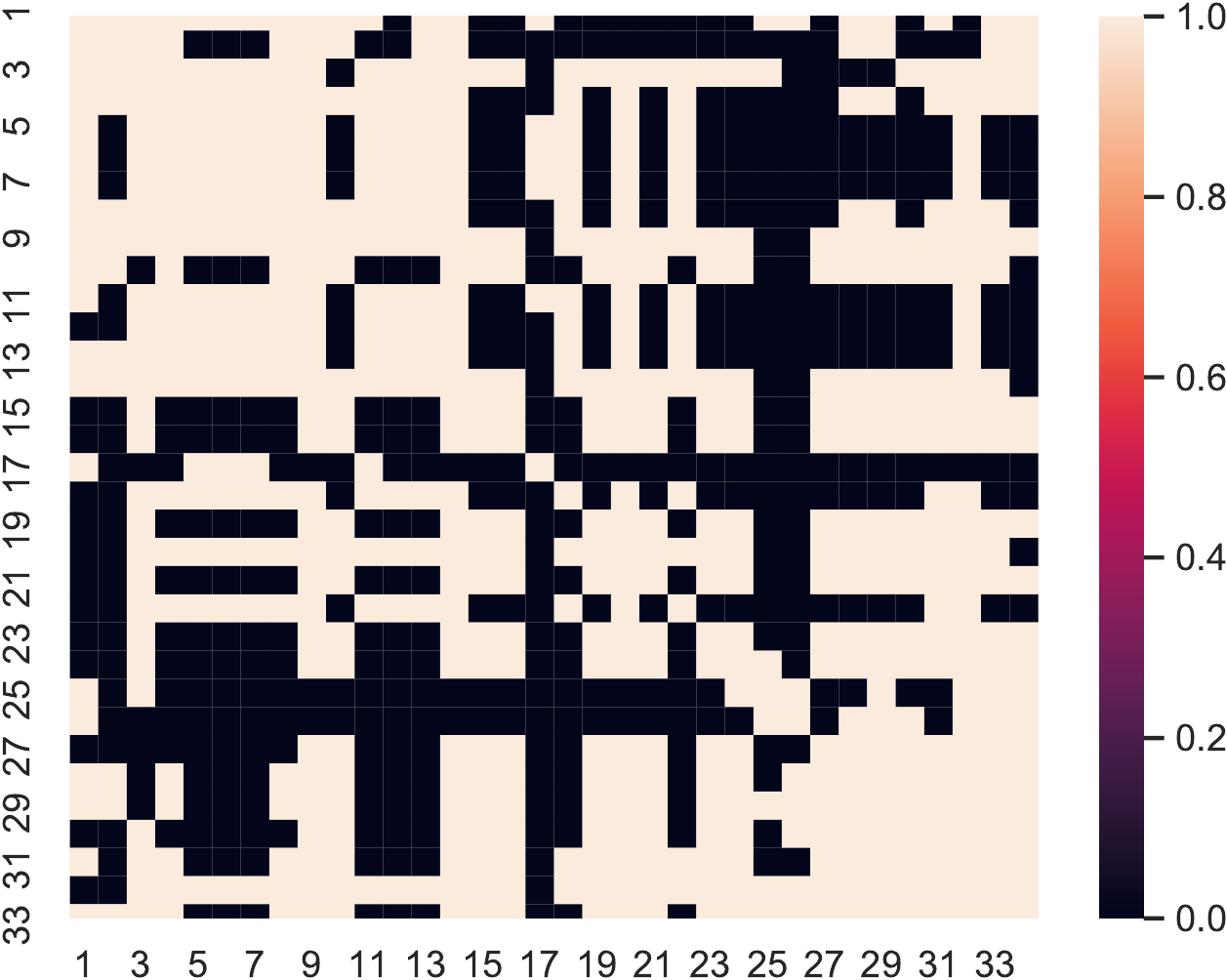}}}
	\qquad
	\subfloat[width=0.30\textwidth][]{{\includegraphics[width=0.29\textwidth]{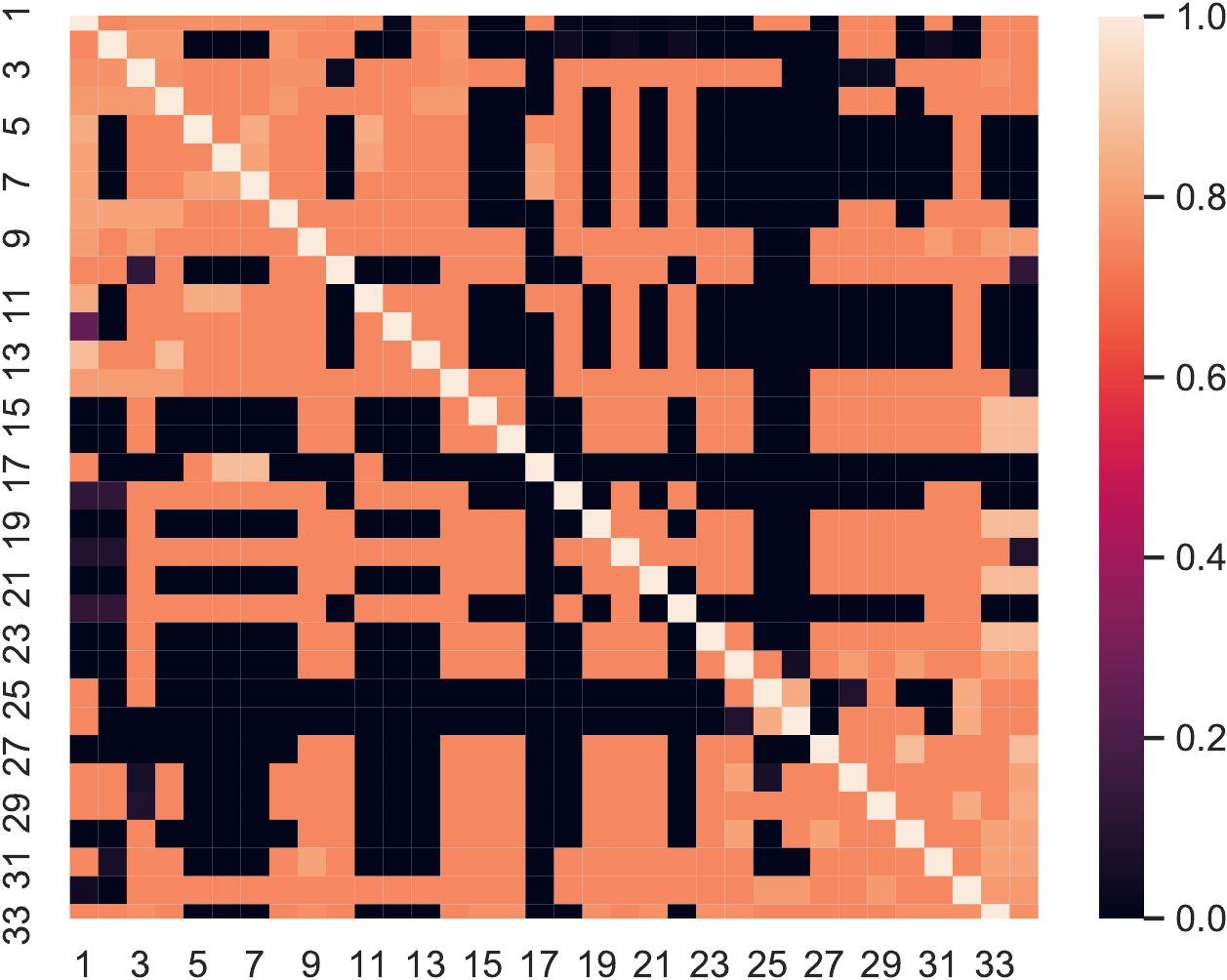}}}
	\qquad
	\subfloat[width=0.30\textwidth][]{{\includegraphics[width=0.29\textwidth]{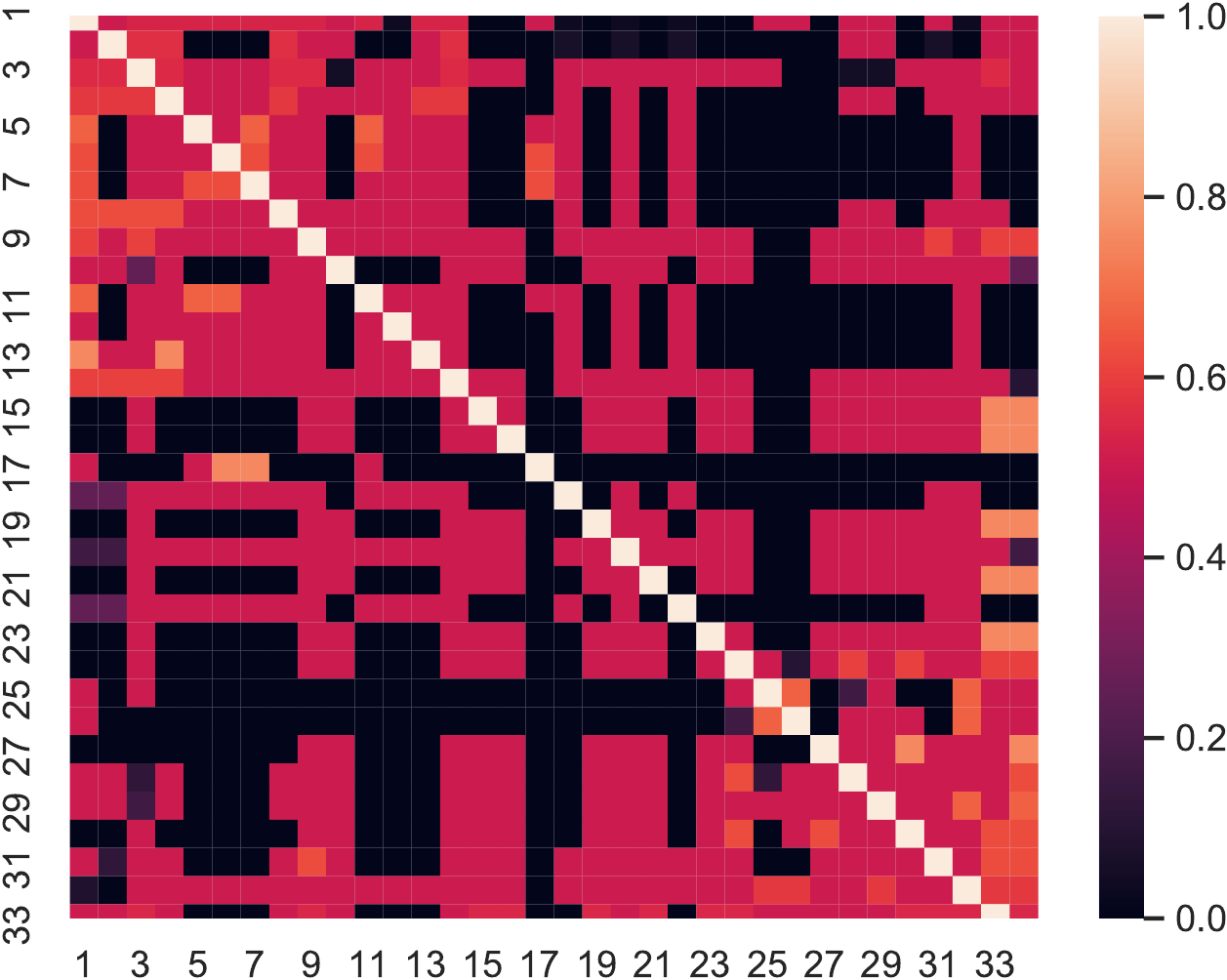}}}
	\\
	\subfloat[width=0.29\textwidth][]{{\includegraphics[width=0.29\textwidth]{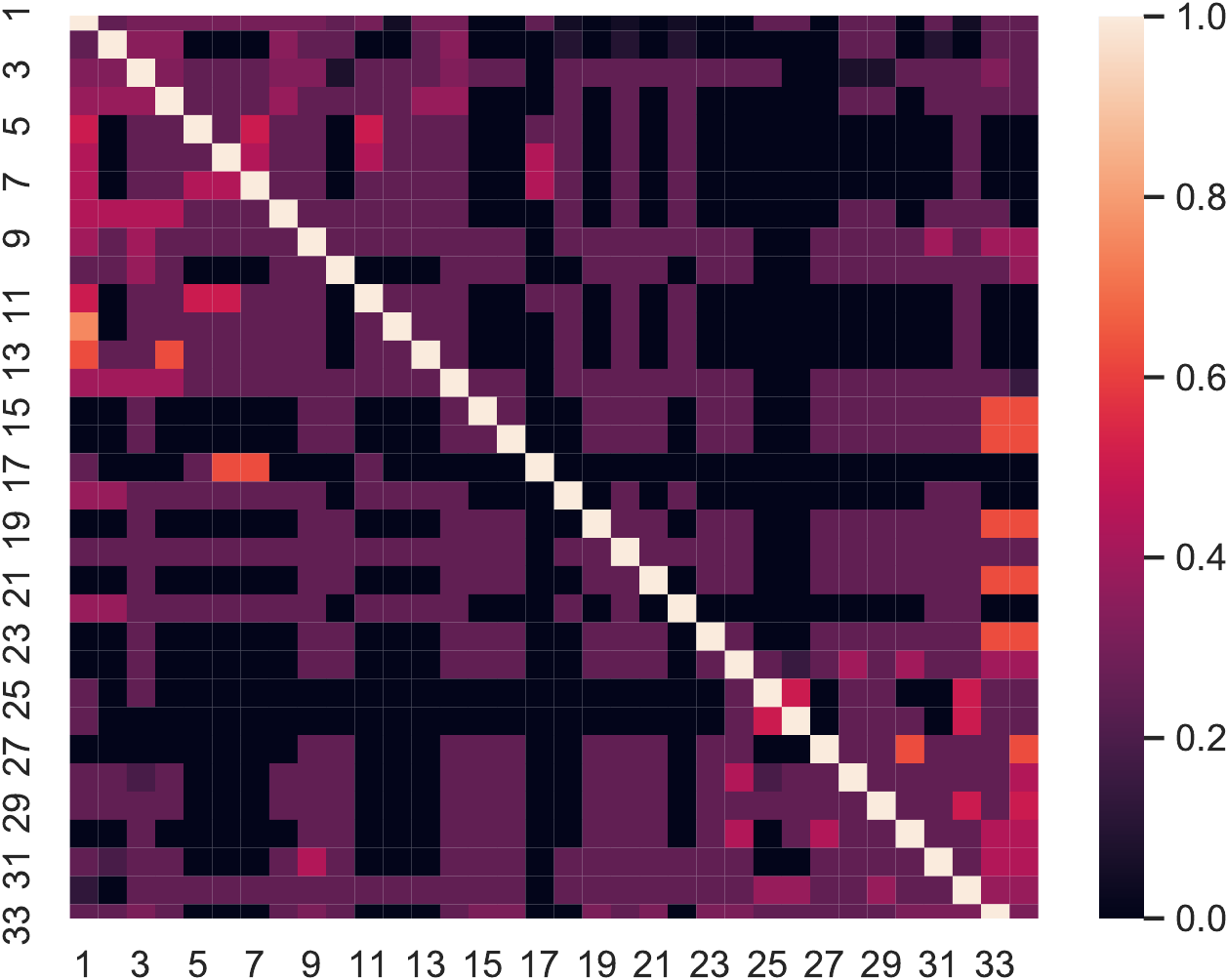}}}
	\qquad
	\subfloat[width=0.29\textwidth][]{{\includegraphics[width=0.29\textwidth]{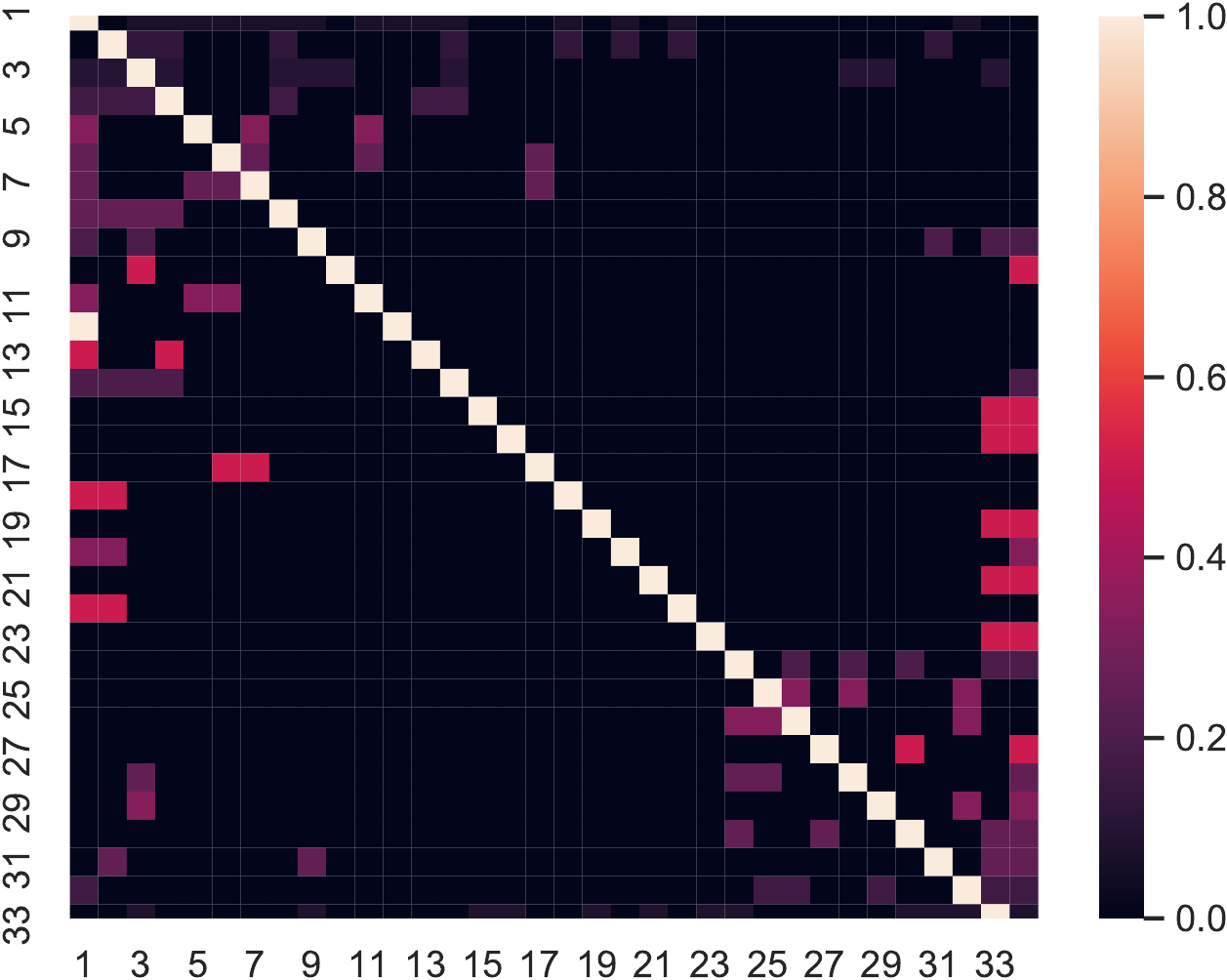}}}
	\caption[Heatmap representation for the Zachary's karate network.]
	{\textbf{Heatmap representation of the Zachary's karate social club.} The $S$ matrix of the Borgia algorithm with different $\alpha$ values. \textbf{a}  $\alpha=0.0$. \textbf{b} $\alpha=0.25$ \textbf{c} $\alpha=0.5$. \textbf{d} $\alpha=0.75$. \textbf{e}  $\alpha= 1.0$.}
	\label{fig:heatmap_zachary_plato}
\end{figure}

\begin{figure}
	\centering
	\subfloat[width=0.30\textwidth][]{{\includegraphics[width=0.29\textwidth]{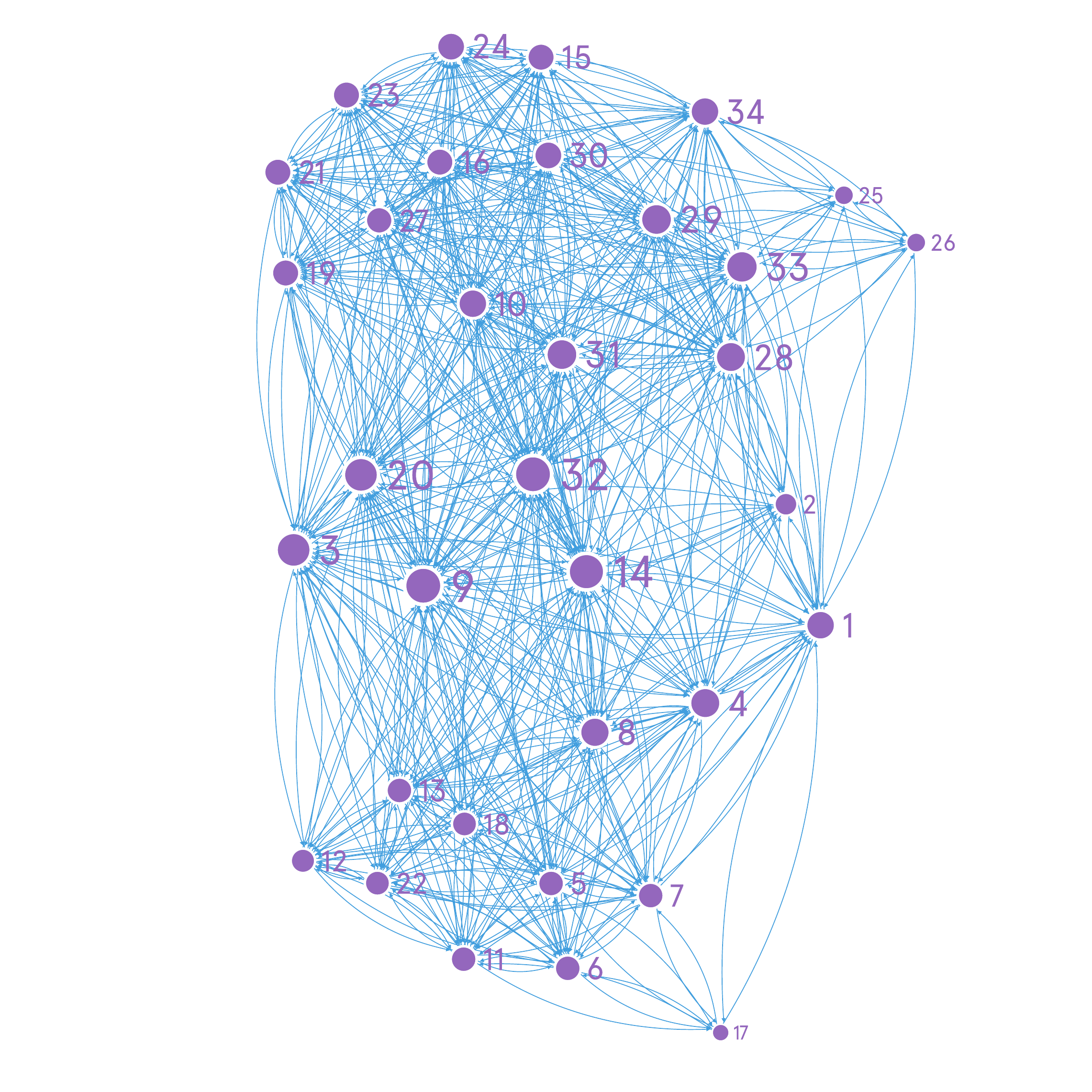}}}
	\qquad
	\subfloat[width=0.30\textwidth][]{{\includegraphics[width=0.29\textwidth]{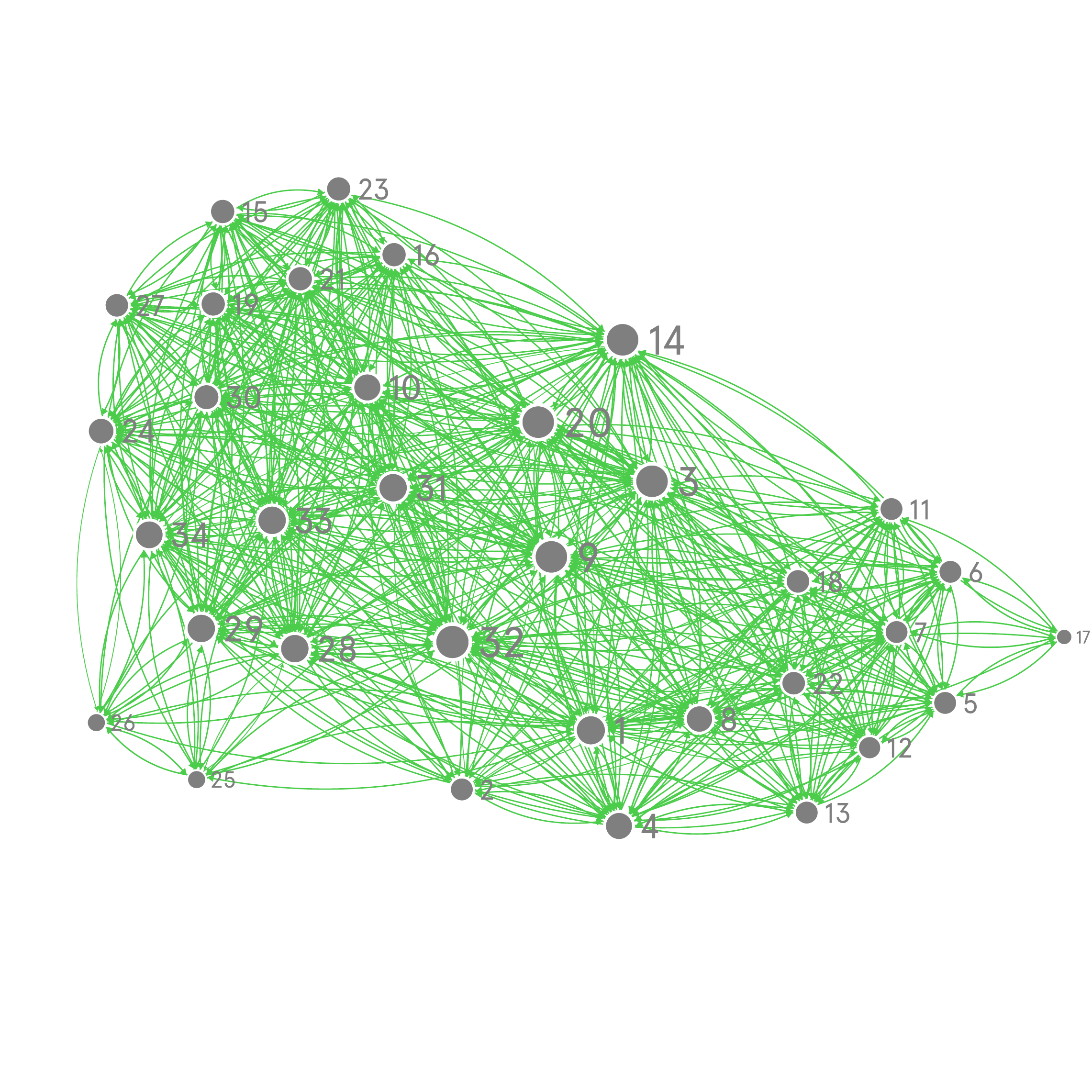}}}
	\qquad
	\subfloat[width=0.30\textwidth][]{{\includegraphics[width=0.29\textwidth]{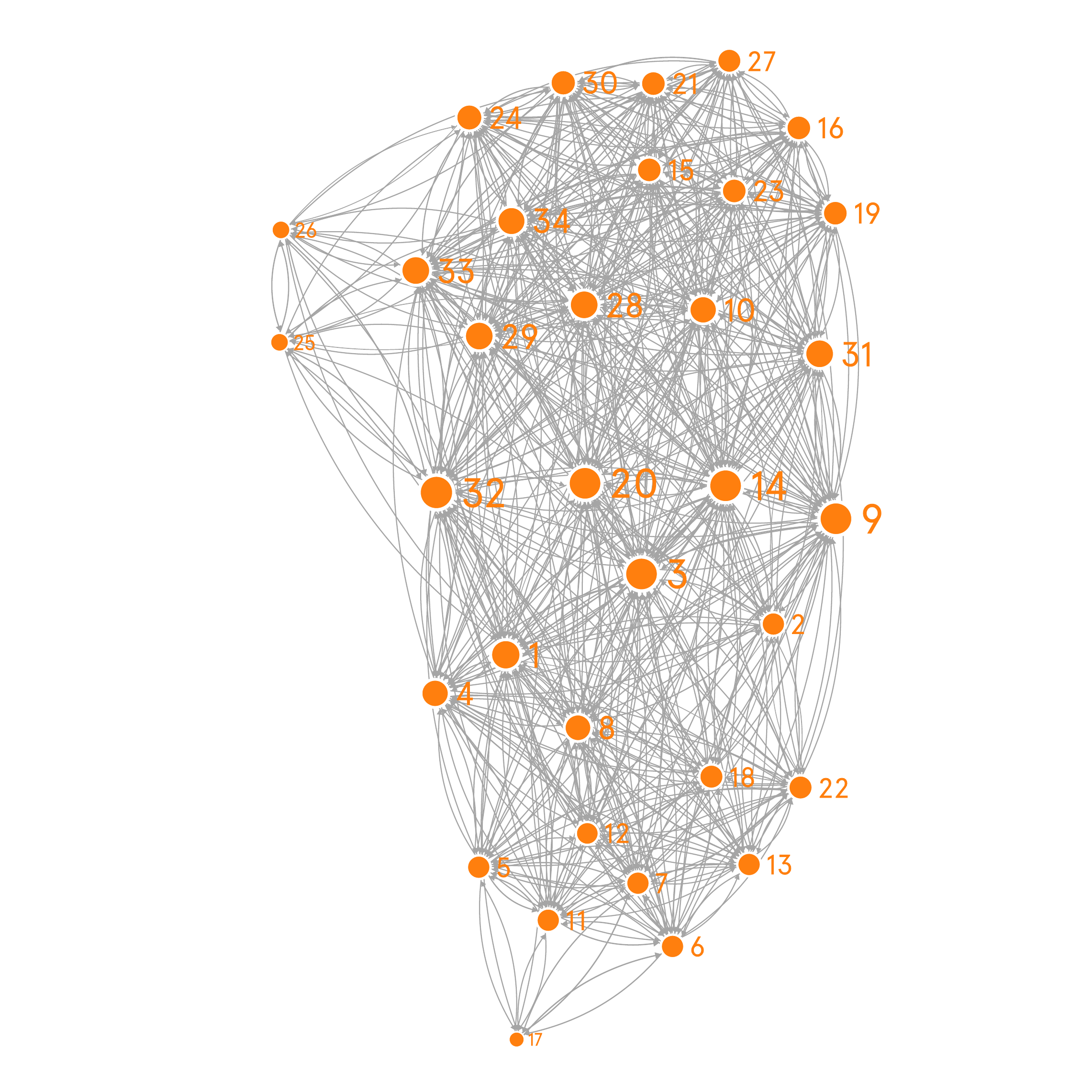}}}
	\\
	\subfloat[width=0.29\textwidth][]{{\includegraphics[width=0.29\textwidth]{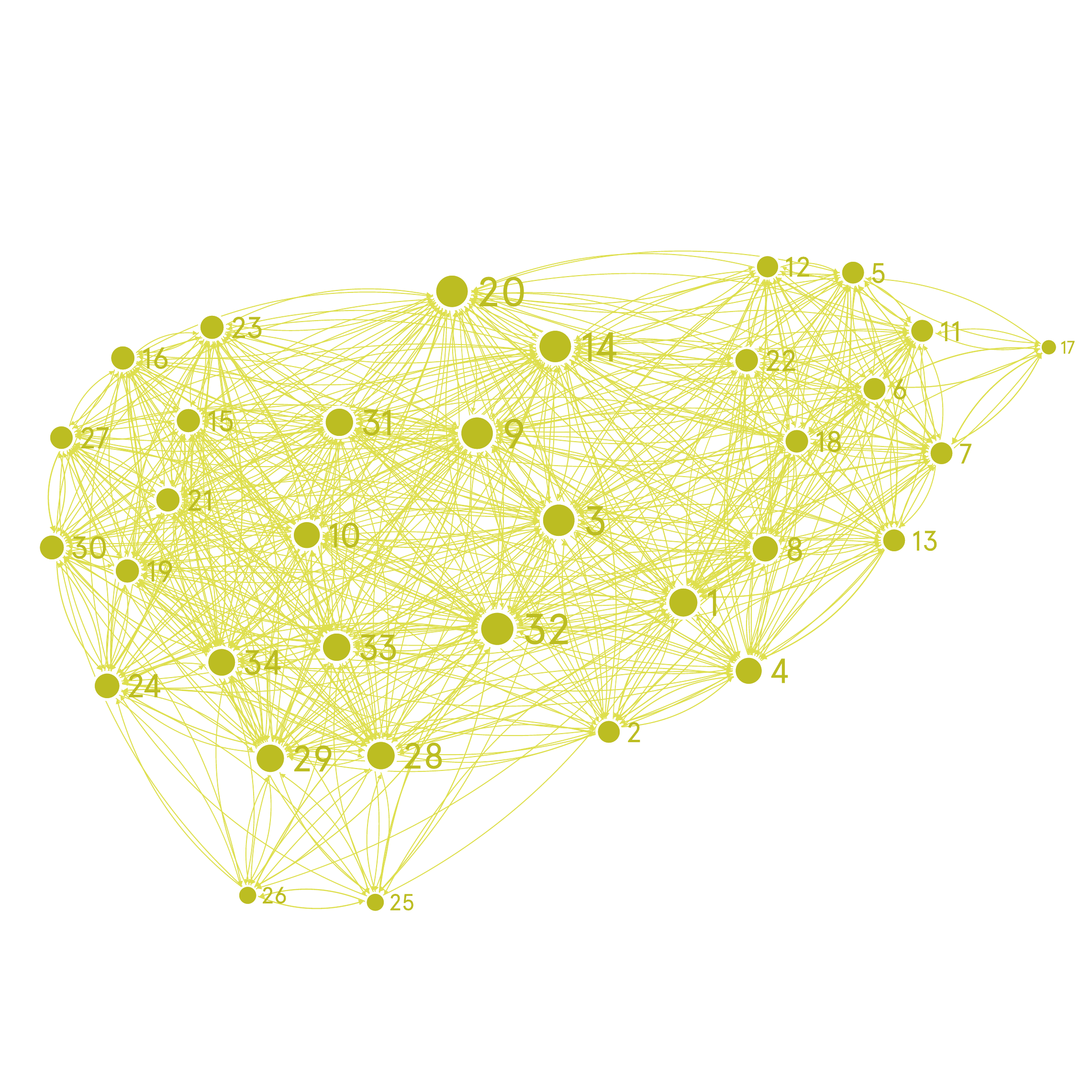}}}
	\qquad
	\subfloat[width=0.29\textwidth][]{{\includegraphics[width=0.29\textwidth]{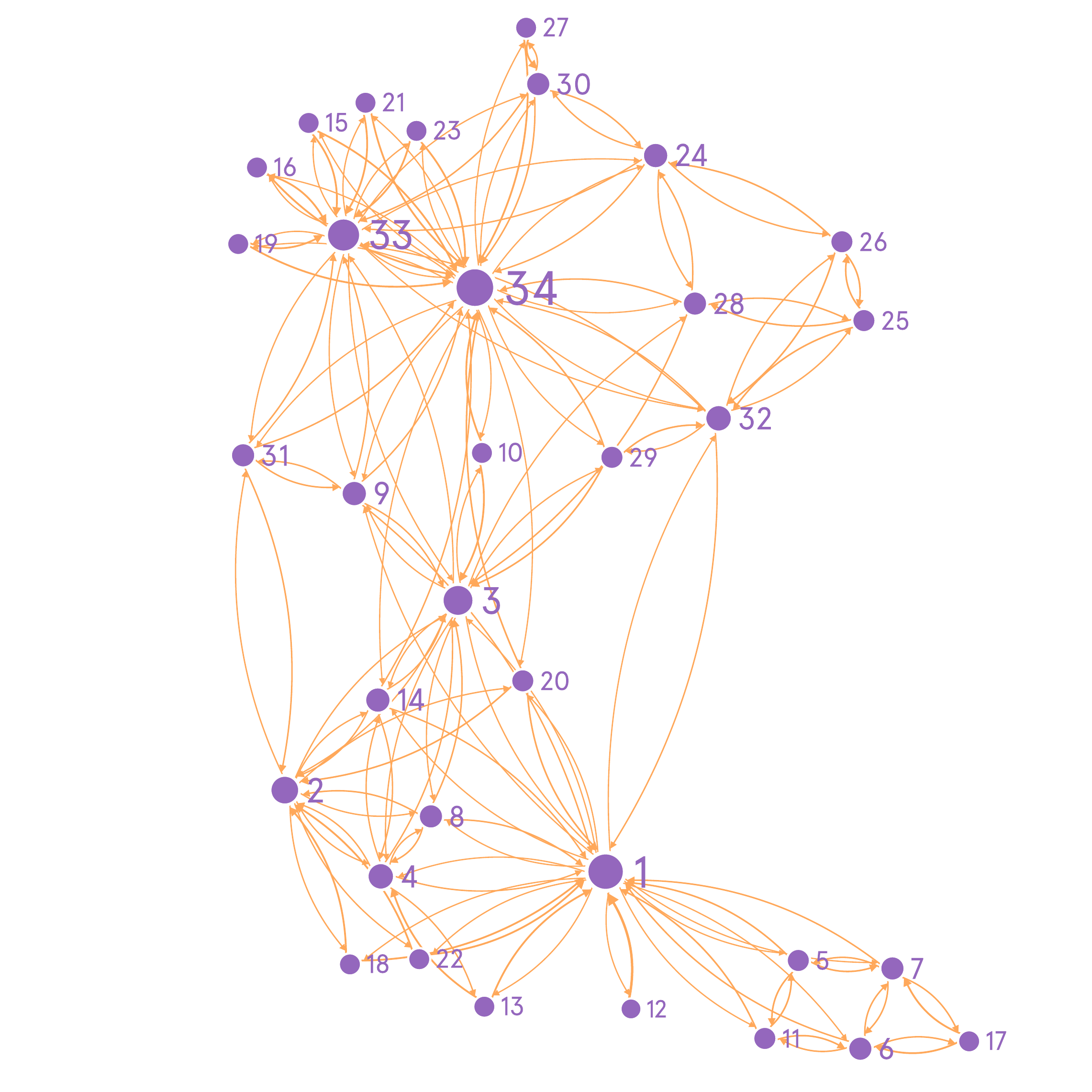}}}
	\caption[Network representation of the Zachary's karate network.]
	{\textbf{Network representation of the Zachary's karate social club Affinity Matrix.} \textbf{a}  $\alpha=0.0$. \textbf{b} $\alpha=0.25$ \textbf{c} $\alpha=0.5$. \textbf{d} $\alpha=0.75$. \textbf{e}  $\alpha= 1.0$.}
	\label{fig:networks_zachary_plato}
\end{figure}

In Figure \ref{fig:heatmap_zachary_plato} we show the corresponding heatmap for each $\alpha$ and in Figure \ref{fig:networks_zachary_plato} we show the corresponding network representations.

\begin{figure}[ht]
	\centering
	\subfloat[width=0.45\textwidth][]{\includegraphics[width=0.42\textwidth]{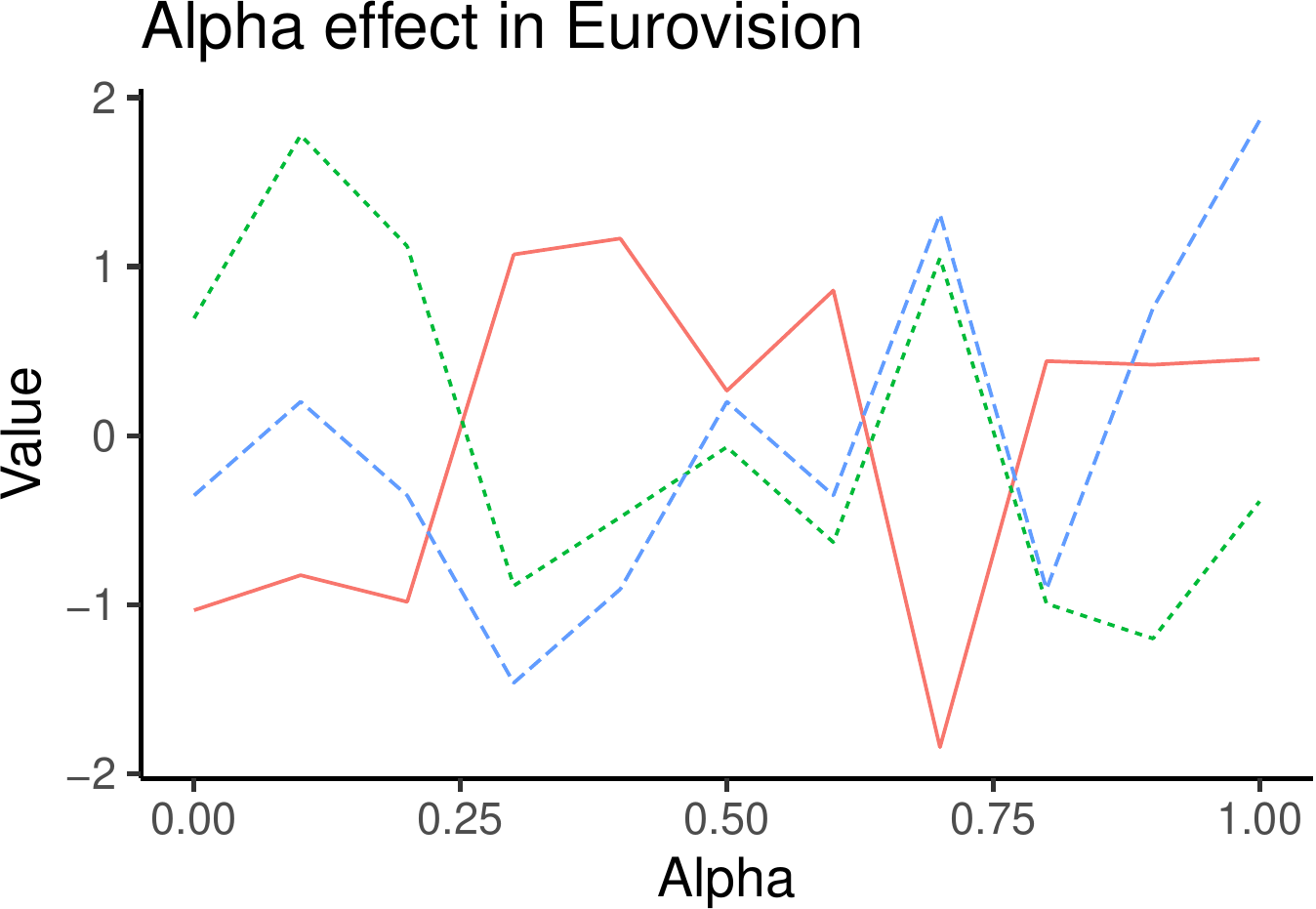}} 
	\subfloat[width=0.45\textwidth][]{\includegraphics[width=0.59\textwidth]{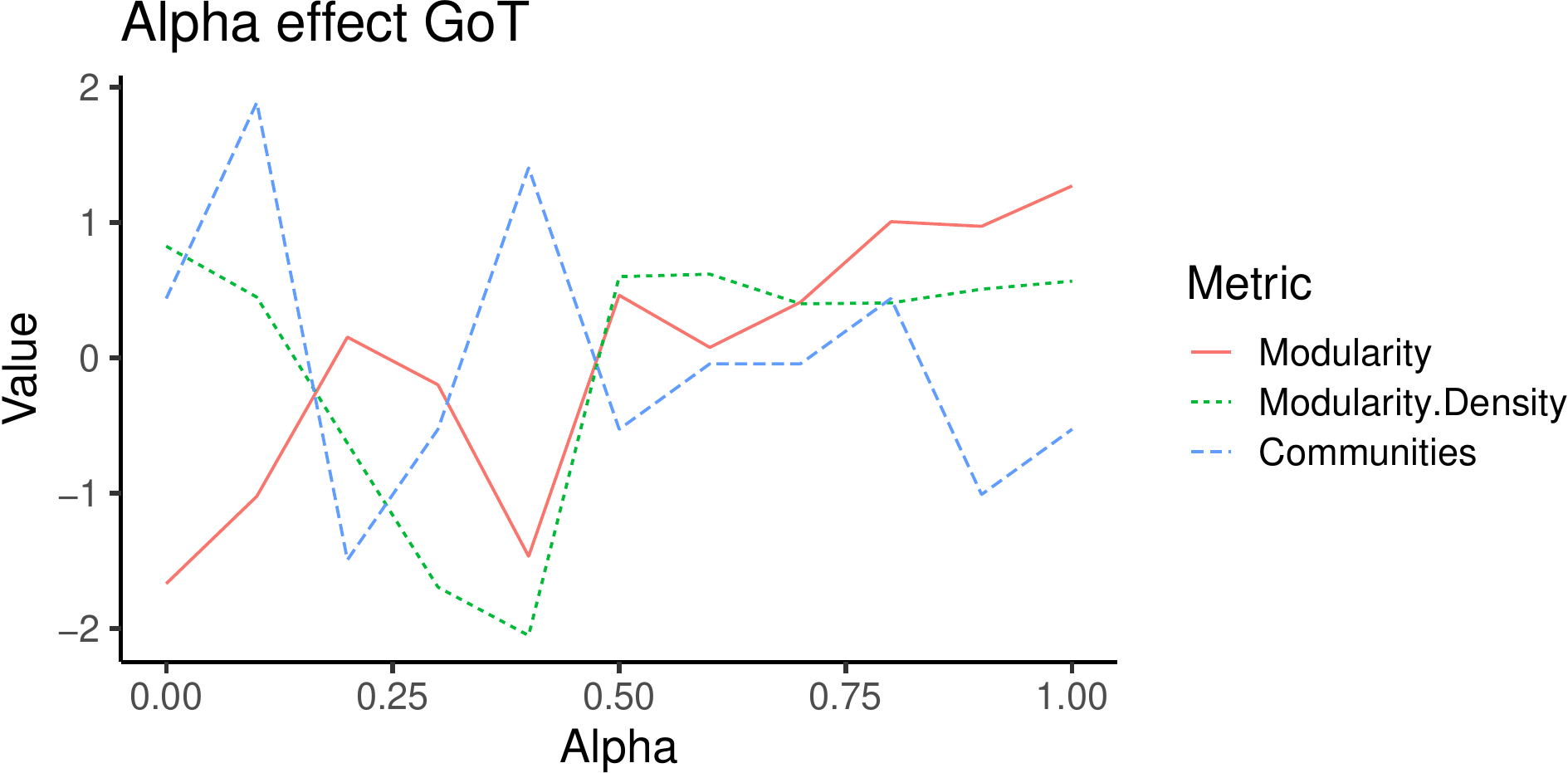}}
	\caption[Affinity combination study]{\textbf{Affinity combination study} \textbf{a.} Modularity values and number of communities for different combinations of Best Friend and Common Friend affinities in the Eurovision voting network with parameters \cfuture{$c=0, p=3$}. \textbf{b.} Modularity values and number of communities for different combinations of Best Friend and Common Friend affinities in the \textit{GoT} social network with parameters \cfuture{$c=0, p=3$}. We have standardized the results for each metric.}
	\label{tab:alpha_got}
\end{figure}

In Fig. \ref{tab:alpha_got} We check the effects of different $\alpha$ in the modularity value for the \textit{GoT} and the Eurovision voting phase networks using a standard greedy expanse parameter and \cfuture{the exponent $c$}. 

When it comes to the number of communities detected, there is no clear effect \cfuture{of} the affinity used. It \cfuture{seems} that using the Common Friend affinity in Eurovision, we get bigger communities, but this does not hold up in the \textit{GoT} network.

By measuring modularity and modularity density, it seems that a lower $\alpha$ benefits the modularity density, whilst a bigger value results in a better modularity. There is no $\alpha$ value that obtains the best  results in both measures, but values in the interval $[0.5, 1]$ apparently give a good result in terms of both metrics and the number of communities obtained. 

\subsubsection{Greedy expanse parameter selection}
\label{sec:greedy}
\begin{figure}
	\centering
	\subfloat[width=0.45\textwidth][]{\includegraphics[width=0.42\textwidth]{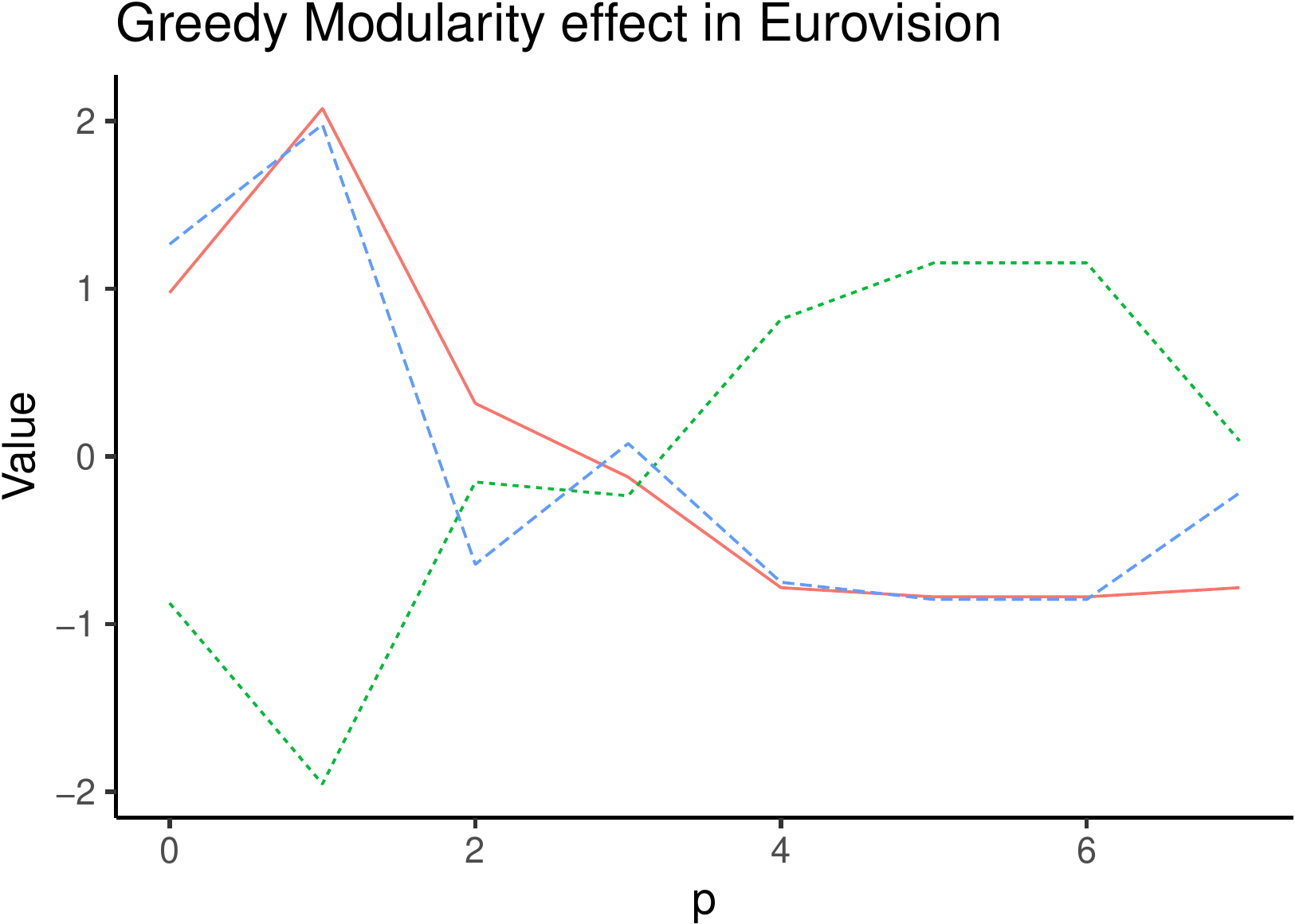}} 
	\subfloat[width=0.45\textwidth][]{\includegraphics[width=0.56\textwidth]{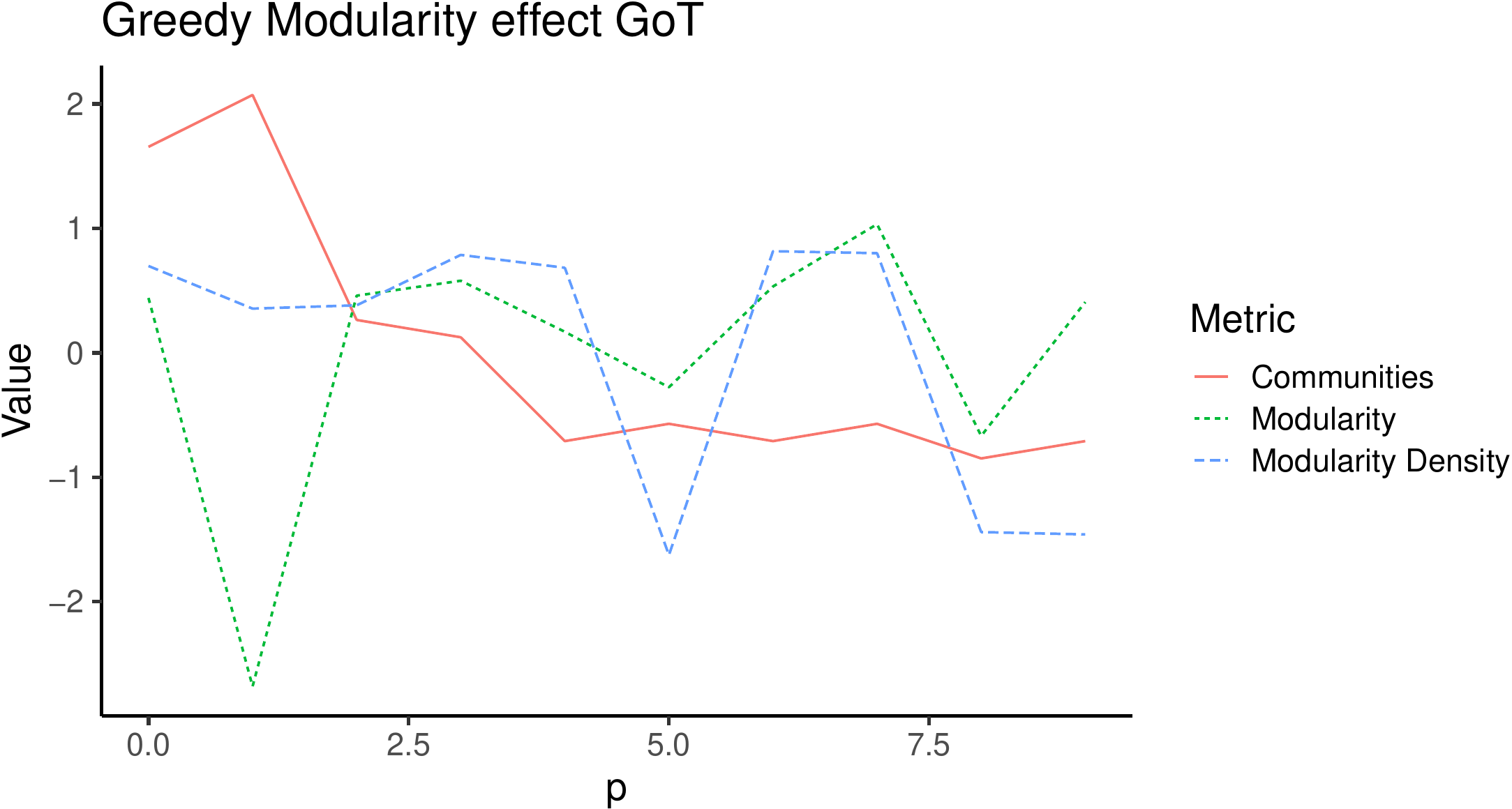}}
	\caption[Greedy parameter modularity study]{\textbf{Greedy parameter modularity study}. Modularity \cite{newman2006modularity} and Modularity Density \cite{li2008quantitative} values and number of communities for different greedy expanse penalization parameter in the Eurovision voting network (\textbf{a}) and \textit{GoT} (\textbf{b}). We have used the mean for three different $\alpha$ ($0.5, 0.75, 1.0$) in each $p$.}
	\label{tab:greedy_euro}
\end{figure}

We use the greedy expanse parameter, $p$, to limit the influence of big particles in the algorithm. The ideal value of the parameter might depend on the topology of the network and the rest of the chosen parameters. If $p$ is too big, then the smaller communities will move too quickly, and if it is too small, the bigger communities will negate the local interactions of the smaller particles. In Fig. \ref{tab:greedy_euro} we have studied how our algorithm performs using different $p$ and different $\alpha$ values for the affinity function. 

The chosen $p$ value has a significant effect in the final result. Generally speaking, a correct value  should be $>2$, since it seems that the number of communities obtained in the $\le2$ cases are abnormally high. The tendency in the modularity values is not so clear alongside $p$.

\subsubsection{Delta parameter}
\label{sec:delta}
The delta parameter delimits the maximal magnitude of movement for an actor in each iteration. In the classical gravitational algorithm, this parameter is a fixed number. However, our experimental tests have revealed that for two actors to collide for the first time we need much more iterations than for the rest of the collisions. We have tackled \cfuture{with} this problem by setting a different delta for the first iteration, such that \cfuture{it} warrants a collision in the first iteration. To do so, we calculate the distances 
to collide for each pair of actors and then, set the delta value as the minimum of them.

The higher the delta, the less accurate the gravitational simulation is. \cfuture{The reason for this is that} we only use this form of delta calculation in the the first iteration. In this moment, actors are far from each other, which makes the attraction forces very weak. 

\begin{figure}
	\centering
	\includegraphics[width=0.7\textwidth]{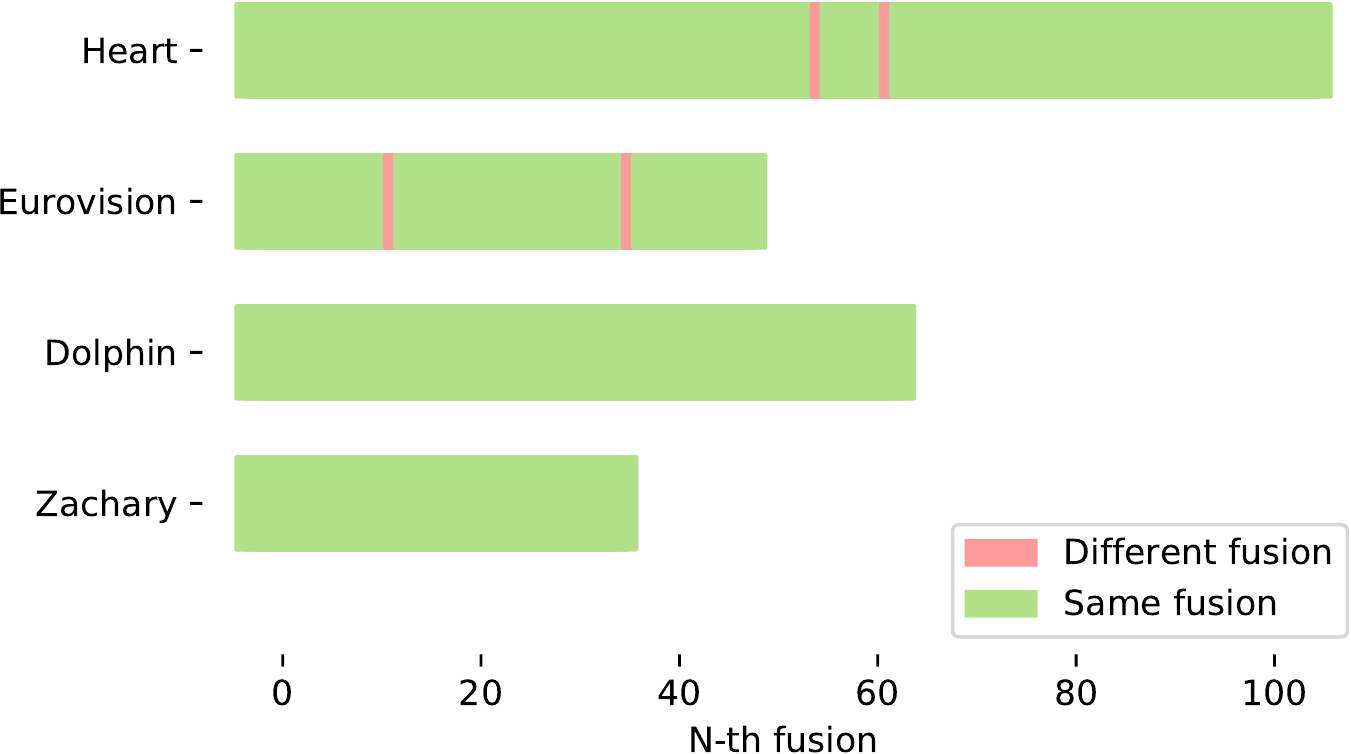}
	\caption[Differences between the dynamic and static delta]{\textbf{Differences in the dynamic and static delta}}
	\label{fig:dy_exec}
\end{figure}

\begin{figure}
	\centering
	\subfloat[width=0.49\textwidth][]{
		\includegraphics[width=0.49\textwidth]{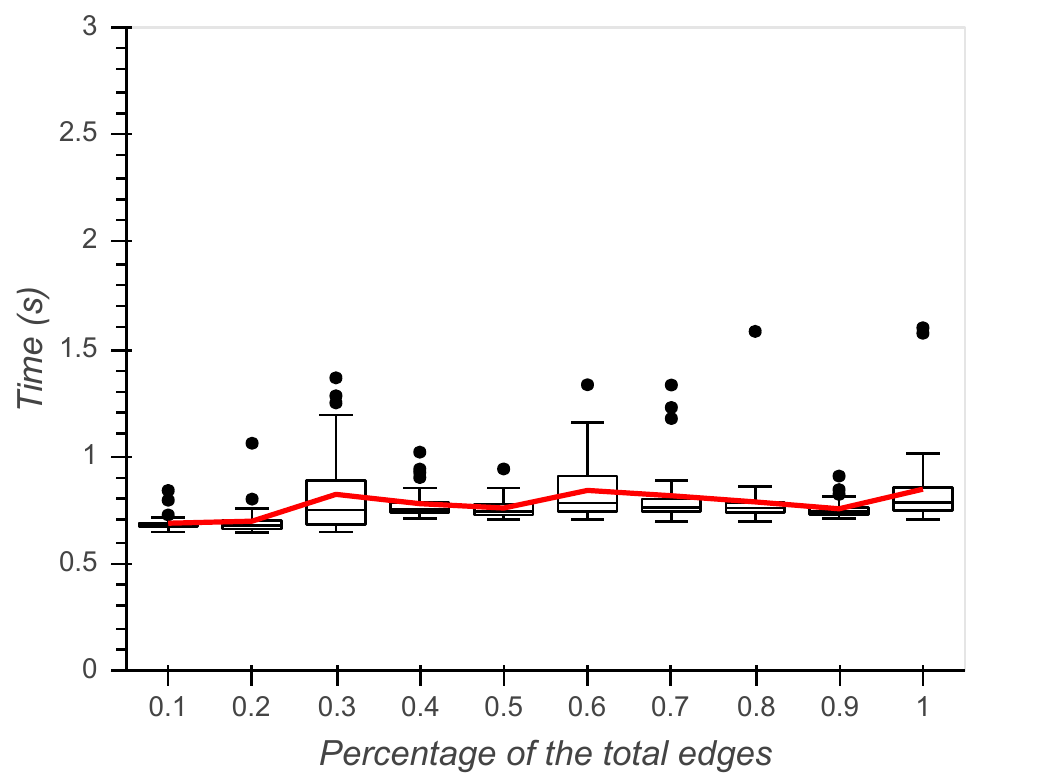}
	}
	\subfloat[width=0.49\textwidth][]{
		\includegraphics[width=0.49\textwidth]{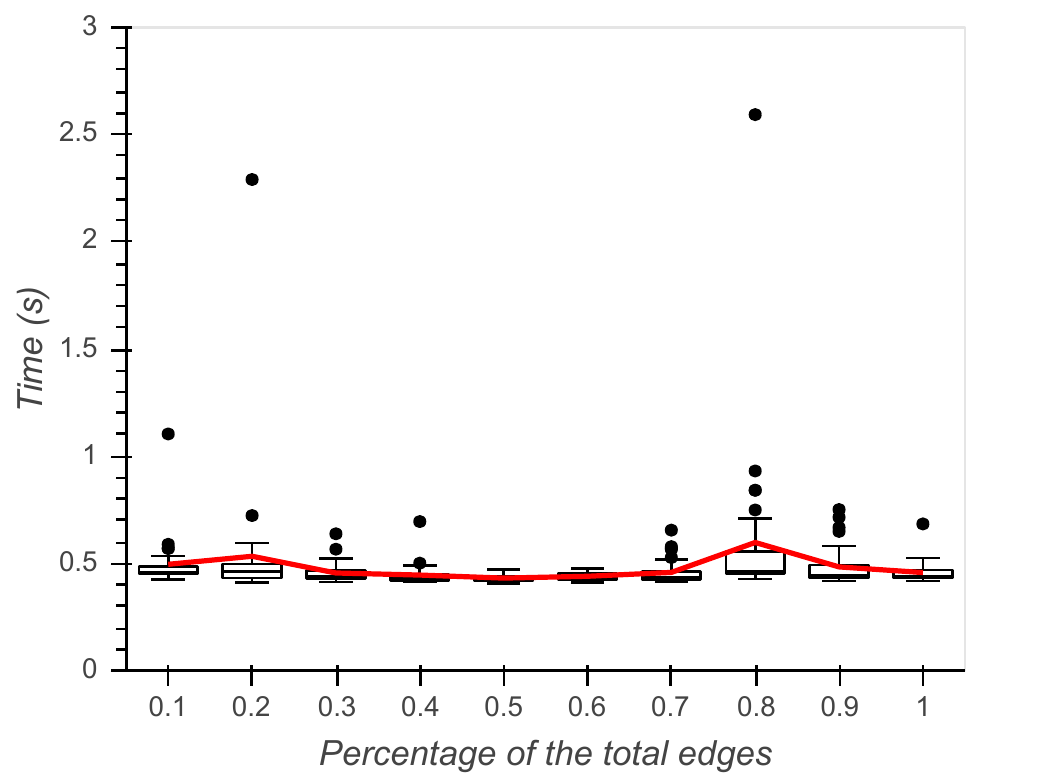}
	}
	\caption[Execution times for the Borgia Algorithm]
	{\textbf{Execution times for the Borgia Algorithm.} \textbf{a, b.} Execution time for different number of edges in the full Eurovision network \cite{BibEntry2019Jul} (2369 total edges). The red line shows the mean of 30 executions of the algorithm, alongside the quartile distribution, for different network sizes. These networks have been obtained by randomly sampling the desired percentage of edges from the original graph. While (a) shows the times for the static delta, (b) shows the times for the dynamic delta.}
	\label{fig:tiempos}
\end{figure}

\cfuture{In Fig. \ref{fig:dy_exec} we have studied the differences for four different datasets when using the static or the dynamic delta. There is no difference in two of them, while in the other two, only two fusions were different. The final result in all cases was not affected. However, changes in the execution time are consistently better (Fig. \ref{fig:tiempos}). It is important to note, however, that the simulation process in the Borgia Clustering can take longer execution time than other community detection algorithms.}


\section{The scalability problem} \label{sec:scale}

Social groups present lots of emergent properties when the number of people \cfuture{composing} them increases. A group of fifty people is not just ten times a group of five, because human interaction does not follow a linear behaviour. In the case of the Borgia Clustering algorithm, we have used what we call ``The Early Roman policy"\cite{fulford1992territorial} which aims at curbing the communities' tendency to grow  as their size increases. In this context, we have implemented this policy by adding a ``greedy expanse" penalization parameter, $p$, to correctly model the growth of social groups. This parameter penalizes ``greedy" nodes that join too quickly with others due to their high attraction force. By setting a high $p$, they become slower, avoiding other nodes to join with one another so fast.

However, there is more than one way to cope with this additional complexity. Apart from the greedy expanse penalization, we have also proposed an additional a set of policies to treat the nodes labelled as too big (formed by the fusion of many actors) to be treated in the same way as smaller ones, , which emerge as the result of different phenomena that still exist or have existed in human societies:

\begin{itemize}
	\item Naive or linear policy: no difference between big and small communities.
	\item French or authoritarian policy: based on the famous French absolutist monarchies \cite{cosandey2002absolutisme}. Only the affinities for the biggest actor in the node are taken into account.
	\item Early Roman policy: based on the end of the expansion of the Roman Empire \cite{fulford1992territorial}. Communities receive a penalization in their attraction power according to their size. This would be the case of the ``greedy expanse" penalization parameter used in the experimental section.
	\item Late Roman policy: based on the Diocletian tetrarchy \cite{rees2004diocletian}. When a community is too big, it is divided in half depending on its members' affinities. 
	\item Greek policy: based on the organization of the ancient Greek world \cite{hansen2006polis}. When two actors are about to form a big node, they are frozen instead and they cannot have any more attractive interactions between them.
	\item Aristocratic policy: based on the natural hierarchy formed in many societies  \cite{gledhill1995state}. When a community is too big, it is treated as a smaller community formed only by the top-k most important members of the community. 
\end{itemize}

\section{Results for real-life networks}
\label{sec:results}
In this section we present the results for the community detection in three different social networks: \textit{GoT}, Eurovision and the word association network for \textit{Heart of Darkness}. \cfuture{The cut for each dendrogram has been obtained by using the best configuration according to the formula in Eq. \ref{eq:cut_dendro}}.

\subsection{Game of Thrones}

\textit{Song of Ice and Fire} is a popular fantasy book in the saga written by George R.R. Martin which, to this day, is constituted by five books. This series of novels presents a numerous set of interesting characters, most of them members of one of the dynasties that rule the fantasy world of Westeros.

We used the data in \cite{Beveridge2018May}, that counts co-occurrences of each character in the original text, and then we created the adjacency matrix.

To compute the algorithm on \textit{Song of Ice Fire} network we used $\alpha=0.5$, $p=3$ and $c=0$.

\begin{figure}
	\centering
	\subfloat[][]{\includegraphics[width=0.7\linewidth]{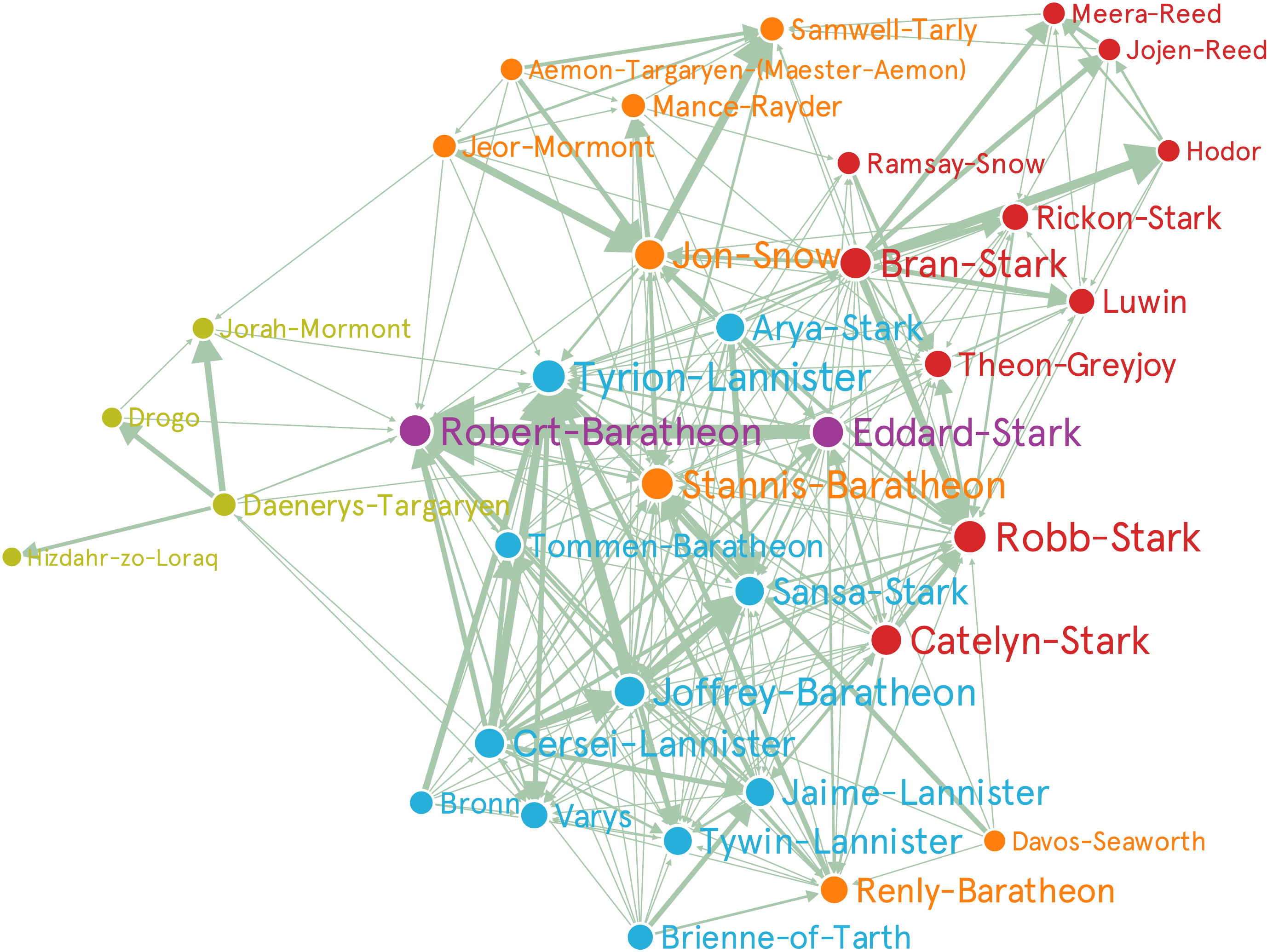}}\\
	\subfloat[][]{\includegraphics[width=0.7\linewidth]{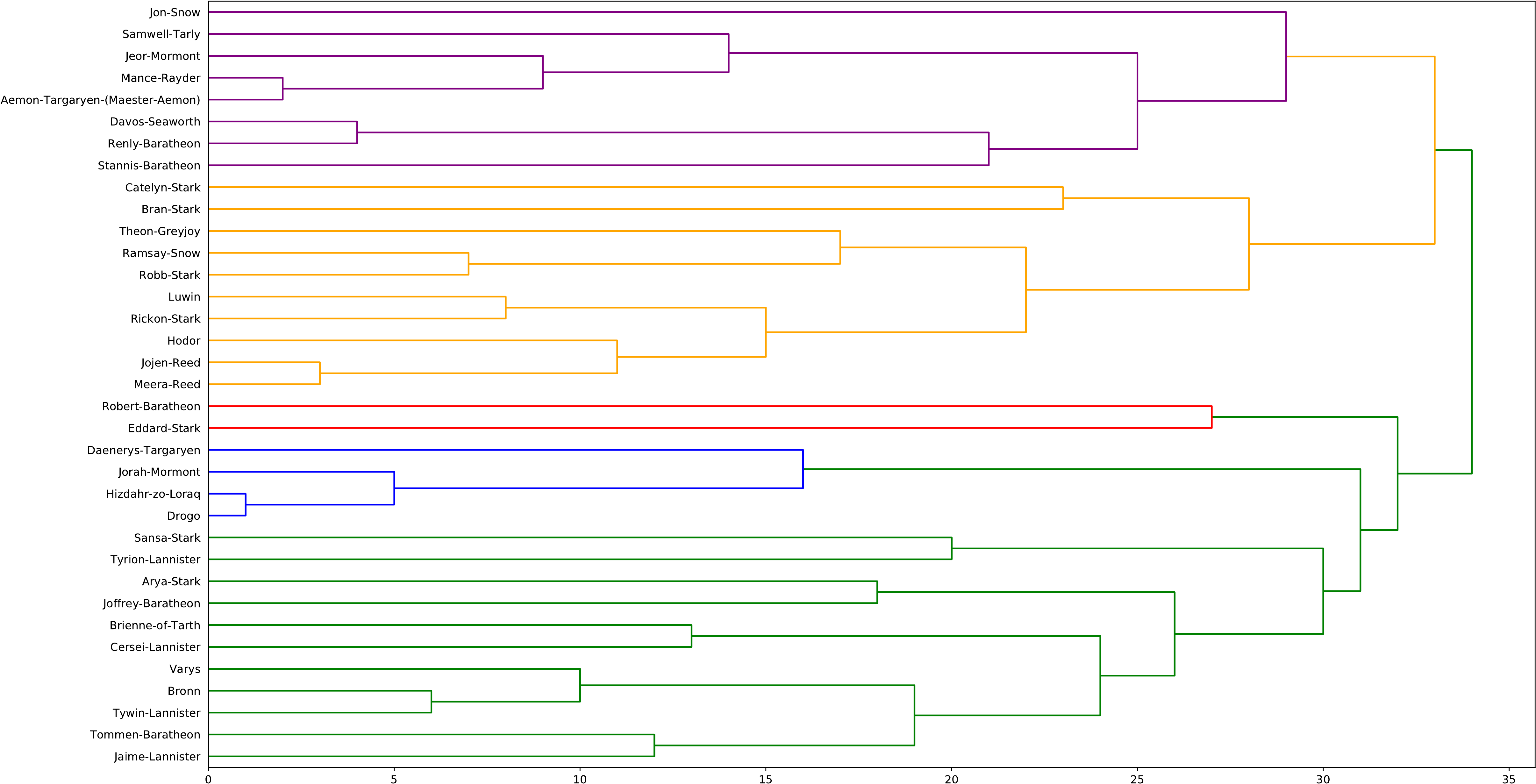}}
	\caption[Community detection in the \textit{Game of Thrones}.]
	{\textbf{Community detection in the \textit{GoT} network}. \textbf{a}. The network formed \cfuture{by} the top-35 most important characters in the book. Each of the five communities is marked with a different colour. \textbf{b}. The dendrogram formed during the execution of the algorithm. Each community is marked with same colours as in (a).}
	\label{fig:got_dendro}
\end{figure}

The result  in Figures \ref{fig:got_dendro} shows how the characters are grouped around the different sub-plots in the book. If we look at the dendrogram, we can see that the characters whose plot is more ``stable" (they do not change much from places or acquaintances) form or join communities faster than those who play an important role in different places or events in the books.

\subsection{Eurovision song contest}

\begin{figure}[ht]
	\centering
	\subfloat[width=0.48\linewidth][]{\includegraphics[width=0.48\linewidth]{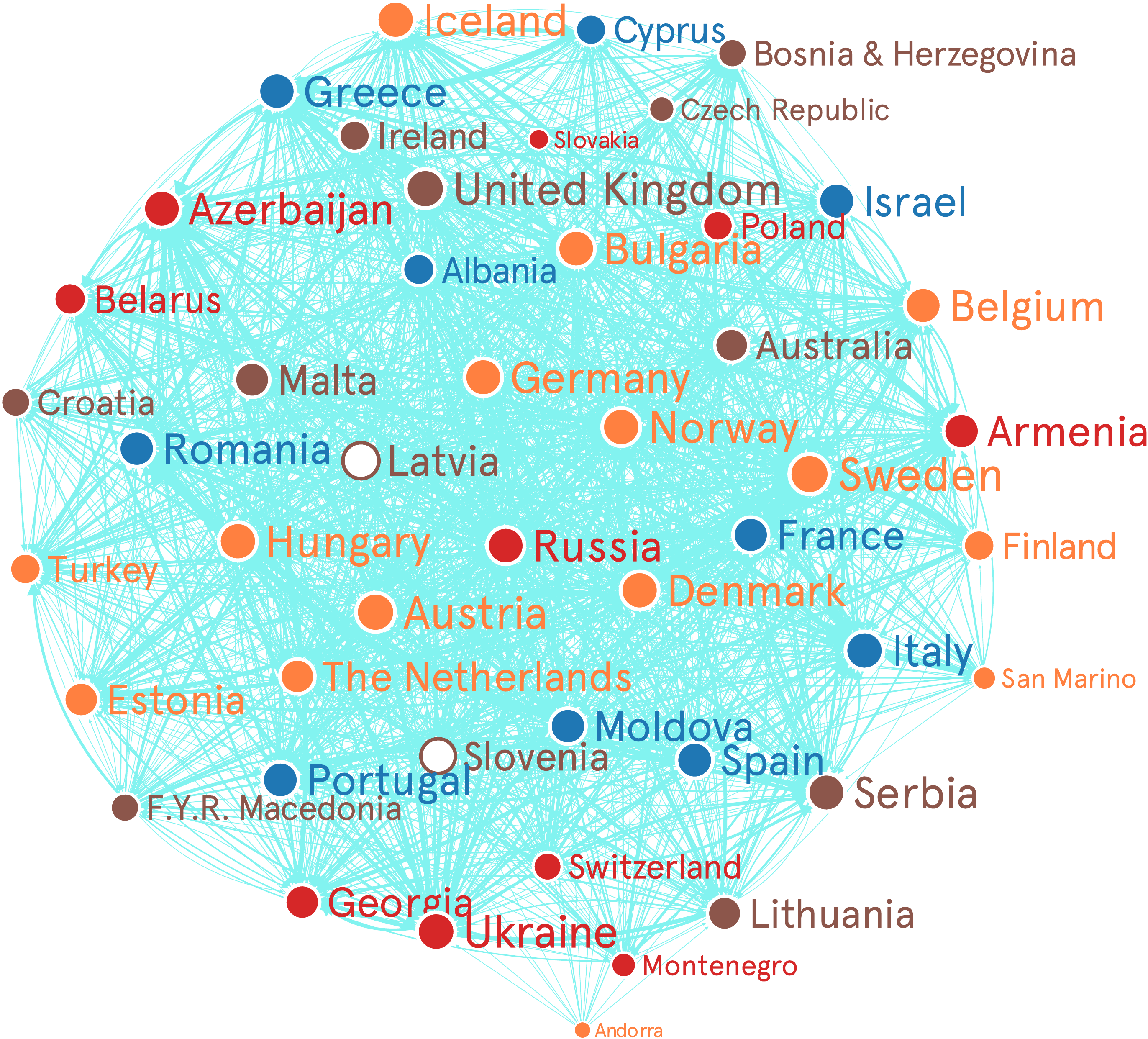}}
	\subfloat[width=0.48\linewidth][]{\includegraphics[width=0.48\linewidth]{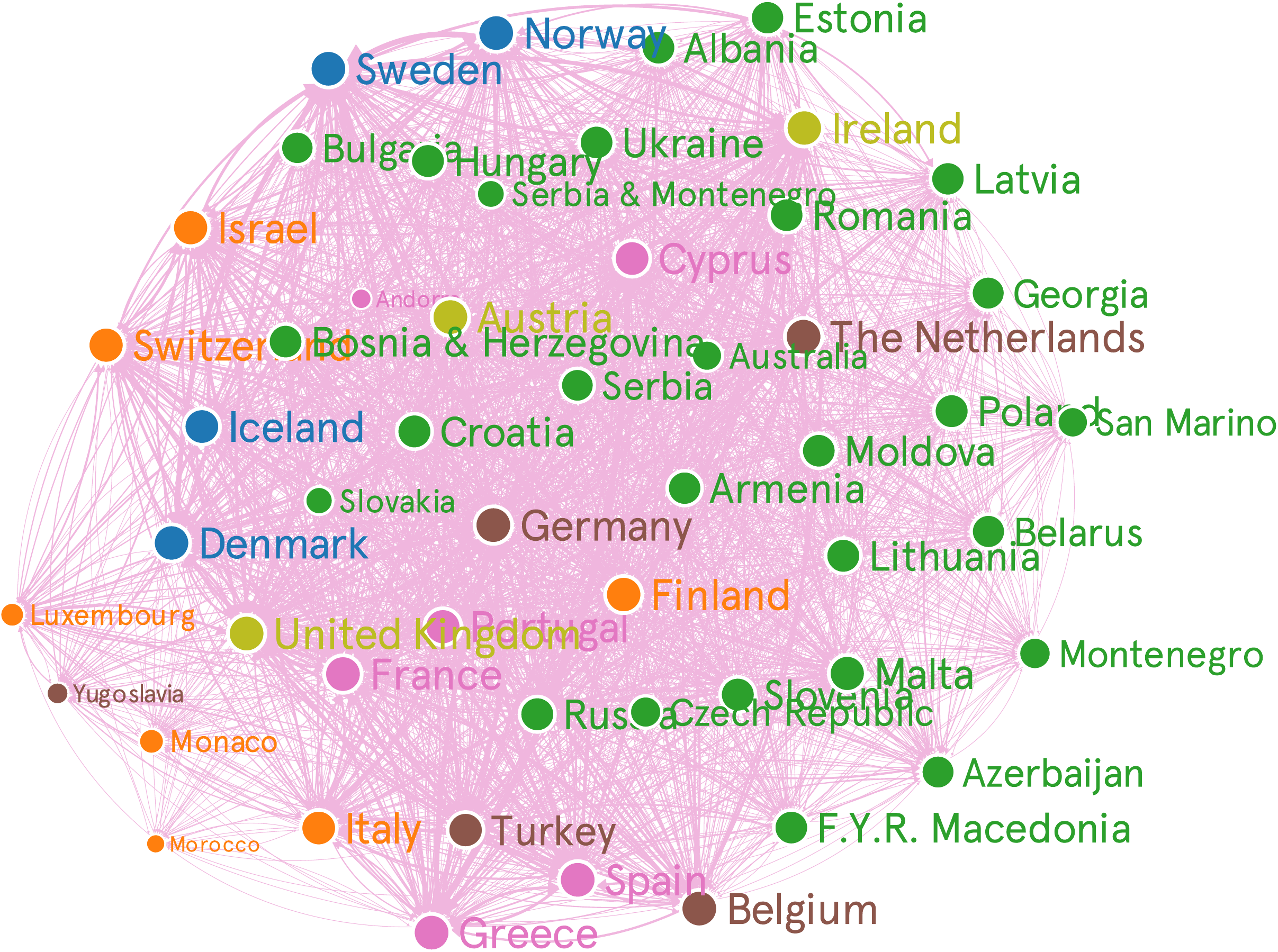}}
	\caption[Networks and community detection in Eurovision contest]
	{\textbf{Networks and community detection in Eurovision contest. } \textbf{a}. Last decade of voting record in Eurovision finals. \textbf{b} Whole voting record in Eurovision finals. }
	\label{fig:eurovision_nets}
\end{figure}

Eurovision is a musical contest where 42 countries compete against each other in order to get the highest score. The participants should judge and give points to the rest according to the quality of the musical performance. This is, of course, only the theory, since in practice it is noticeable how cultural and geographical proximities play an important role in the outcome of the contest. From the dataset in \cite{BibEntry2019Jul}, we have taken not only the last decade of voting, but also both the whole record of finals voting to study how countries historically behave when they are voting. 
These networks are interesting to us because they have very high density (Fig. \ref{fig:eurovision_nets}), as they are almost fully connected, and because we actually have semantic information about each actor. This means that we can interpret the resulting communities without relying too much on numerical measures.

In this case, we use the same parameter selection as in the previous case; $\alpha=1$, $p=3$ and $c=0$.

\begin{figure}[ht]
	\centering
	\subfloat[width=0.48\linewidth][]{\includegraphics[width=0.48\linewidth]{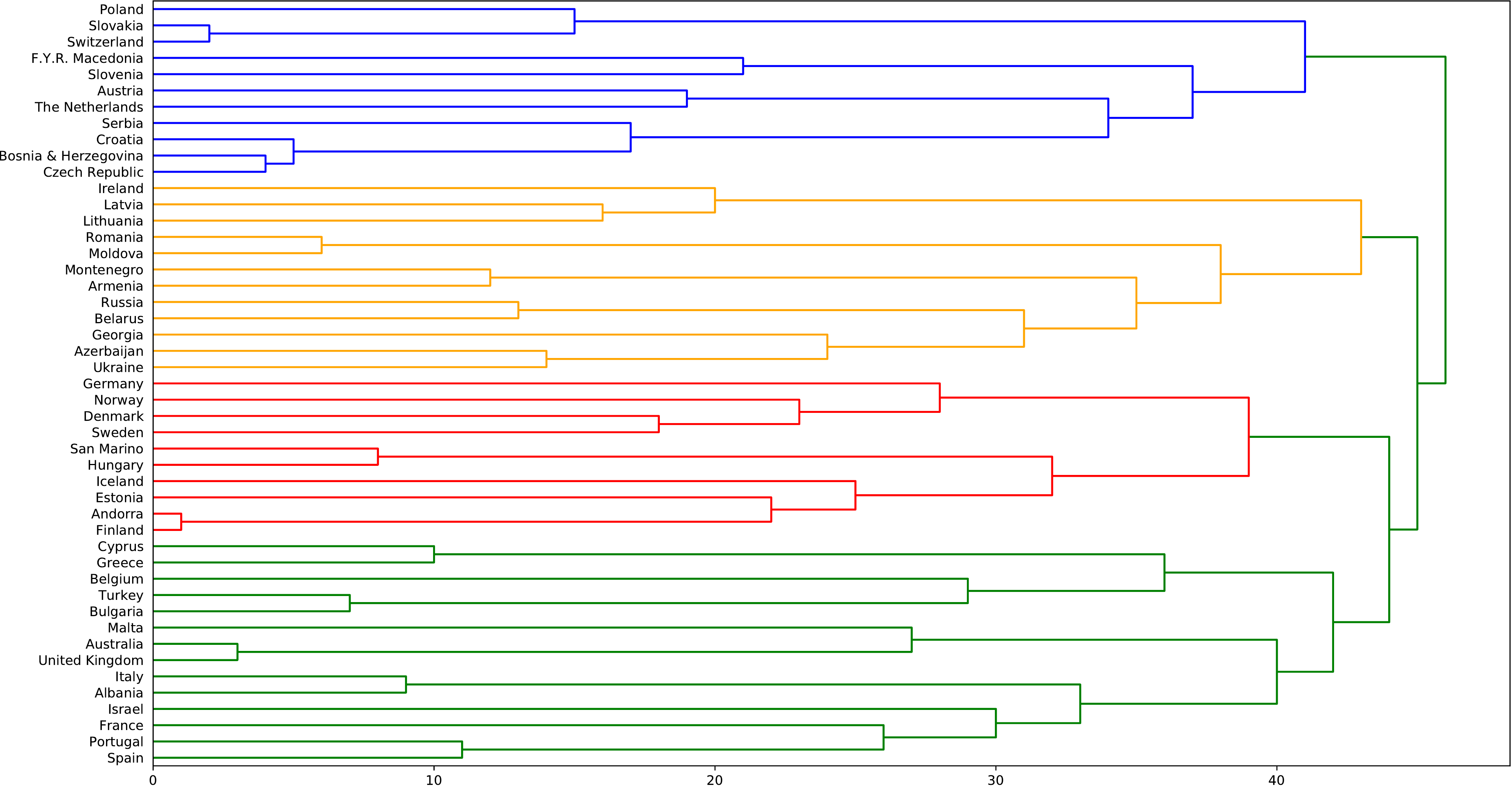}}
	\subfloat[width=0.48\linewidth][]{\includegraphics[width=0.48\linewidth]{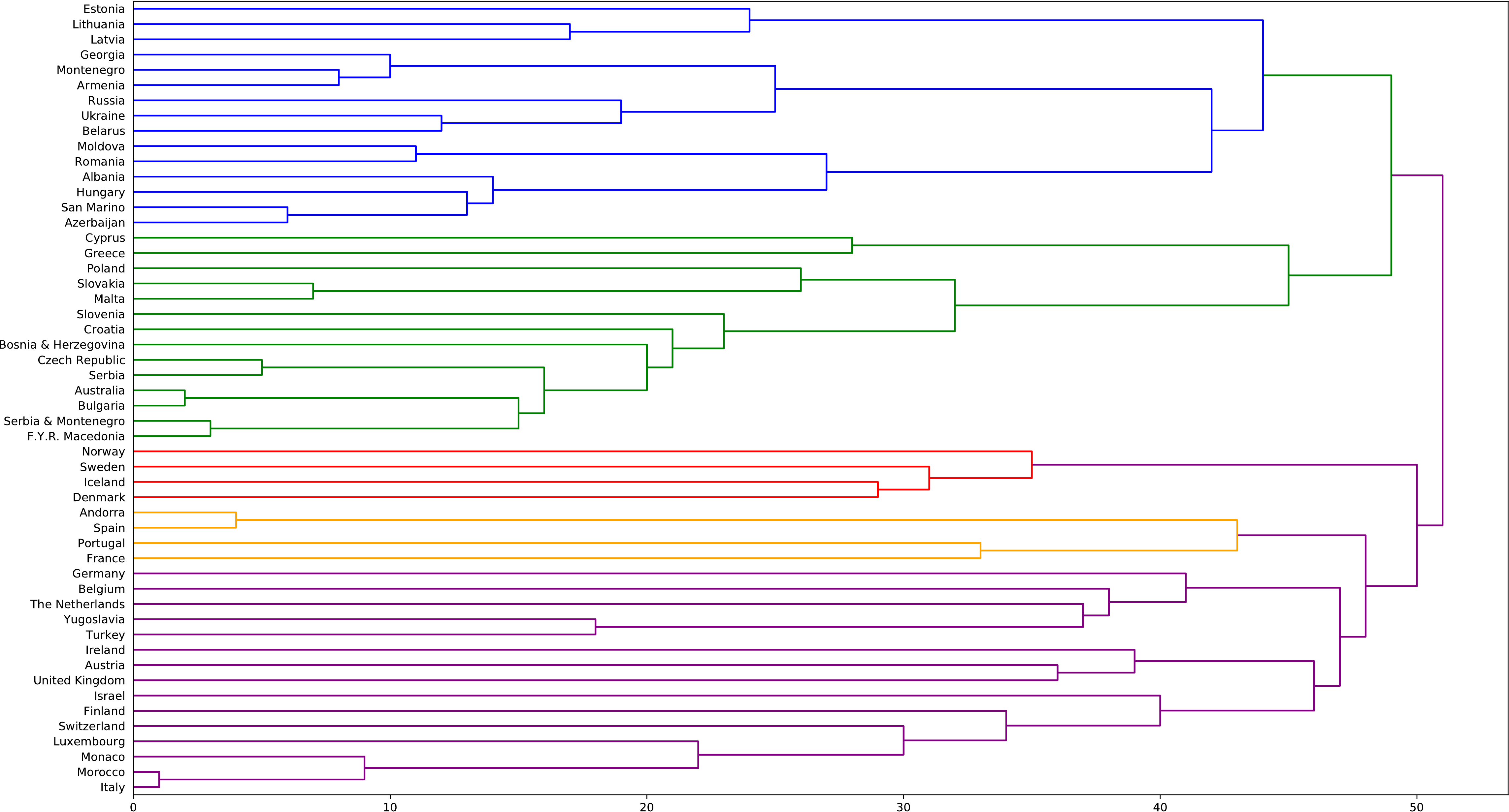}}
	\caption[Dendrograms for Eurovision in contest]
	{\textbf{Dendrogram and community detection for the last decade of voting record in Eurovision finals.} \textbf{a}. Dendrogram and community detection for the last decade of voting record in Eurovision finals. \textbf{b} Dendrogram community detection for the whole voting record in Eurovision finals. }
	\label{fig:eurovision_always}
\end{figure}

In the first case, the last decade of Eurovision (Fig. \ref{fig:eurovision_nets}a, Fig. \ref{fig:eurovision_always}a), we have obtained four different communities. At a local level, it seems that strong cultural and geographical \cfuture{components} influence the voting process, e.g Greece and Cyprus, Spain and Portugal. These components are more relevant in some communities than in others. There is a community consisting almost exclusively of countries of Slavic Europe, other has mainly Mediterranean countries, etc. However, in our time cultural bias is not so pronounced when compared to the whole historic record of voting. 

If we take into account the whole voting record since 1975, we obtain five communities (Fig. \ref{fig:eurovision_nets}b, Fig. \ref{fig:eurovision_always}b). Two of them have a clear cultural resemblance between their members: the Nordic countries, and the Spain-France-Portugal-Andorra axis. There \cfuture{are} also two Eastern-Europe groups and a Western one (with some exceptions in them that could be a consequence of a ``migration" effect \cite{spierdijk2006geography}).

\begin{table}
	\centering
	\scalebox{0.65}{
		\subfloat[]{
			\begin{tabular}{lrrrrr}
				\toprule
				{} &  Mod. &  M. density &       ARI &       NMI &  $C$ \\
				\midrule
				Greedy Mod.   &    0.3871 &            0.1868 &  0.5684 &  0.4540 &          3 \\
				Girvan-Newman       &    \textbf{0.4156} &            0.2258 &  0.4070 &  0.3529 &          4 \\
				Label prop.   &    0.3956 &            0.2139 &  0.6841 &  0.5750 &          3 \\
				L. eigenvector &    0.4012 &            0.1969 &  0.4351 &  0.4172 &          4 \\
				Lovaine             &    0.4138 &            \textbf{0.2296} &  0.4292 &  0.3592 &          4 \\
				Borgia C.   &    0.3693 &            0.1797 &  \textbf{0.8822} &  \textbf{0.8324} &          2 \\
				\cfuture{Grav. Clus.}   & 0.0342   &    0.0555       &  0.0325 & 0.0757  &    3    \\
				\bottomrule
			\end{tabular}
		}
		
		\subfloat[]{
			\begin{tabular}{rrrrr}
				\toprule
				Mod. &  M. density &       ARI &       NMI &  $C$ \\
				\midrule
				0.4954 &            0.1659 &  0.4658 &  0.4149 &          4 \\
				\textbf{0.5193} &            0.2011 &  0.4505 &  0.4404 &          5 \\
				0.4684 &            0.1829 &  0.492&  0.4418 &          4 \\
				0.4911 &            0.1809 &  0.3211 &  0.3439 &          5 \\
				0.5175 &            \textbf{0.2032} &  0.3140 &  0.3397 &          5 \\
				0.3787 &            0.1362 &  \textbf{1.0000} &  \textbf{1.0000} &          2 \\
				-0.0010   &    0.0422       & 0.0091  & 0.0115  &     4   \\
				\bottomrule
			\end{tabular}
		}
	}
	\\
	\scalebox{0.65}{
		\subfloat[]{
			\begin{tabular}{lrrrrr}
				\toprule
				{} &  Mod. &  M. density &       ARI &       NMI &  $C$ \\
				\midrule
				Greedy Mod.   &    0.5564 &            0.2744 &  0.4844 &  0.5584 &          6\\
				Girvan-Newman       &    0.5996 &            0.4321 &  0.7781 &  0.8014 &         10\\
				Label prop.   &    0.5974 &            0.4027 &  0.7445 &  0.7701 &          9\\
				L. eigenvector &    0.4926 &            0.2593 &  0.4640 &  0.5611 &          8\\
				Lovaine             &    \textbf{0.6044} &            0.4167 &  0.7070 &  0.7551 &          9\\
				Borgia C.   &    0.6005 &            \textbf{0.4909} &  \textbf{0.8966} &  \textbf{0.8978} &         12\\
				\cfuture{Grav. Clus.}   &  0.2319  &    0.0848        & 0.1202  &  0.2899 &   7     \\
				
				\bottomrule
			\end{tabular}
		}
		
		\subfloat[]{
			\begin{tabular}{rrrrr}
				\toprule
				Mod. &  M. density &       ARI &       NMI &  $C$ \\
				\midrule
				0.5019 &            0.1739 &  0.6378 &  0.4929 &          4 \\
				0.5168 &            0.1989 &  \textbf{0.6823} &  0.4875 &          5 \\
				0.5106 &            0.1874 &  0.6701 &  0.5088 &          4 \\
				0.4671 &            0.1439 &  0.5466 &  0.4428 &          4 \\
				\textbf{0.5267} &            \textbf{0.1991} &  0.6463 &  0.4576 &          5 \\
				0.4994 &            0.1665 &  0.6685 &  \textbf{0.5649} &          3 \\
				0.0053 &    0.0646        & 0.0234  & 0.0203  &   2     \\
				\bottomrule
			\end{tabular}
		}
	}
	\caption[Results for the Borgia Clustering.]
	{\textbf{Results for the Borgia Clustering.} 
		\textbf{a.} Zachary Karate club network. Parameters: $\alpha=0.7, p=3, G(x, y)=1$.
		\textbf{b.} Dolphin network. Parameters: $\alpha=0.7, p=3, G(x, y)=1$.
		\textbf{c.} Football network. Parameters: $\alpha=1, p=0, G(x, y)=1$.
		\textbf{d.}  Polbooks club network. Parameters: $\alpha=1.0, p=0, G(x, y)=1$.
	}
	\label{tab:pool_boks}
\end{table}

\subsection{Heart of Darkness}

\emph{Heart of Darkness} is the famous novel written by Joseph Conrad in 1899. The story narrates the abuse committed by the Belgium King during the late 19th century. Besides its humanistic interest, this novel presents a mysterious character, Kurtz, which is particularly interesting for our research. This omnipresent individual, seen as a demigod by the natives who constantly refer to him, only makes an appearance in the final stages of the book. 
\cfuture{We have computed the word association network in the same way as that in section \ref{sec:affinities}. Each node corresponds to a word in the original text and each edge is the corresponding affinity between two entities, using as $C$ the original number of co-occurrences for each pair of words in the same paragraph.}

\begin{figure}[ht]
	\centering
	\subfloat[][]{\includegraphics[width=0.5\textwidth]{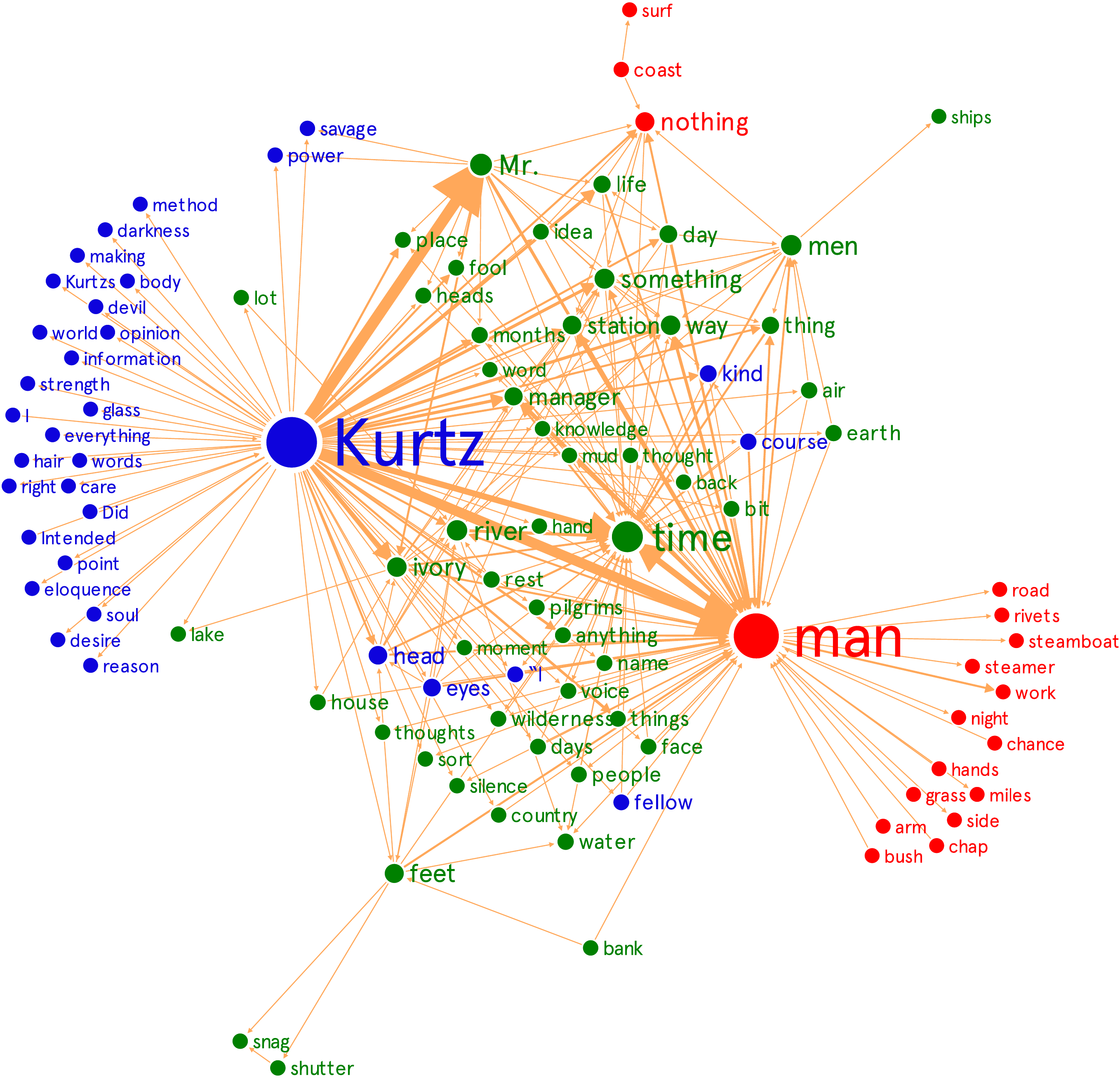}}\\
	\subfloat[][]{\includegraphics[width=0.5\textwidth]{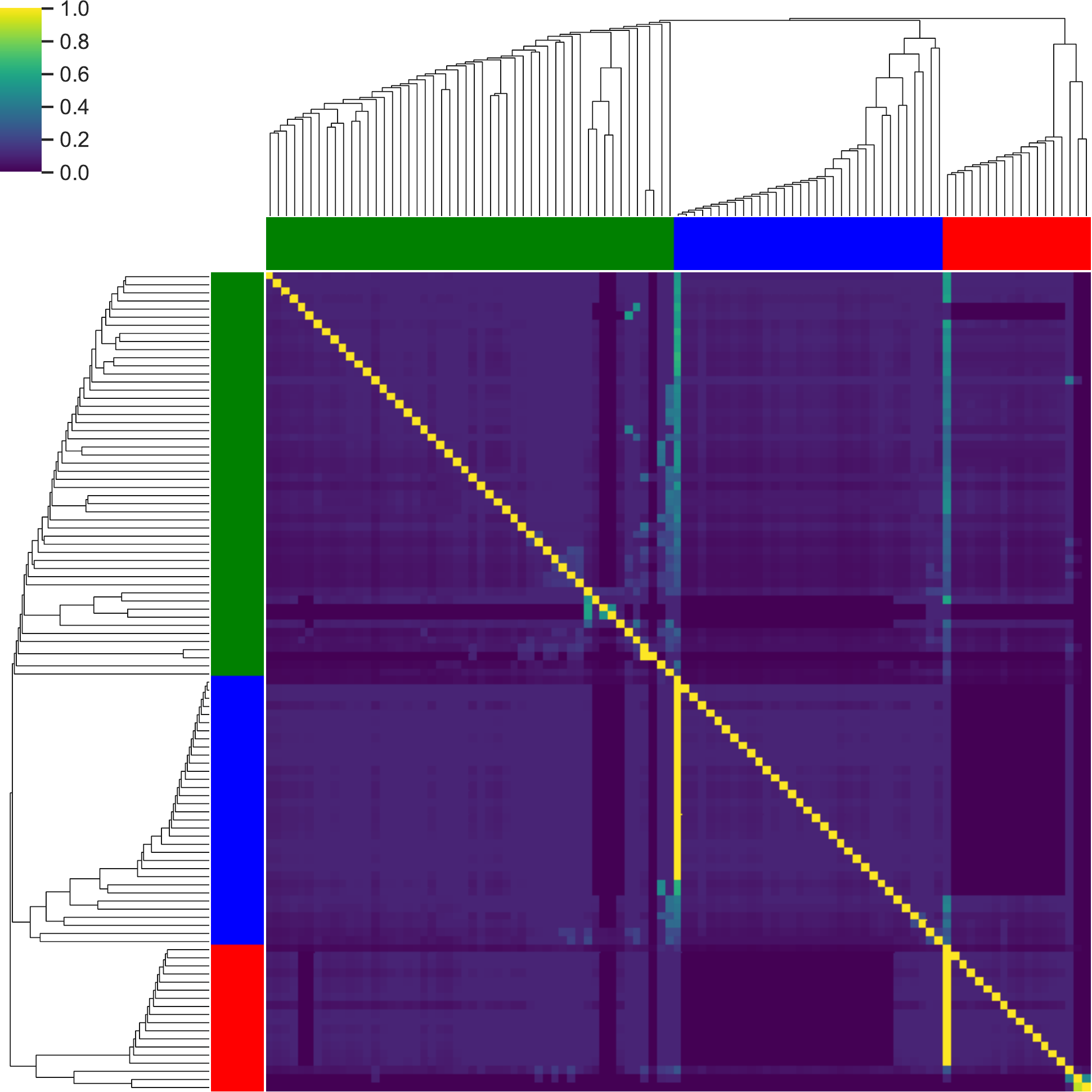}}
	\caption[Community detection in the \textit{Heart of Darkness}]
	{\textbf{Community detection in \textit{Heart of Darkness}.} \textbf{a}. The network resulting from the top-200 most repeated names in the book. Each one of the five communities is marked with a different colour. \textbf{b}. $A_C$ heatmap and the dendrogram formed during the execution of the algorithm.}
	\label{fig:dendrogramaheart}
\end{figure}

We can see the result using Borgia Clustering community detection in Figure \ref{fig:dendrogramaheart}. We obtain three different communities. The most notable one is the ``Kurtz" community. It is characterized by the number of words that induce a moral bias in the reader about this character, such as: ``Darkness'', ``Devil'', or ``Power''. There are also some other important concepts linked to this character: ``Knowledge", ``Desire'', ``Reason'', ``Eloquence'', etc. All of these show the fascination that his figure inspires in the natives that the first-person narrator and protagonist encounters through the journey. The ``man" community and the ``time" are both semantically more heterogeneous, although the latter one  is mainly composed of terms that describe the place in which the action takes place. 

\section{Comparison \cfuture{with} other community detection algorithms}
\label{sec:comparison}
Finally, to benchmark our community detection algorithm, we have chosen datasets with ground truth labels: the famous Zachary's karate club social network \cite{zachary1977information}, politics books \cite{adamic2005political}, that contains the number of co-purchases of different books about US politics, football network \cite{Girvan2002Jun}, that represents the number of matches between each pair of teams, and Dolphins \cite{lusseau2003bottlenose} \cfuture{which} is a network that registers the frequency each pair of dolphins played together. 

To test the quality of our solution and to study the best parameter selection in the Borgia Clustering algorithm, we have compared it \cfuture{to} other community detection algorithms: \cfuture{Girvan-Newman \cite{girvan2002community}, Newman greedy modularity optimization \cite{newman2004fast}, the Lovaine algorithm \cite{Blondel_2008}, using the eigenvalues of matrices to detect communities \cite{newman2006finding}, and label propagation \cite{raghavan2007near}. The reported results for the label propagation method is the median of five executions due to its stochastic nature.} We have used the Normalized Mutual Information (NMI) \cite{strehl2002cluster} and Random Adjusted Index (ARI) \cite{rand1971objective} to compare the results against ground truth labels.  We have also compared Modularity \cite{newman2006modularity} and Modularity Density \cite{li2008quantitative} for each solution. 

\begin{figure}[ht]
	\includegraphics[width=1\linewidth]{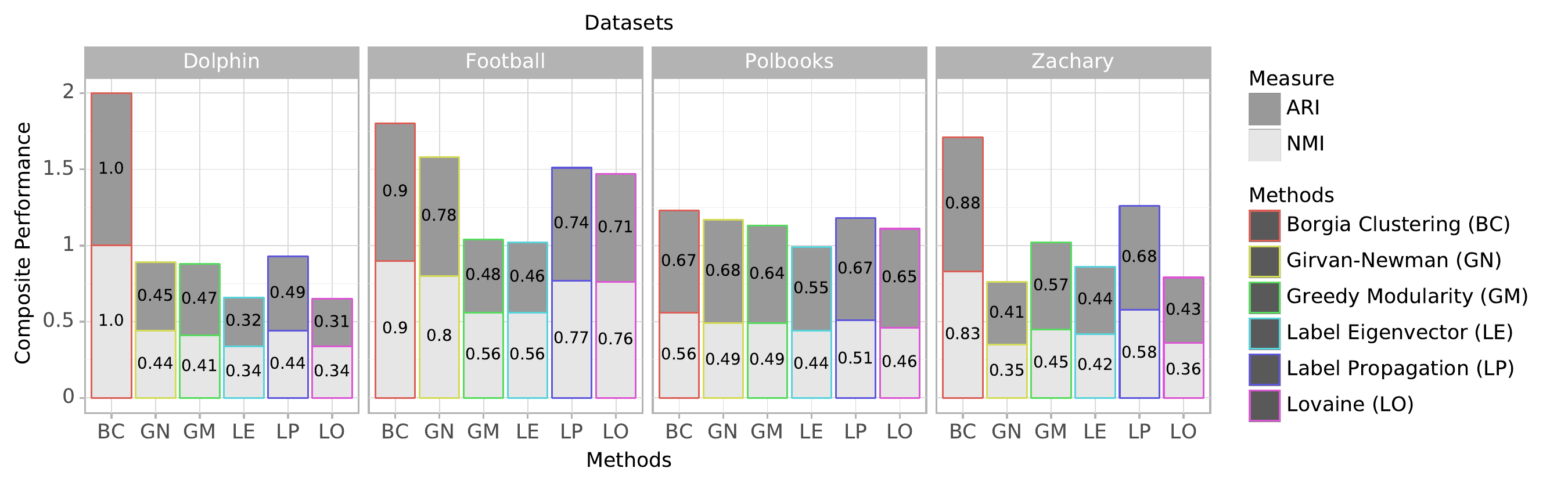}
	\caption[Comparison of different Community Detection Algorithms]
	{\textbf{Comparison of different Community Detection Algorithms.} We have compared the Borgia Clustering algorithm against three modularity optimization methods: Girvan-Newman, Newman greedy modularity optimization, and the Lovaine algorithm. We have also compared it against using eigenvalues of matrices to detect communities and label propagation.}
	\label{fig:comparison}
\end{figure}

\cfuture{The} results are shown in Figure \ref{fig:comparison} using the Borgia Clustering algorithm in the ARI and NMI measures. A  comparison including modularity and modularity density can be found in Table \ref{tab:pool_boks}. Borgia Clustering does not use at all the concept of modularity, so it was expected to perform worse \cfuture{with respect to this index} compared to the methods that actually optimize this metric: Greedy Modularity, Girvan-Newman and Lovaine.

\cfuture{In the same table, we also show the results for the traditional gravitational algorithm that was developed originally for clustering problems \cite{Wright}. As expected, this algorithm performs poorly compared to the others, as it was not \cfuture{specifically designed} for this problem. Using the affinity matrices instead of the adjacency matrices showed no improvement for this algorithm either.}

\section{Conclusions and future work} \label{sec:conclusions}

In this work we have proposed a new set of functions based on human nature to represent actors and we have shown the effects that different affinities have on the same real-work networks. We have discussed the importance of scale in networks, and we have applied our ideas to develop a new human-based community detection algorithm, the Borgia Clustering, based on the centre-Italian wars of the 15th century. Using this algorithm, we have obtained good results compared to the most used community detection algorithms up-to-date. \cfuture{We think that this algorithm can offer significant improvements in very dense networks, when modularity is not a reliable measure to optimize. Also, this algorithm always converges to the same value, so there is no need to run it multiple times to obtain a valid result. Furthermore, due to the intuitive principles that the algorithm stands on, we think that results obtained with it can be easily interpretable.}

 \cfuture{Future research will aim at applying Borgia Clustering in large scale networks, studying the constraints and requirements to perform the community detection process efficiently, in terms of memory and execution time. We also intend to explore the effects of affinity functions in dynamic environments, where networks change along time \cite{dynamic2010}, and exploit them with the Borgia Clustering algorithm. }


\section{Acknowledgements}

Javier Fumanal Idocin's and Humberto Bustince's research has been supported by the project TIN2016-77356-P (AEI/FEDER,UE).

Oscar Cordón's research was supported by the Spanish Spanish Ministry of Science, Innovation and Universities under grant EXASOCO (PGC2018-101216-B-I00), including , European Regional Development Funds (ERDF).

Amparo Alonso-Betanzos' research has been financially supported in part by the Spanish Ministerio de Econom\'ia y Competitividad (research project TIN2015-65069-C2-1-R), by European Union FEDER funds and by the Conseller\'ia de Industria of the Xunta de Galicia (research project GRC2014 /035).

M. Minárová's research has been funded by the project work was supported by the project APVV-17-0066.



\clearpage
\bibliographystyle{elsarticle-num}
\bibliography{socialbib}

\begin{thebibliography}{10}
\expandafter\ifx\csname url\endcsname\relax
  \def\url#1{\texttt{#1}}\fi
\expandafter\ifx\csname urlprefix\endcsname\relax\def\urlprefix{URL }\fi
\expandafter\ifx\csname href\endcsname\relax
  \def\href#1#2{#2} \def\path#1{#1}\fi

\bibitem{scott1988social}
J.~Scott, Social network analysis, Sociology 22~(1) (1988) 109--127.

\bibitem{wasserman1994social}
S.~Wasserman, K.~Faust, et~al., Social network analysis: Methods and
  applications, Vol.~8, Cambridge university press, 1994.

\bibitem{borgatti2009network}
S.~P. Borgatti, A.~Mehra, D.~J. Brass, G.~Labianca, Network analysis in the
  social sciences, Science 323~(5916) (2009) 892--895.

\bibitem{horvath2011weighted}
S.~Horvath, Weighted network analysis: applications in genomics and systems
  biology, Springer Science \& Business Media, 2011.

\bibitem{BENITEZANDRADES2020154}
J.~A. Benítez-Andrades, I.~García-Rodríguez, C.~Benavides,
  H.~Alaiz-Moretón, A.~Rodríguez-González, Social network analysis for
  personalized characterization and risk assessment of alcohol use disorders in
  adolescents using semantic technologies, Future Generation Computer Systems
  106 (2020) 154 -- 170.

\bibitem{fischer1982dwell}
C.~S. Fischer, To dwell among friends: Personal networks in town and city,
  University of Chicago Press, 1982.

\bibitem{wellman1988social}
B.~Wellman, S.~D. Berkowitz, Social structures: A network approach, Vol.~2, CUP
  Archive, 1988.

\bibitem{stopczynski2014measuring}
A.~Stopczynski, V.~Sekara, P.~Sapiezynski, A.~Cuttone, M.~M. Madsen, J.~E.
  Larsen, S.~Lehmann, Measuring large-scale social networks with high
  resolution, PloS one 9~(4) (2014).

\bibitem{sanchezmar2015}
N.~S{\'a}nchez-Maro{\~{n}}o, A.~Alonso-Betanzos, O.~Fontenla-Romero,
  C.~Brinquis-N{\'u}{\~{n}}ez, J.~G. Polhill, T.~Craig, A.~Dumitru,
  R.~Garc{\'i}a-Mira, An agent-based model for simulating environmental
  behavior in an educational organization, Neural Processing Letters 42~(1)
  (2015) 89--118.

\bibitem{2019cascades}
C.~Gunaratne, C.~Senevirathna, C.~Jayalath, N.~Baral, W.~Rand, I.~Garibay, A
  multi-action cascade model of conversation, in: 5th International Conference
  on Computational Social Science, 2019.

\bibitem{lobel2015information}
I.~Lobel, E.~Sadler, Information diffusion in networks through social learning,
  Theoretical Economics 10~(3) (2015) 807--851.

\bibitem{McAuley2014DiscoveringSC}
J.~McAuley, J.~Leskovec, Discovering social circles in ego networks, ACM
  Transactions on Knowledge Discovery from Data 8~(1) (2014) 4.

\bibitem{SLESS2018217}
L.~Sless, N.~Hazon, S.~Kraus, M.~Wooldridge, Forming k coalitions and
  facilitating relationships in social networks, Artificial Intelligence 259
  (2018) 217 -- 245.

\bibitem{YU2018312}
W.~Yu, S.~Li, Recommender systems based on multiple social networks
  correlation, Future Generation Computer Systems 87 (2018) 312 -- 327.

\bibitem{LUO20191023}
X.~Luo, C.~Jiang, W.~Wang, Y.~Xu, J.-H. Wang, W.~Zhao, User behavior prediction
  in social networks using weighted extreme learning machine with distribution
  optimization, Future Generation Computer Systems 93 (2019) 1023 -- 1035.

\bibitem{DELGADO2002171}
J.~Delgado, Emergence of social conventions in complex networks, Artificial
  Intelligence 141~(1) (2002) 171 -- 185.

\bibitem{Girvan2002Jun}
M.~Girvan, M.~E.~J. Newman, {Community structure in social and biological
  networks}, Proceedings of the National Academy of Sciences 99~(12) (2002)
  7821--7826.

\bibitem{zhou2003distance}
H.~Zhou, Distance, dissimilarity index, and network community structure,
  Physical review E 67~(6) (2003) 061901.

\bibitem{palla2005uncovering}
G.~Palla, I.~Der{\'e}nyi, I.~Farkas, T.~Vicsek, Uncovering the overlapping
  community structure of complex networks in nature and society, Nature
  435~(7043) (2005) 814.

\bibitem{girvan2002community}
M.~Girvan, M.~E. Newman, Community structure in social and biological networks,
  Proceedings of the National Academy of Sciences 99~(12) (2002) 7821--7826.

\bibitem{pizzuti2008ga}
C.~Pizzuti, Ga-net: A genetic algorithm for community detection in social
  networks, in: International conference on parallel problem solving from
  nature, Springer, 2008, pp. 1081--1090.

\bibitem{MA2020533}
T.~Ma, Q.~Liu, J.~Cao, Y.~Tian, A.~Al-Dhelaan, M.~Al-Rodhaan, Lgiem: Global and
  local node influence based community detection, Future Generation Computer
  Systems 105 (2020) 533 -- 546.

\bibitem{newman2006modularity}
M.~E. Newman, Modularity and community structure in networks, Proceedings of
  the national academy of sciences 103~(23) (2006) 8577--8582.

\bibitem{newman2004fast}
M.~E. Newman, Fast algorithm for detecting community structure in networks,
  Physical review E 69~(6) (2004) 066133.

\bibitem{Blondel_2008}
V.~D. Blondel, J.-L. Guillaume, R.~Lambiotte, E.~Lefebvre, Fast unfolding of
  communities in large networks, Journal of Statistical Mechanics: Theory and
  Experiment 2008~(10) (oct 2008).

\bibitem{danon2005comparing}
L.~Danon, A.~Diaz-Guilera, J.~Duch, A.~Arenas, Comparing community structure
  identification, Journal of Statistical Mechanics: Theory and Experiment
  2005~(09) (2005) P09008.

\bibitem{newman2004finding}
M.~E. Newman, M.~Girvan, Finding and evaluating community structure in
  networks, Physical review E 69~(2) (2004) 026113.

\bibitem{Wright}
W.~Wright, Gravitational clustering, Pattern Recognition 9~(3) (1977) 151 --
  166.

\bibitem{beliakov2016practical}
G.~Beliakov, H.~B. Sola, T.~C. S{\'a}nchez, A practical guide to averaging
  functions, Vol. 329, Springer, 2016.

\bibitem{Milgram1967TheSW}
S.~Milgram, The small world problem, Psychology today 2~(1) (1967) 60--67.

\bibitem{friedkin1991theoretical}
N.~E. Friedkin, Theoretical foundations for centrality measures, American
  Journal of Sociology 96~(6) (1991) 1478--1504.

\bibitem{hu2005efficient}
M.~Jacomy, T.~Venturini, S.~Heymann, M.~Bastian, Forceatlas2, a continuous
  graph layout algorithm for handy network visualization designed for the gephi
  software, PLOS ONE 9~(6) (2014) 1--12.

\bibitem{italy_before}
Kayac,
  \href{https://commons.wikimedia.org/wiki/File:Grandi_Casate_Italiane_nel_1499.png}{Italian
  lordships in 1499}, [Online; accessed 20. Mar. 2019] (2013).
\newline\urlprefix\url{https://commons.wikimedia.org/wiki/File:Grandi_Casate_Italiane_nel_1499.png}

\bibitem{catalan2008principe}
J.~Catal{\`a}n~Deus, El pr{\'\i}ncipe del renacimiento, Debate 26 (2008) 90.

\bibitem{merton1968matthew}
R.~K. Merton, The {M}atthew effect in science: The reward and communication
  systems of science are considered, Science 159~(3810) (1968) 56--63.

\bibitem{GUPTA1991431}
M.~Gupta, J.~Qi, Theory of t-norms and fuzzy inference methods, Fuzzy Sets and
  Systems 40~(3) (1991) 431 -- 450.

\bibitem{AGOP19}
J.~Armentia, I.~Rodr{\'i}­guez, J.~Fumanal~Idocin, H.~Bustince,
  M.~Min{\'a}rov{\'a}, A.~Jurio, Gravitational clustering algorithm
  generalization by using an aggregation of masses in newton law, in:
  R.~Hala{\v{s}}, M.~Gagolewski, R.~Mesiar (Eds.), New Trends in Aggregation
  Theory, Springer International Publishing, Cham, 2019, pp. 172--182.

\bibitem{zachary1977information}
W.~W. Zachary, An information flow model for conflict and fission in small
  groups, Journal of Anthropological Research 33~(4) (1977) 452--473.

\bibitem{mathbeveridge2017Aug}
mathbeveridge, \href{https://github.com/mathbeveridge/asoiaf}{{asoiaf}},
  [Online; accessed 26. Sep. 2019] (Aug 2017).
\newline\urlprefix\url{https://github.com/mathbeveridge/asoiaf}

\bibitem{BibEntry2019Jul}
\href{https://www.kaggle.com/datagraver/eurovision-song-contest-scores-19752018}{{Eurovision
  Song Contest scores 1975-2018}}, [Online; accessed 2. Jul. 2019] (Jul 2019).
\newline\urlprefix\url{https://www.kaggle.com/datagraver/eurovision-song-contest-scores-19752018}

\bibitem{li2008quantitative}
Z.~Li, S.~Zhang, R.-S. Wang, X.-S. Zhang, L.~Chen, Quantitative function for
  community detection, Physical review E 77~(3) (2008) 036109.

\bibitem{fulford1992territorial}
M.~Fulford, Territorial expansion and the roman empire, World Archaeology
  23~(3) (1992) 294--305.

\bibitem{cosandey2002absolutisme}
F.~Cosandey, R.~Descimon, L'absolutisme en France: histoire et historiographie,
  Editions du Seuil, 2002, (In French).

\bibitem{rees2004diocletian}
R.~Rees, Diocletian and the Tetrarchy, Edinburgh University Press Edinburgh,
  2004.

\bibitem{hansen2006polis}
M.~H. Hansen, Polis: an introduction to the ancient Greek city-state, Oxford
  University Press, 2006.

\bibitem{gledhill1995state}
J.~Gledhill, B.~Bender, M.~T. Larsen, State and society: the emergence and
  development of social hierarchy and political centralization, Vol.~4,
  Psychology Press, 1995.

\bibitem{Beveridge2018May}
A.~Beveridge, M.~Chemers, The game of game of thrones: Networked concordances
  and fractal dramaturgy, in: Reading Contemporary Serial Television Universes,
  Routledge, 2018, pp. 201--225.

\bibitem{spierdijk2006geography}
L.~Spierdijk, M.~Vellekoop, Geography, culture, and religion: Explaining the
  bias in eurovision song contest voting, Enschede: Department of Applied
  Mathematics, University of Twente (2006) 33.

\bibitem{adamic2005political}
L.~A. Adamic, N.~Glance, The political blogosphere and the 2004 us election:
  divided they blog, in: Proceedings of the 3rd international workshop on Link
  discovery, ACM, 2005, pp. 36--43.

\bibitem{lusseau2003bottlenose}
D.~Lusseau, K.~Schneider, O.~J. Boisseau, P.~Haase, E.~Slooten, S.~M. Dawson,
  The bottlenose dolphin community of doubtful sound features a large
  proportion of long-lasting associations, Behavioral Ecology and Sociobiology
  54~(4) (2003) 396--405.

\bibitem{newman2006finding}
M.~E. Newman, Finding community structure in networks using the eigenvectors of
  matrices, Physical review E 74~(3) (2006) 036104.

\bibitem{raghavan2007near}
U.~N. Raghavan, R.~Albert, S.~Kumara, Near linear time algorithm to detect
  community structures in large-scale networks, Physical review E 76~(3) (2007)
  036106.

\bibitem{strehl2002cluster}
A.~Strehl, J.~Ghosh, Cluster ensembles---a knowledge reuse framework for
  combining multiple partitions, Journal of Machine Learning Research 3~(Dec)
  (2002) 583--617.

\bibitem{rand1971objective}
W.~M. Rand, Objective criteria for the evaluation of clustering methods,
  Journal of the American Statistical Association 66~(336) (1971) 846--850.

\bibitem{dynamic2010}
D.~{Greene}, D.~{Doyle}, P.~{Cunningham}, Tracking the evolution of communities
  in dynamic social networks, in: 2010 International Conference on Advances in
  Social Networks Analysis and Mining, 2010, pp. 176--183.

\end{thebibliography}

\end{document}